\documentclass[pre,notitlepage,twocolumn,superscriptaddress]{revtex4-1}
\usepackage{graphicx}

\usepackage{eucal}
\usepackage[dvips]{epsfig}
\usepackage{amssymb}
\usepackage{bm}
\usepackage{amsmath}
\usepackage{color}

\newcommand{\be}{\begin{eqnarray}}
\newcommand{\ee}{\end{eqnarray}}
\newcommand{\bse}{\begin{subequations}}
\newcommand{\ese}{\end{subequations}}

\newcommand{\bnum}{\begin{enumerate}}
\newcommand{\enum}{\end{enumerate}}

\newcommand{\bit}{\begin{itemize}}
\newcommand{\eit}{\end{itemize}}

\newcommand{\bc}{\begin{cases}}
\newcommand{\ec}{\end{cases}}

\newcommand{\bpm}{\begin{pmatrix}}
\newcommand{\epm}{\end{pmatrix}}

\newcommand{\bvm}{\begin{vmatrix}}
\newcommand{\evm}{\end{vmatrix}}

\newcommand{\bs}{\boldsymbol}

\newcommand{\ovl}{\overline}

\newcommand{\gc}{\gamma}
\newcommand{\gd}{\delta}
\newcommand{\eps}{\epsilon}

\newcommand{\gt}{\theta}
\newcommand{\gr}{\rho}
\newcommand{\gs}{\sigma}

\newcommand{\gvf}{\varphi}

\newcommand{\p}{\partial}
\newcommand{\f}{\frac}

\newcommand{\lan}{\langle}
\newcommand{\ran}{\rangle}

\newcommand{\csp}{\;,\qquad}

\newcommand{\bX}{{\bf X}}

\newcommand{\bF}{{\bf F}}
\newcommand{\bN}{{\bf N}}

\newcommand{\bP}{{\bf P}}
\newcommand{\bI}{{\bf I}}

\newcommand{\bx}{{\bf x}}
\newcommand{\bn}{{\bf n}}
\newcommand{\bu}{{\bf u}}

\newcommand{\bw}{{\bf w}}

\newcommand{\br}[1]{\left( #1 \right)}
\begin{document}
\title{Derivation of a hydrodynamic theory for mesoscale dynamics in microswimmer suspensions}

\author{Henning Reinken}
\email{henning.reinken@itp.tu-berlin.de}
\affiliation{Institute for Theoretical Physics, Technische Universit\"at Berlin,
Hardenbergstr. 36, D-10623, Berlin, Germany} 

\author{Sabine H. L. Klapp}
\affiliation{Institute for Theoretical Physics, Technische Universit\"at Berlin,
Hardenbergstr. 36, D-10623, Berlin, Germany} 

\author{Markus B\"ar}
\affiliation{Department of Mathematical Modelling and Data Analysis,
Physikalisch-Technische Bundesanstalt Braunschweig und Berlin, Abbestr. 2-12,
10587 Berlin, Germany}

\author{Sebastian Heidenreich}
\affiliation{Department of Mathematical Modelling and Data Analysis,
Physikalisch-Technische Bundesanstalt Braunschweig und Berlin, Abbestr. 2-12,
10587 Berlin, Germany}

\date{January 31, 2018}

\begin{abstract} 
In this paper we systematically derive a fourth-order continuum theory capable 
of reproducing mesoscale turbulence in a three-dimensional suspension of microswimmers. We start from overdamped Langevin equations for a 
generic microscopic model (pushers or pullers), which include hydrodynamic interactions 
on both, small length scales (polar alignment of neighboring swimmers) and large length scales, where the 
solvent flow interacts with the order parameter field.
The flow field is determined via the Stokes equation supplemented by an {ansatz} 
for the stress tensor. 
In addition to hydrodynamic interactions, we allow for nematic pair interactions 
stemming from excluded-volume effects. The results here substantially extend and generalize earlier findings [Phys. Rev. E \textbf{94}, 020601(R) (2016)], in which we derived a two-dimensional hydrodynamic theory. From the corresponding mean-field Fokker-Planck equation combined with a self-consistent closure scheme, we derive nonlinear field equations for the polar and the nematic order parameter, involving gradient terms of up to fourth order.
We find that the effective microswimmer dynamics depends on the coupling between solvent flow and orientational order. For very weak coupling corresponding to a high viscosity of the suspension, the dynamics of mesoscale turbulence can be described by a simplified model containing only an effective microswimmer velocity. 
\end{abstract}


\maketitle

\section{Introduction}
Self-organized structures of actively driven constituents are fascinating 
non-equilibrium phenomena which are fundamental for the development of life. Due 
to the large number of constituents in such systems, continuum theories provide 
a useful tool to describe and analyse how the microscopic details of system 
constituents and the interactions between them give rise to large scale 
collective structures observed in experiments. These range from large scale vortex structures \cite{Sumino_Nature, Schaller_Nature,Nishiguchi2015,Rabani2013}, 
dynamical clustering \cite{buttinoni2013dynamical,peruani2012collective}, giant 
number fluctuations \cite{Ramaswamy,sw10} and actively driven phase separations 
\cite{schwarz2012phase,wittkowski2014scalar,buttinoni2013dynamical,
speck2014effective} to bacterial 
turbulence \cite{dombrowski2004self,bratanov2015new,oza2016generalized, 
heidenreich2014numerical}.
For recent reviews on active matter, see \cite{Ramaswamy2010,Romanczuk2012,Marchetti2013,Elgeti2015,Bechinger2016,Zoettl2016,Menzel2015}.
Even if the interactions between living microswimmers are complex and 
characterized by many details, the observed 
collective patterns often show a generalized behavior. Therefore, the continuum 
description of such systems uses symmetry arguments and extra terms for the 
activity driving the system out of equilibrium. 

Due to the fascinating nature of collective phenomena in active systems, many 
efforts have been made to find general models for large collections of active moving particles. The models proposed in recent years range from 
kinetic and continuum theories based on microscopic models to 
purely phenomenological approaches 
\cite{peshkov2012nonlinear,peshkov2012continuous,MarchettiP,MarchettiQ,
grossmann2012active,menzel2016dynamical,Shelley1,Shelley2,Toner_Tu,Grossmann2014,Grossmann2015}. 
In most of the systems the individual microswimmer exhibits an axis defining a 
direction that breaks the rotational symmetry. The density distribution of a 
large collection of microswimmers then depends on the position and the 
orientation of the individuals. Depending on the symmetry of the distribution, 
the system can either exhibit polar (e.g., in polar active fluids) or nematic order (e.g., for active 
nematics). Examples for nematically ordered matter are microtubuli-kinesin mixtures
\cite{sanchez2012spontaneous} or swimming bacteria in a lyotropic liquid 
crystal \cite{zhou2014living,genkin2017topological}. Due to the symmetry of the 
orientational distribution, active nematics are usually described by a second rank 
tensor (nematic order parameter) coupled to a vector field describing the fluid in 
which 
the microswimmers are swimming 
\cite{MarchettiQ,hemingway2016correlation,doostmohammadi2016stabilization}. In 
this paper we consider an active fluid consisting of polar moving active particles 
like bacterial suspensions \cite{Shelley1,Shelley2,Toner_Tu}. 

In a recent approach, a phenomenological model for the collective dynamics of 
\textit{B. subtilis} suspensions has been proposed \cite{PNAS}.  The idea was to 
extend the seminal Toner-Tu equation \cite{Toner_Tu} introduced  solely on 
symmetry arguments as the continuum counterpart of the Vicsek model 
\cite{vicsek1995novel} for moving active particles (e.g., to describe flocks of 
flying birds). The Toner-Tu equation exhibits polar symmetry and is able to 
describe the transition 
from a disordered state to an ordered (polar)
state that corresponds to the collective movement of active individuals. For 
constant density (as expected 
in dense suspensions) the Toner-Tu equation has two homogeneous fixed points, 
but shows no heterogeneous 
dynamics as observed for bacterial suspensions. The reason is a positive 
parameter of the Laplacian that damps 
excited spatial modes. By allowing negative values, excited modes increase
and the homogeneous state becomes unstable. Such a theory 
then has to be extended to higher (fourth) order derivatives for stability reasons. 

Such higher order derivative equations have been frequently used in the theory of 
pattern formation. A simple theory modeling the formation of patterns at a 
certain length scale is the Swift-Hohenberg equation. This scalar fourth order 
theory exhibits a Turing instability and shows, e.g., stripes, labyrinth or hexagonal 
patterns in two dimensions. 
A combination of the characteristic features of the Swift-Hohenberg with the 
Toner-Tu equations yields a surprisingly rich phenomenological description of 
bacterial turbulence. In particular, a fourth order field theory for the divergence-free collective velocity $\bw$, i.e.,
\be
\label{GSHE}
\left(\p_t +\lambda_0 \bw \cdot \nabla \right) \bw =&& -\nabla p + \lambda_1
\nabla|\bw|^2 - \alpha \bw - \beta |\bw|^2 \bw \notag \\&&+ \Gamma_2 \nabla^2
\bw + \Gamma_4 \nabla^4 \bw, 
\ee
is able to model the main features of mesoscale turbulence \cite{PRL, 
NJP, PNAS, bratanov2015new,Zhang2009}. The ratio of the phenomenological 
parameters $\Gamma_2$ and $\Gamma_4$ determines the characteristic scale, the 
typical vortex size $\Lambda$, in bacterial suspensions, viz., $\Lambda = 2 \pi 
\sqrt{2 \Gamma_4/\Gamma_2}$. By adjusting these parameters and the parameter 
$\lambda_0$ to experimental data, the dynamics of the observed bacterial 
suspensions can be described quantitatively. 

The phenomenological model describes the dynamics of an effective velocity and 
does not distinguish between orientation and the dynamics of the surrounding fluid. 
However, recent experiments of \textit{B. subtilis} suspensions and 
simulations of Wioland, Lushi and coworkers have shown the importance of the
solvent hydrodynamics for the collective dynamics \cite{Lushi14}. In an earlier 
rapid communication we introduced a two-dimensional field theory including the solvent 
flow dynamics \cite{heidenreich2016hydrodynamic}.

The purpose of the present article is to derive a continuum model from a 
generic microscopic model with polar and nematic alignment interactions 
combined with an effective treatment of hydrodynamic interactions with the 
ambient fluid for two and three spatial dimensions. In particular, the theory resulting in this paper is capable to describe the dynamics of three-dimensional microswimmer suspensions. 

The next section introduces 
our microscopic model by a set of overdamped Langevin equations. We then derive 
the related Fokker-Planck equation and by projection onto the orientational moments we obtain an infinite hierarchy of moment equations.
Using appropriate closure conditions to truncate the 
hierarchy finally yields the continuum equations. 

In our previous publication \cite{heidenreich2016hydrodynamic} we sketched the 
derivation of a 2D field theory from  a microscopic model. The 
focus was to show that the typical length scale in mesoscale turbulence depends 
on details of the microswimmers themselves.
In this paper we give an elaborate derivation of a 2D and 3D field theory from 
a microscopic model that in addition includes nematic interactions. 
We show that the microscopic parameters have a strong influence on the collective dynamics of mesoscale turbulence. 
In particular, we identify a coupling parameter characterizing the flow's 
response to the activity. In the limit of very weak coupling, the system can be 
adequately described by the dynamics of only one order parameter and we recover 
the phenomenological Eq.~(\ref{GSHE}) for the effective microswimmer velocity.
Finally, we present 3D simulations focusing on the limit of weak coupling.

\section{Microscopic theory}
\label{theory_micro}
\subsection{Microswimmer model}
\label{model}
\begin{figure}
   \begin{center}
   \begin{tabular}{c}
   \includegraphics [width = 9.0 cm]{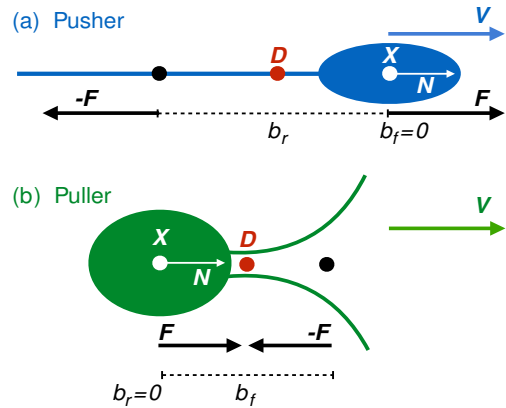}
   \end{tabular}
   \end{center}
   \caption
   { \label{Swimmer_Sketch} 
(Color online) Schematic of a model microswimmer of type "pusher" (a) or "puller" (b). The 
center
of hydrodynamic stress (X) is located in front (pusher) or behind (puller)
the location of the force dipole (D). The distances between X and
the acting forces is given by $b_f$ and $b_r$, respectively. The orientation
of the swimmer's long axis, 
${\bf N}$ defines the direction of the velocity, ${\bf v}$.}
   \end{figure}

Active suspensions are comprised of millions of microswimmers moving in a 
Newtonian solvent fluid such as water. The swimming mechanism is based on
non-reciprocal movements in time, which cause a flow of the
surrounding fluid. The characteristics of the flow
depend on the type of swimmer considered (e.g., bacterial swimmer, algae cell or 
chemically driven nano-rod). However, at large distances and on time scales 
beyond a typical stroke period,
the flow can often be approximately described by a moving 
force dipole resulting from two close point forces $\bF$ with opposite 
directions. 
One distinguishes between pushers and pullers. A pusher particle
 (such as, e.g., \textit{Escherichia Coli} bacteria~\cite{Knut}) pumps the solvent fluid from 
its sides towards the rear and the front,
 as described by two forces pointing away from one another. In contrast, for 
pullers (e.g., \textit{Chlamydomonas Reinhardtii})  the forces point towards each other.
 An illustration of these situations is given in Fig.~\ref{Swimmer_Sketch} which 
also introduces our particle model. 

 Each microswimmer is represented by a
 cylindrical rod moving along the direction of its long axis, described
 by the vector $\bN$. 
 The front force, $\bF_f=f_0 {\bf N}$ (with $f_0$ being positive or negative for 
a pusher or puller, respectively) acts 
at  position ${\bf X}_f ={\bf X}+b_f {\bf N}$, where $b_f>0$ and ${\bf X}$
is the center of hydrodynamic stress, i.e., the point where the hydrodynamic 
net torque on a rigid body vanishes \cite{Happel}. 
We will later use ${\bf X}$ to describe the position of the swimmer in space.
The rear force, $\bF_r=-f_0 {\bf N}=-\bF_f$ acts 
at  position ${\bf X}_r ={\bf X}-b_r {\bf N}$ (with $b_r>0$). 
The entire rod is then moving with velocity ${\bf v}=v_0\bN$ (with $v_0$ being 
the self-propulsion velocity) similarly to self-propelled rod models 
\cite{ginelli2010large, yang2010swarm, wensink2008aggregation,Peruani2006,Grossmann2016}. 

With these assumptions, the force density in the solvent generated by a \textit{single} microswimmer is given by 
\be
\label{dforce}
{\bf f} (\bx, t)
&=&
 \bF_r \gd(\bx- \bX_r) +  \bF_f \gd(\bx- \bX_f).
\ee
We will later make use of this expression to calculate the active stress tensor 
${\boldsymbol \sigma}^{\mathrm{a}}$ 
resulting from the forcing of the fluid by a collection of swimmers. The active 
stress is an
important ingredient of the Stokes equation (valid at low Reynolds numbers) for 
the fluid velocity
${\bf u}$.
\subsection{Microscopic equations of motion}
\label{micro_eqs}
We now consider an ensemble of $\sigma=1,\ldots,S$ identical microswimmers of 
the type introduced in Fig.~\ref{Swimmer_Sketch}. 
The aim in this section is to formulate equations of motion incorporating, on 
the one hand,
direct conservative (e.g., repulsive) interactions between the swimmers, and, on 
the other hand,
indirect (hydrodynamic) interactions induced by the solvent flow.
Following our earlier paper~\cite{heidenreich2016hydrodynamic} we take into 
account these interactions on a coarse-grained level of description.
Various studies (see, e.g., Refs.~\cite{najafi2004simple, 
pooley2007hydrodynamic,avron2005pushmepullyou})
have proposed a microscopic modeling of the interaction between swimmer and 
solvent flow.
However, due to the detailed nature of these models they are not convenient
for the derivation of a continuum theory targeting the collective behavior of a
large number of microswimmers.

\begin{figure}
   \begin{center}
   \includegraphics [width = 7.0 cm]{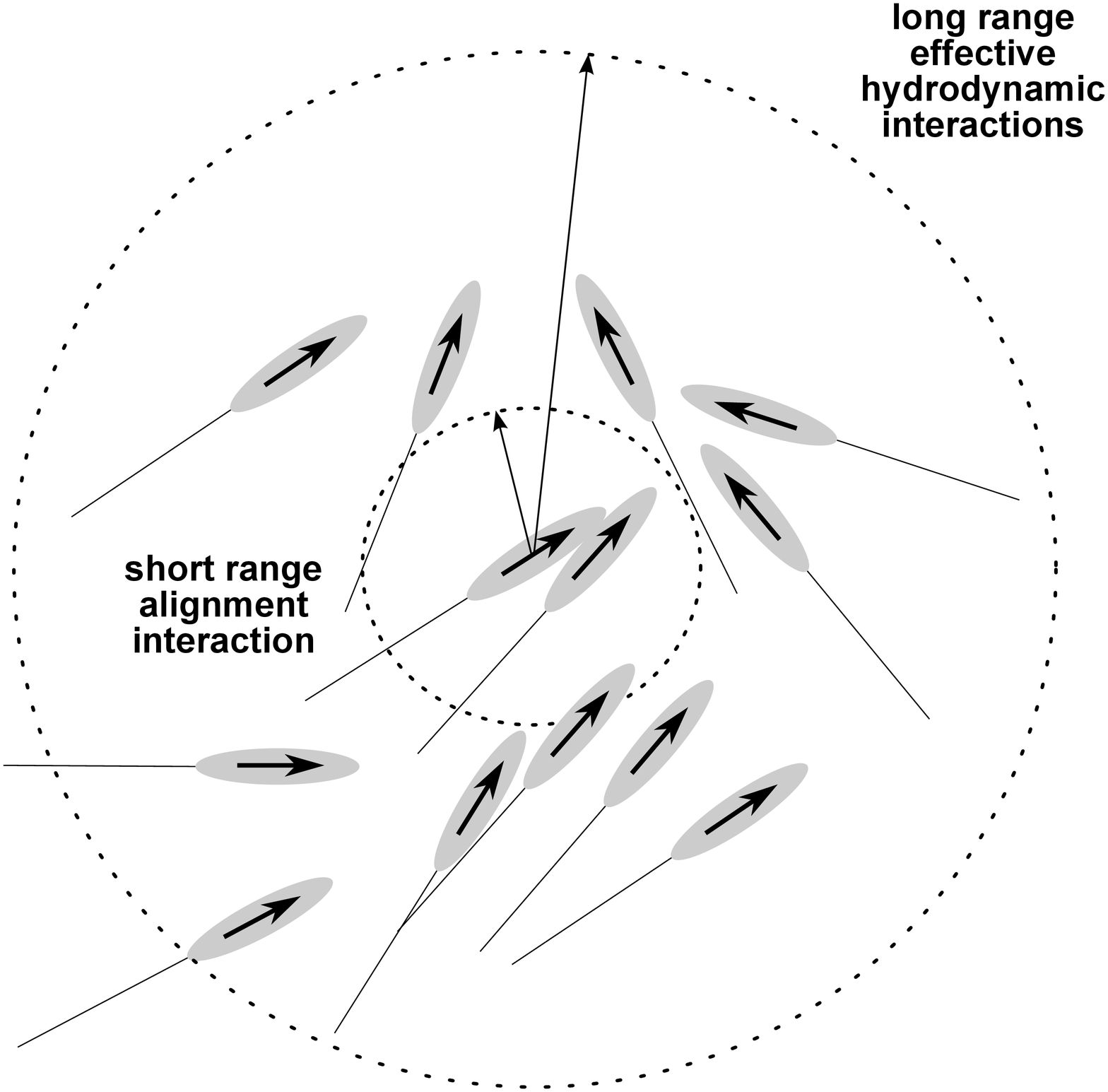}
   \end{center}
   \caption
   { \label{interactions}  Schematic of the two types of interactions 
   incorporated in our model: short-range interactions favoring
   alignment (inner circle), and long range hydrodynamic interactions (outer 
ring).}
   \end{figure}

Within our coarse-grained description, the interactions are grouped into two 
contributions, see Fig.~\ref{interactions} for an illustration.
The first contribution comprises all short-range effects. Specifically, it 
summarizes the repulsive (rod-rod) pair interactions as well as the near-field 
hydrodynamic interactions
between swimmers $\nu$ and $\mu$ by an anisotropic pair potential (to be 
specified below), which couples the orientations $\bN^\nu$ and $\bN^\mu$ at 
distances up to a fixed range.
The second contribution accounts for far-field hydrodynamic effects;
these are incorporated by terms which couple the orientation ${\bf N}^\nu$ with 
the (averaged) solvent flow ${\bf u}$. 
The latter is a collective quantity in the sense that it 
involves the stroke-averaged far fields generated by all the
individuals (see Sec.~\ref{field_equations}).

\subsubsection{Pair interactions}
In the following we first specify the short-range pair interactions. 
Experimentally one observes that bacteria and 
other microswimmers locally tend
to swim along the same direction \cite{sw10}. Within our description
this corresponds to an activity-driven polar alignment of the orientation
vectors $\bN^\nu$, $\bN^\mu$ of neighboring particles. Note that this is a simplified model describing only the aligning effect of near-field hydrodynamic interactions and neglecting any higher order tensorial components. For a classification of active systems see~\cite{Marchetti2013}. Specifically, we use the 
ansatz
\be
\label{phi_act}
\Phi^\mathrm{act}\left(\bN^\nu,\bN^\mu,r_{\mu\nu}\right)
&=&-\f{\gc_0 v_0}{2}\bN^\nu \cdot\bN^\mu \,\Theta(\eps_a-r_{\mu\nu}),
\ee
where $\gamma_0>0$ is the magnitude of the interaction, $r_{\mu\nu}=|\bs X^\mu 
-\bs X^\nu|$ is the particle distance,
and $\Theta(x)$ is the theta-function (with $\Theta(x)=1$, $x\geq 0$, and zero 
otherwise). 
The latter expresses the fact that the range of the polar interaction is 
restricted to distances
smaller than $\epsilon_a$. 
The appearance of the self-propulsion velocity $v_0$
in the prefactor reflects the "active" nature of this interaction.
However, even at $v_0=0$
(dead microswimmer) there is a steric interaction between
the bodies. On a mean-field (Maier-Saupe) level, these
lead to a passive nematic potential 
\cite{maier1958einfache, maier1959einfache,maier1960einfache}
\be
\label{phi_pass}
\Phi^\mathrm{pass}\left(\bN^\nu,\bN^\mu,r_{\mu\nu}\right)
&=&-\f{\gc_1}{4}\left(\bN^\nu \cdot\bN^\mu\right)^2 \,\Theta(\eps_p-r_{\mu\nu}),\notag \\
\ee
where $\gamma_1>0$. Both potentials are of square-well type, i.e., they are 
independent of the distance
within the range of the respective interaction ranges.
This choice reflects our focus on reorientation effects; in fact,
the impact of direct steric collisions on the translational motion is neglected 
in our model.
Realistically, 
the ranges $\epsilon_a$, $\epsilon_p$ should be comparable to the 
microswimmers' length. For convenience we choose  $\epsilon_a= 
\epsilon_p = \epsilon$.
The total potential function stemming from short-range interactions is then 
given by the pair sums 
of Eqs.~(\ref{phi_act}) and (\ref{phi_pass}),
\be
\label{phi_total}
\Phi = \sum_{\mu,\nu}\left(\Phi^\mathrm{act} 
\left(\bN^\nu,\bN^\mu,r_{\mu\nu}\right)
+\Phi^\mathrm{pass}\left(\bN^\nu,\bN^\mu,r_{\mu\nu}\right)\right).\notag \\
\ee

Our approach of including the far-field hydrodynamic effects will be specified 
in the next paragraphs.

\subsubsection{Langevin equations}
\label{sec: Langevin}
To model the (overdamped) translational and orientational
motion of the ensemble of microswimmers we
employ the Langevin equations
\begin{subequations}
\begin{eqnarray}
\label{LE}
\frac{d}{dt} {\bf X}^\sigma(t)  &=& v_0  {\bf N}^\sigma(t) +  {\bf u} ({\bf
X}^\sigma,t) + \sqrt{2 D}\; \boldsymbol{\xi}^\sigma(t),   \\
\label{LE1}
\frac{d}{dt}  {\bf N}^\sigma(t) &=& 
{\bs\Omega ({\bf X}^\sigma,t)\cdot {\bf N}^\sigma}(t) +{\boldsymbol \Pi}({\bf
N}^\sigma) \cdot \left[a_0 \boldsymbol \Sigma( {\bf
X}^\sigma,t) \cdot {\bf N}^\sigma(t)  \right. \notag \\
&&\left. -  {\nabla}_{{\bf
N}^\sigma} \Phi({\bf N}) +
\tau^{-1/2} \;{\boldsymbol  \eta}^\sigma(t)  \right],
\end{eqnarray}
\end{subequations}
where the random functions $\boldsymbol{\xi}^\sigma(t)$ and 
$\boldsymbol{\eta}^\sigma(t)$ denote Gaussian white noise
mimicking diffusion effects in the translational and orientational motion (we 
neglect
shape-induced effects on the noise). 
The translational dynamics (\ref{LE}) is determined by the
self-propulsion (of magnitude $v_0$) in direction of ${\bf N}$, as well as by 
the advection ${\bf u}$.
Here, $\bf u$ is the average flow field to be calculated via the averaged Stokes 
equation (see Eq.~(\ref{SE-b}) below). Thus, we adopt the mean-field-like 
assumption that the fluctuating field ${\bf u}$ can be replaced by its average.

The orientational dynamics (\ref{LE1}) is determined, first,
by the short-range anisotropic interactions $\Phi$ defined in 
Eq.~(\ref{phi_total}). 
Second, Eq.~(\ref{LE1}) takes into account the far-field hydrodynamic effects
via the coupling terms involving the vorticity 
${\bs\Omega}=\f{1}{2}\left[\left(\nabla{\bf u}\right)^\top - \nabla \bs 
u\right]$, which describes the torque of the solvent flow,
and the strain rate tensor ${\boldsymbol \Sigma} =\f{1}{2}\left[\left(\nabla{\bf 
u}\right)^\top + \nabla \bs u \right]$ describing the effect of a flow gradient.
Finally, there appears the projector $\bs \Pi ({\bf N}^\sigma) ={\bf I}-{\bf N}^\sigma{\bf N}^\sigma$ (with
${\bf I}$ being the unit matrix), which conserves the length of the orientation
vector ${\bf N}^\sigma$. The entire
Eq.~\eqref{LE1} is interpreted as a Stratonovich stochastic differential 
equation
with rotational relaxation time $\tau$.

We note that the coupling terms $ \boldsymbol \Omega \cdot {\bf N}^\sigma + \bs \Pi \cdot a_0 \boldsymbol 
\Sigma \cdot {\bf N}^\sigma$ 
in Eq.~(\ref{LE1}) 
remind of Jeffrey's theory for the oscillatory tumbling motion of
elongated passive particles \cite{Jeffery,HinchLeal,PedleyKessler}.
In that theory, the magnitude of the coupling depends 
on the aspect (length-to-width) ratio $r$ of the particles via the 
shape parameter
\be
\label{shape_eq}
a_0=(r^2-1)/(r^2+1).
\ee
Because the Jeffrey equation has been proposed for passive rods,
the validity of Eq.~(\ref{shape_eq}) for active particles is not immediately 
clear.
Indeed, Rafa{\"i} and co-workers \cite{Peyla2010} have 
experimentally shown  that the Jeffrey tumbling period 
\cite{Jeffery} of \textit{Chlamydomonas } algae depend on their swimming 
conditions. Further, it has been found \cite{Jibuti2017}
that the tumbling period in a shear flow of living cells can be 
larger or lower than the tumbling period of the corresponding dead cells. 
Nevertheless, one may expect a dependence of the shape 
parameter $a_0$ on the swimmer size $\ell$ also for active 
particles, thus we stick to Eq.~(\ref{shape_eq}).

\subsection{Far-field hydrodynamic interactions}

We distinguish between near-field and far-field hydrodynamic interactions, the former already specified by the effective polar interaction potential above (see Eq.~\ref{phi_act}). The latter are taken into account by the coupling of Eqs.~(\ref{LE}) and (\ref{LE1}) to the average flow field ${\bf u}$.

\subsubsection{Stokes equation}

In the limit of low Reynolds numbers, ${\bf u}$ is determined by the averaged Stokes equation
\begin{equation}
0=-{\nabla}  p + \mu^* \nabla^2 {\bf u}  +
\nabla \cdot \lan {\boldsymbol \sigma}\ran ,
\label{SE-b} 
\end{equation}
combined with the incompressibility condition (appropriate at large densities)
 \begin{equation}
\label{SE1} 
0=\nabla \cdot {\bf u}.
\end{equation}
In Eq.~(\ref{SE-b}), $p$ is the hydrodynamic pressure, 
$\lan{\boldsymbol\sigma}\ran$ is the averaged stress tensor (to be discussed in 
the next paragraph), and $\mu^*$ is the effective bulk viscosity.
The latter includes contributions from the solvent as well
as from the embedded colloids or bacteria. As an approximation,
we here employ the Batchelor-Einstein
relation for passive, spherical particles \cite{Hinch10,Einstein,Batchelor}. 
The latter is given as 
$\mu^*=\mu_0(1+k_1 c + k_3 c^2)$ where $\mu_0$ is the "bare" solvent viscosity 
and $c=V_0 n$ is the volume fraction
(with $n$ being the number of particles per unit volume $V_0=4 \pi a^3/3$
for a sphere of radius $a$). The coefficients are $k_1 = 5/2$ in three
dimensions \cite{Hinch10, Einstein} and $k_1=2$ in two dimensions
\cite{Heines08}, respectively. The second order coefficient can be similarly calculated for passive spherical 
colloidal particles \cite{Hinch10}. For active microswimmer suspensions there is an additional term $k_2$ with a negative sign that emerges within the derivation given later \cite{heidenreich2016hydrodynamic}.

Using the values for the passive case for $k_1$ and $k_3$ the effective bulk viscosity of \textit{Chlamydomonas 
Reinhardtii}  (puller) suspensions is well 
described \cite{Peyla2010,heidenreich2016hydrodynamic}. 
On the other hand for \textit{B. subtilis} or \textit{Escherichia Coli} (pusher) 
suspensions the value of $k_3$ has to be fitted to be in agreement 
with experimental observations \cite{heidenreich2016hydrodynamic}.

An alternative to the Batchelor-Einstein relation is the Krieger-Dougherty equation valid for dense suspensions~\cite{Krieger1959}, assuming a different functional form of the effective viscosity in terms of the volume fraction. In the scaling we propose later, the explicit form of the effective viscosity does not enter. In this work we thus refrain from a detailed discussion on the topic.

The Stokes equation \eqref{SE-b} is formulated in three spatial dimensions,
and this is the situation for which we later present numerical results.
Likewise, we assume that the orientation vectors ${\bf N}^\sigma$ can have
all directions on the unit sphere.
However, the theory may also be adapted for a 
quasi-two dimensional system where one spatial dimension is very small compared 
to the 
others (thin film); this is achieved by integrating over the small dimension. 
Depending on boundary conditions the integration procedure may introduce some 
extra boundary terms,
such as the Brinkmann equation for no slip conditions \cite{Brinkmann49}. 

\subsubsection{Stress tensor}
We now consider in more detail the stress tensor $\bs \sigma$, whose average 
appears in the Stokes equation~(\ref{SE-b}). This ensures that
the properties of the interacting microswimmer suspension are fed back into the 
flow field $\bf u$.
In the present system, $\lan{\bs \sigma}\ran$ consists of an active (a) and 
passive (p) contribution, $\lan{\boldsymbol \sigma}\ran=\lan{\boldsymbol 
\sigma}^\mathrm{a}\ran+\lan{\boldsymbol \sigma}^\mathrm{p}\ran$. Here, we 
discuss first the microscopic, non-averaged stress tensor (the averaging will be 
done later in Sec.~\ref{sec: active stress}).

The active part ${\boldsymbol \sigma}^\mathrm{a}$ is determined by forces 
exerted by the
microswimmers onto the solvent fluid.
Specifically, one has
\be
\nabla \cdot {\boldsymbol \sigma}^\mathrm{a}= {\bf f},
\ee
where ${\bf f}$ is the force density related to the $\sigma=1,\ldots,S$ force 
dipoles.
From Eq.~(\ref{dforce}) we find that each particle $\sigma$ gives a contribution
\be
{\bf f}^\gs (\bx, t)
&=&\notag
 \bF^\gs_r \gd(\bx- \bX^\gs_r) +  \bF^\gs_f \gd(\bx- \bX^\gs_f)
 \\
&=&\notag -f_0 \bs N^\gs
\left[ \gd(\bx- {\bf X}^\gs +b_r {\bf N}^\gs)  \right. \\
&&\left. -  \gd(\bx- {\bf X}^\gs -b_f 
{\bf
N}^\gs)\right].
\ee
where we have used that $\bF^\gs_r=-\bF^\gs_f=-f_0{\bf N}$,
as well as the corresponding expressions for ${\bf X}_{r}$ and ${\bf 
X}_f^{\sigma}$
(see Fig.~\ref{Swimmer_Sketch}).

Following earlier studies~\cite{Ramaswamy,hatwalne2004rheology}, we expand the force density in a Taylor series for small lengths $b_r$ and $b_f$ up to fourth order, analogous to a multipole expansion~\cite{Baskaran2009,Spagnolie2012}, yielding
\be
\label{stress_Taylor}
{\bf f}^\gs
\approx\notag
-f_0 \bf N^\gs &&
\biggl[ 
(b_r+b_f) {\bf N}^\gs \cdot \nabla  +
\f{(b_r^2-b_f^2)}{2!} {\bf N}^\gs  {\bf N}^\gs :\nabla \nabla 
\\
\notag&&
+ \f{(b_r^3+b_f^3)}{3!} {\bf N}^\gs  {\bf N}^\gs{\bf N}^\gs :\cdot\nabla \nabla 
\nabla \\
&&+
\f{(b_r^4-b_f^4)}{4!} {\bf N}^\gs  {\bf N}^\gs {\bf N}^\gs{\bf N}^\gs::\nabla 
\nabla\nabla\nabla
 \notag
\\
&&+ \dots \biggr]\, \delta^\sigma,
\ee 
where we introduced the abbreviation $\delta^\sigma = \gd(\bx- {\bf X}^\gs)$.
Using $ {\bf f}^\gs= \nabla \cdot \bs \gs^\mathrm{a}_\gs, \notag$
where $\bs \gs^\mathrm{a}_\gs$ is the active stress related to particle 
$\sigma$,
and performing a sum over all particles we find that the
total active stress is given 
(up to an irrelevant constant) by
\begin{eqnarray}
\label{e:active_stress}
{\boldsymbol \sigma^\mathrm{a}}({\bf x}, t)
 &\approx& 
 - f_0\biggl[\zeta_1
 \sum_\sigma {\bf N}^\sigma {\bf N}^\sigma\delta^\sigma +
 \notag\\
&&
\qquad\;
\zeta_2\;\nabla\cdot  
\sum_\sigma {\bf N}^\sigma {\bf N}^\sigma {\bf N}^\sigma\delta^\sigma +
\notag  \\
&&
\qquad\;
\zeta_3\; {\nabla} {\nabla} :  
 \sum_\sigma {\bf N}^\sigma {\bf N}^\sigma {\bf N}^\sigma {\bf 
N}^\sigma\delta^\sigma +
\, 
 \notag\\
&&
\qquad\;
\zeta_4\; \nabla \nabla \nabla : \cdot 
\sum_\sigma {\bf N}^\sigma {\bf N}^\sigma {\bf N}^\sigma {\bf N}^\sigma  {\bf 
N}^\sigma\delta^\sigma \biggr],\notag\\
\end{eqnarray}
where we have truncated the expansion (\ref{stress_Taylor}) after the 
fifth-order term in ${\bf N}^\sigma$.
The coefficients of the lower-order terms are given by
\begin{subequations}
\be
\zeta_1 &=& b_f + b_r 
\csp \; \;\; \;\; \;\;
\zeta_2 = \f{1}{2}(b_r^2-b_f^2),  \\
\zeta_3 &=&  \f{1}{3!}(b_r^3+b_f^3)
\csp 
\zeta_4 = \f{1}{4!}(b_r^4-b_f^4).
\ee
\end{subequations}

The passive stress, $\bs \sigma^\mathrm{p}$, is determined by the relation
\cite{Dhont2002},
 \be
 \label{e:pass_stress}
 \nabla \cdot \bs \sigma^\mathrm{p} = - \sum_\sigma \bigg\langle \int_{\partial 
V_j 
 }dS \delta(r-r') {\bf f}^h(r') \bigg\rangle,
 \ee
where ${\bf f}^h(r')$ is the force per unit area that the solvent exerts on the 
surface element, and $\partial V_j$  is the surface area of the rod.

To summarize, our microscopic model is defined by the (overdamped) Langevin 
equations~(\ref{LE}) and \eqref{LE1}
for the positions ${\bf X}^\sigma$ and orientations ${\bf N}^\sigma$, 
respectively,
combined with the Stokes Eq.~\eqref{SE-b}
for the flow field ${\bf u}$. The latter includes the active and passive stress 
from  the
microswimmers according to Eqs.~\eqref{e:active_stress} and Eq. 
\eqref{e:pass_stress}.

Similar equations of motions have been derived on the basis of the slender body 
theory by Shelley 
and Saintillan \cite{Shelley1,Shelley2}. The major difference to their work
concerns the form of alignment interactions, as well as the fact that we 
consider higher order contributions to the 
active stress. 
\section{Continuum theory}
Starting from the microscopic theory presented in Sec.~\ref{theory_micro} we now
derive continuum equations for suitable order parameter fields describing the 
dynamics
on the mesoscale. To this end, we first set up the Fokker-Planck (FP) equation 
(Sec.~\ref{Fokker}) 
corresponding to the Langevin equations \eqref{LE} and~\eqref{LE1}. The FP 
equation then
generates order parameter equations by projection. In the second step, we 
combine these order parameter equations with a suitably rewritten
equation for the flow field (Sec.~\ref{field_equations}).

\subsection{Fokker-Planck equation}
\label{Fokker}

\subsubsection{Probability densities and order parameters}

We consider the dynamics of the 
one-particle probability density function
\be
\label{e:def_PDF}
\mathcal{P}( {\bf x} ,\bn, t) =  
\frac{1}{S} \sum_{\sigma=1}^S  
\bigg\langle
\delta( {\bf x } - {\bf X}^\sigma(t))\,\delta( {\bf n } - {\bf N}^\sigma(t)) 
\bigg\rangle, \quad
\ee  
where $\lan\,\cdot\,\ran$ denotes an average with respect to the Gaussian white 
noise processes as well as an ensemble average. Based on $\mathcal{P}( {\bf x} 
,\bn, t)$, 
we can define the marginal position density $\mathcal{P} (\bx,t)$ and the local 
concentration field $\gr(\bx,t)$ via
\be
\label{rho_def}
\gr(\bx,t)&=& S\, \mathcal{P} ({\bf x},t) 
= S\int d\bn\; \mathcal{P} ({\bf x},{\bf n},t).
\ee
From this we obtain the (conditional) orientational density
\be
\mathcal{P} ({\bf n},t|{\bf x},t)=
\f{\mathcal{P} ({\bf x},{\bf n},t) }{\mathcal{P} (\bx,t)} 
=
\f{S\mathcal{P} ({\bf x},{\bf n},t) }{\gr(\bx,t)},
\ee
which describes the probability that a particle has orientation ${\bf n}$ given
that it is located at position ${\bf x}$.

With $\mathcal{P}( {\bf x} ,\bn, t)$ or $\mathcal{P} ({\bf n},t|{\bf x},t)$ we 
can now define
the moments of the orientation ${\bf n}$. Up to third order, the moments
are given as
\begin{subequations}
\label{moments}
\be
\ovl{\bn} (\bx,t)
&\equiv&
 \int d\bn\; \mathcal{P} ({\bf n},t|{\bf x},t)\; \bn 
 \quad\; \notag\\
&=& \f{S}{\gr}\int d\bn\; \mathcal{P} ({\bf x},{\bf n},t)\; \bn,\\
\ovl{\bn\bn} (\bx,t)
&\equiv&
 \int d\bn\; \mathcal{P} ({\bf n},t|{\bf x},t)\; \bn \bn
\;\;\notag\\
&=&\f{S}{\gr}\int d\bn\; \mathcal{P} ({\bf x},{\bf n},t)\; \bn  \bn,\\ 
\ovl{\bn\bn\bn} (\bx,t) 
&\equiv &
 \int d\bn\; \mathcal{P} ({\bf n},t|{\bf x},t)\;\notag \bn \bn\bn\\
 &=&
\f{S}{\gr} \int d\bn\; \mathcal{P} ({\bf x},{\bf n},t)\; \bn \bn \bn.  
\ee
\end{subequations}
From the above definitions it follows that the moments may be considered as 
expectation values (averages)
over the fluctuating microscopic degrees of freedom. 
These expectation values directly lead to the definition of the two
key order parameters of our system, that is, the polarization measuring the net
orientation
\be
\label{polarization}
\bP(\bx,t)\equiv \ovl{\bn} (\bx,t)
\ee
and the ${\bf Q}$-tensor measuring nematic order 
\be
\label{qtensor}
{\bf 
Q}(\bx,t)\equiv(\ovl{\bn\bn})^\text{ST}(\bx,t)=\ovl{\bn\bn}(\bx,t)-\f{\bI}{d}.
\ee
In Eq.~(\ref{qtensor}), the notation $(\ldots)^\text{ST}$ denotes the symmetric 
traceless
part of a tensor, and $d$ is the spatial dimension. For $d=2,3$, the explicit 
expressions
for the symmetric traceless tensors in terms of the full tensors are given by
\begin{subequations}
\label{brackets}
\be
\label{bracket1}
(\ovl{n_in_j})^\text{ST}(\bx,t)&=&\ovl{n_i n_j}-\f{1}{d}\gd_{ij},
\\
(\ovl{n_in_jn_k})^\text{ST}(\bx,t)
&=&
\ovl{n_i n_jn_k}-\f{3!}{2d+4} 
(\gd_{ij} \ovl{n_{k}})^\text{SY},\notag\\
\\
(\ovl{n_in_jn_kn_l})^\text{ST}(\bx,t)
&=&
\ovl{n_i n_jn_kn_l}-
\f{4!}{4d+16}(\gd_{ij}\ovl{n_k n_l})^\text{SY} \notag
\\
&&+
\f{4!}{88d+16}(\gd_{ij}\gd_{kl})^\text{SY},\\
(\ovl{n_i n_jn_kn_l n_m})^\text{ST}(\bx,t)
&=& \ovl{n_i n_jn_kn_ln_m} \notag\\
&&-\f{5!}{12 d+72} (\gd_{ij} \ovl{n_{k}n_{l}n_{m}})^\text{SY} \notag\\
&&+ \f{5!}{120 d+144} 
(\gd_{ij} \gd_{kl}\ovl{n_{m}})^\text{SY},\notag\\
\ee
\end{subequations}
and so on, where $(\ldots)^\text{SY}$ denotes the symmetric part of a tensor and 
is calculated by the sum over permutations yielding $k!$ terms,
\be
\label{bracket2}
(a_{i_1\ldots i_k})^\text{SY}=\f{1}{k!}\sum_{\pi(i,j,\ldots)} 
a_{\pi(i_1)\ldots\pi(i_k)}.
\ee

To simplify the calculation of the moments and order parameters, it is useful to 
expand the orientational (or full probability) density
in terms of a basis set of orthogonal angle-dependent functions. 
For axially symmetric objects (such as the rods considered here), this can be 
done
by using either spherical harmonics $Y_{\ell m}(\gt,\gvf)$ \cite{McCourt}, or, 
alternatively,
irreducible symmetric traceless tensors 
$(\bn\ldots\bn)^\text{ST}$~\cite{Turzi2011, Kroeger2008, hess2015tensors}.
Here we follow the second route. The expansion of the orientational
density then reads
\be\label{e:conditional_expansion}
\mathcal{P} ({\bf n},t|{\bf x},t)&=&\f{1}{V_d}\sum_{k=0}^\infty
\f{(2k+1)!!}{k!}\times
\notag\\&&
(n_{i_1}\cdots n_{i_k})^\text{ST}\;
(\ovl{n_{i_1}\cdots n_{i_k}})^\text{ST}(\bx,t), 
\ee
where $k!!=k(k-2)(k-4)\ldots$ and $V_d$ is the surface volume of the 
$d$-dimensional unit sphere. In the expansion~(\ref{e:conditional_expansion}),
the (symmetric traceless) tensorial expectation values 
$(\ovl{\bn\ldots\bn})^\text{ST}(\bx,t)$ play the role of coefficients
of the irreducible tensors. 

\subsubsection{Mean-field FP equation}
The Fokker-Planck equation for the probability 
distribution $\mathcal{P}( {\bf x},\bn, t)$, which corresponds to the 
Langevin equations~\eqref{LE} and~\eqref{LE1},  can be derived by applying 
standard techniques 
\cite{Risken, Jacobs}. One obtains
\begin{eqnarray}
\label{FP}
\partial_t \mathcal{P}({\bf x},{\bf n},t) &=&  
 -  {\nabla}\cdot \left[
 \left(v_0   {\bf n} + {\bf u}\right)
 \mathcal{P} \right]
 + D  {\nabla}^2\mathcal{P}
\notag \\
&&
- {\nabla}_{\bf n} \cdot
\left\{\left[\bs \Omega \cdot \bn + a_0 {\boldsymbol \Pi}(\bn) \cdot {\bs
\Sigma} \cdot {\bf n} -\frac{1}{\tau} \bn 
\right] \mathcal{P} 
\right\} \notag \\
&&+ \frac{1}{2 \tau} {\nabla}_{\bf n} {\nabla}_{\bf n} : 
\left[{\boldsymbol \Pi}(\bn)\cdot {\boldsymbol \Pi(\bn)}^\top
\;\mathcal{P}\right] + \mathcal{C}^{(2)}[\Phi]. \notag \\
\end{eqnarray}
Note that ${\boldsymbol \Pi}(\bn)\cdot {\boldsymbol \Pi(\bn)}^\top={\boldsymbol 
\Pi}(\bn)\cdot {\boldsymbol \Pi(\bn)}={\boldsymbol
\Pi}(\bn)$. 
The first line in Eq.~\ref{FP} contains the translational drift and diffusion term, respectively. The second line arises due to orientational drift, the first term in the third line corresponds to orientational diffusion.
Further, the last term in Eq.~(\ref{FP}), $\mathcal{C}^{(2)}[\Phi]$, represents 
the "interaction integral"
stemming from the anisotropic pair potentials defined in Eqs.~(\ref{phi_act}) 
and (\ref{phi_pass}).
Its explicit expression reads
\be
\label{C2}
\mathcal{C}^{(2)}[\Phi] 
&=& (S-1)\;
{\nabla}_{\bf n} \cdot \left\{ \int d{\bf n}'\int d{\bx'}\; {\boldsymbol
\Pi(\bn)} \cdot \right. \notag \\ 
&& \left[ {\nabla}_\bn  
 \phi^\mathrm{act}\left(\bn,\bn',|\bx-\bx'|\right)\right. \notag \\
&& \left. \left. + {\nabla}_\bn  
 \phi^\mathrm{pass}(\bn,\bn',|\bx-\bx'|) \right]
 \;\mathcal{P}^{(2)}({\bf x},{\bf n};\bx ',{\bf n}',t)
\right\}. \notag\\
&\equiv&\mathcal{C}^{(2)}[\Phi^\mathrm{act}]+\mathcal{C}^{(2)}[\Phi^\mathrm{pass
}].
\ee
As seen from in Eq.~(\ref{C2}), $\mathcal{C}^{(2)}[\Phi]$ and 
thus, the entire FP equation for the one-particle density, 
couples to the two-particle
distribution function, $\mathcal{P}^{(2)}({\bf x},{\bf n};\bx ',{\bf n}',t)$. 
This coupling initiates
a hierarchy of distribution functions which has to be truncated by a closure 
approximation. Here we use a mean field (factorization) approximation for 
$\mathcal{P}^{(2)}$
(see Appendix~\ref{A:MF}) and focus on the limit $S\gg 1$, yielding
\be
\label{C2_act}
\mathcal{C}^{(2)}[\Phi^\mathrm{act}]
 &\approx&
- \gc_0 v_0
{\nabla}_{\bf n} \cdot  \left\{
{\boldsymbol \Pi(\bn)} \;\mathcal{P}({\bf x},{\bf n},t)
\cdot  \bs J[\gr\bP]
\right\},\notag \\
\ee 
where
\be
\label{J_act}
 \bs J[\gr\bP]=
 A_d (\gr\bP)+ B_d \nabla^2 (\gr\bP) + 
C_d (\nabla^2)^2 (\gr\bP)\notag. \\
 \ee
Note that $ \bs J[\gr\bP]$ still depends on $\bf x$ and $t$.
The index $d$ labels the spatial dimension (which coincides
with that of the orientation vectors). For $d=2$ one finds (see Appendix~\ref{A:MF})
\be 
\label{A2}
A_2 =   \pi\eps^2  \csp  B_2 &=&
\f{\pi\eps^4}{8}  \csp  C_2 =
\f{\pi\eps^6}{192},
\ee
whereas for $d=3$
\be
\label{A3}
A_3 = \f{4\pi\eps^3}{3} \csp B_3 &=&  \f{2\pi\eps^5}{15}  \csp C_3 = 
\f{\pi\eps^7}{210}.
\ee 
An analogous mean-field treatment of the 
passive contribution yields (see Appendix~\ref{A:MF})
\be
\label{C2_pass}
\mathcal{C}^{(2)}[\Phi^\mathrm{pass}]
 &\approx&
- \gc_1 {\nabla}_{\bf n} \cdot  \left\{
{\boldsymbol \Pi(\bn)} \;\mathcal{P}({\bf x},{\bf n},t)
\cdot  \bs J[\gr {\bf Q}]\cdot \bn
\right\}, \notag
\\
 \ee 
 where
 \be
\label{J_pass}
 \bs J[\gr {\bf Q}]&=&
 A_d( \gr  {\bf Q}) + B_d \nabla^2 (\gr  {\bf Q} )
 + C_d (\nabla^2)^2 (\gr {\bf Q}). 
 \ee
 
Inserting the mean-field approximations (\ref{C2_act}) and (\ref{C2_pass})
into Eq.~\eqref{FP} we obtain the FP equation in its final form,
 \begin{eqnarray}
\label{FPE}
\partial_t \mathcal{P}({\bf x},{\bf n},t) &=&  
 -  {\nabla}\cdot \left[
 \left(v_0   {\bf n} + {\bf u}\right)
\mathcal{P}({\bf x},{\bf n},t) \right] + D \nabla^2 \mathcal{P}
\notag \\
&&- {\nabla}_{\bf n} \cdot
\left\{
\left[ \bs \Omega \cdot \bn + a_0
{\boldsymbol \Pi}(\bn) \cdot \bs \Sigma\cdot {\bf n}
\notag \right. \right. \\
&&- \tau^{-1}\bn +\gc_0 v_0 {\boldsymbol \Pi}(\bn) \cdot 
\bs J[\gr\bP] \notag \\
&& \left.  \left. +\gc_1 {\boldsymbol \Pi}(\bn) \cdot 
\bs J[\gr {\bf Q}] \cdot \bn 
\right] \mathcal{P}({\bf x},{\bf n},t)
\right \}  \notag \\
&&+ \frac{1}{2 \tau} {\nabla}_{\bf n} {\nabla}_{\bf n} : 
\left[{\boldsymbol \Pi}(\bn) \;\mathcal{P}({\bf x},{\bf n},t) \right].
\end{eqnarray}
Equation~(\ref{FPE}) will be our starting point for the derivation of order 
parameter equations.

\subsection{Field equations}
\label{field_equations}
\subsubsection{Order parameter equations}
We start with the evolution equation for the overall density which, according to 
Eq.~(\ref{rho_def}),
may be obtained by multiplying the FP equation (\ref{FPE}) with $S$ and 
integrating over the orientational degree of freedom. This procedure yields
\be
 \label{rho}
\partial_t \gr &=&  
-{\nabla} \cdot \left[\gr \left(v_0   {\bP} + {\bf u}\right) \right]
 + D  {\nabla}^2 \gr,
\ee
where we have used the definition of the polarization, Eq.~(\ref{polarization}).
In a dense suspension, fluctuations of the overall density are typically 
negligible, such that we can set $\rho=const$.
From Eq.~\eqref{rho} then follows immediately $ v_0\nabla \cdot \bP + \nabla 
\cdot \bu = 0$. Using, furthermore,
that the solvent flow is incompressible, i.e., $\nabla \cdot \bu =0$, we find 
that
the polarization is source free, that is,
\be 
\label{divP}
\nabla \cdot \bP = 0.
\ee

The evolution equation for $\bP=\ovl{\bn}$ [see Eqs.~(\ref{polarization}) and 
(\ref{moments})]
is derived by multiplying the FP equation ({\ref{FPE}) with $S\bn$ and 
performing again the orientational average.
Setting $\rho=const$ and dividing by this factor we find
\be
\label{moment1_time}
\partial_t  \bP&=&  
 -  {\nabla}\cdot 
 \left[v_0   \ovl{\bn\bn} + \ovl{{\bf u} \bn}\right]
 + D  {\nabla}^2\bP -\frac{1}{\tau} \bP
\notag \\
&&
+ \ovl{\bs \Omega \cdot \bn} + 
a_0  \ovl{({\bf I}-{\bn\bn}) \cdot {\bs \Sigma} \cdot {\bn}}
 \notag \\
&&+\gc_0 v_0 \gr\ovl{({\bf I}-\bn\bn)\cdot  \bs J[\bP]}\notag \\
&&+ \gamma_1 \gr\ovl {\left( \bI- \bn \bn \right) \cdot  \bs J[{\bf Q}] 
\cdot \bn}, 
\ee
where the quantities $J[\bP]$ and $J[{\bf Q}]$ follow from Eqs.~(\ref{J_act}) 
and (\ref{J_pass}) via division
by $\gr$, that is,
\begin{subequations}
\label{J_act_pass}
\be
\bs J[\bP]&=&
 A_d \bP+ B_d \nabla^2 \bP + 
C_d (\nabla^2)^2 \bP,  \\
 \bs J[{\bf Q}]&=&
 A_d {\bf Q} + B_d \nabla^2 {\bf Q} 
 + C_d (\nabla^2)^2 {\bf Q}. 
 \ee
 \end{subequations}
To simplify the ensemble averages (denoted by overbars) appearing on the 
right side of Eq.~(\ref{moment1_time}), we use the fact that the quantities 
$J[\bP]$, and $J[{\bf Q}]$
are \textit{per se} averaged quantities (since they solely depend on order 
parameters) and can thus be pulled out of the overbars. Recall that the flow 
field ${\bf u}$ and therefore also the tensors ${\bf \Sigma}$ and ${\bf \Omega}$ 
 are considered as averaged quantities as well (see also Eq~(\ref{SE-b}) in 
Sec.~\ref{sec: Langevin}) and thus, the same argument holds.
These manipulations allow us to group together the factors ${\bf n}$. Finally, 
we use that $\ovl{\bn}=\bP$ and express the second moment through the 
${\bf Q}$-tensor [see Eq.~(\ref{qtensor})], i.e., $\overline{{\bf n}{\bf n}} 
={\bf Q} + {\bf I}/d$.
Taken altogether we find
\begin{equation}
\label{polarl}
\begin{split}
(\partial_t + {\bf u}\cdot\nabla) {\bf P} = & \ {\bf \Omega} \cdot {\bf P} + a_0 
{\bf \Sigma} \cdot {\bf P} - v_0 \nabla \cdot {\bf Q} \\
&+ D  \nabla^2{\bf P} -\frac{1}{\tau} {\bf P} - a_0 {\bf \Sigma} : 
\overline{{\bf n}{\bf n}{\bf n}} \\
&+\gamma_0 v_0 \rho \frac{d-1}{d} \boldsymbol{J}[{\bf P}] - \gamma_0 v_0 \rho 
{\bf Q} \cdot \boldsymbol{J}[{\bf P}]\\
&+ \gamma_1 \rho \boldsymbol{J}[{\bf Q}] \cdot {\bf P} - \gamma_1\rho 
\boldsymbol{J}[{\bf Q}] : \overline{{\bf n}{\bf n}{\bf n}}\,.
\end{split}
\end{equation}
As seen from Eq.~\eqref{polarl}, the dynamics of the polarization (i.e., the 
first moment)
involves the second and third moment. In other words, there is again a hierarchy 
problem 
whose treatment requires a closure condition. We stress that this
comes on top of the mean-field approximation of
two-particle distribution function in the FP equation.

Not surprisingly, a similar problem arises in the evolution
equation for the second moment, which is related to the ${\bf Q}$ tensor 
according to Eq.~(\ref{qtensor}).
Multiplying the FP equation by $S{\bf n}{\bf n}$ and integrating yields
\begin{equation}
\label{2moment_1}
\begin{split}
\partial_t (\rho \overline{{\bf n}{\bf n}}) = &- \nabla\cdot \left[\rho (v_0 
\overline{{\bf n}{\bf n}{\bf n}} + \overline{{\bf u}{\bf n}{\bf n}})\right] + D  
\nabla^2(\rho \overline{{\bf n}{\bf n}}) -2 \frac{\rho}{\tau} \overline{{\bf 
n}{\bf n}} \\
&+ 2 \rho \overline{{\bf \Omega} \cdot {\bf n}{\bf n}} + 2\rho\; a_0  
\overline{({\bf I}-{\bf n}{\bf n}) \cdot {\bf \Sigma} \cdot{\bf n}{\bf n}}\\
&+ 2 \gamma_0 v_0 \rho \overline{({\bf I}-{\bf n}{\bf n})\cdot  
\boldsymbol{J}[\rho {\bf P}]{\bf n}} + \frac{\rho}{\tau}({\bf I} - 
\overline{{\bf n}{\bf n}})\\
&+ 2 \gamma_1 \rho \overline{({\bf I}-{\bf n}{\bf n}) \cdot \boldsymbol{J}[\rho 
{\bf Q}] \cdot{\bf n}{\bf n}}\,.
\end{split}
\end{equation}
We now proceed in analogy to our treatment of the equation for the 
polarization: 
First, we divide by $\rho = const$ and pull the quantities $\boldsymbol{J}[{\bf 
P}]$ and $\boldsymbol{J}[{\bf Q}]$
[defined in Eqs.~(\ref{J_act_pass})], as well as ${\bf u}$, ${\bf \Omega}$ and 
${\bf \Sigma}$ out of the overbars, yielding
\begin{equation}
\label{2moment_2}
\begin{split}
\partial_t (\overline{{\bf n}{\bf n}}) = &- \nabla\cdot \left[(v_0 
\overline{{\bf n}{\bf n}{\bf n}} + {\bf u}\overline{{\bf n}{\bf n}})\right] + D  
\nabla^2\overline{{\bf n}{\bf n}}-\frac{3}{\tau} \overline{{\bf n}{\bf n}}  + 
\frac{1}{\tau}{\bf I}\\
&+ 2 {\bf \Omega} \cdot \overline{{\bf n}{\bf n}} + 2 a_0 {\bf \Sigma}\cdot 
\overline{{\bf n}{\bf n}} - 2 a_0  {\bf \Sigma} : \overline{{\bf n}{\bf n}{\bf 
n}{\bf n}}\\
&+ 2 \gamma_0 v_0\rho \boldsymbol{J}[{\bf P}]\overline{{\bf n}} - 2 \gamma_0 
v_0\rho \boldsymbol{J}[{\bf P}] \cdot \overline{{\bf n}{\bf n}{\bf n}}\\
&+ 2 \gamma_1 \rho \boldsymbol{J}[{\bf Q}] \cdot\overline{{\bf n}{\bf n}} - 2 
\gamma_1 \rho  \boldsymbol{J}[{\bf Q}] : \overline{{\bf n}{\bf n}{\bf n}{\bf 
n}}\,.
\end{split}
\end{equation}

The nematic order parameter is related to the second moment via ${\bf Q} = 
\overline{{\bf n}{\bf n}} - {\bf I}/d$. Thus, we take the symmetric traceless 
part of the entire equation~(\ref{2moment_2}), yielding
\begin{equation}
\label{nematic1}
\begin{split}
\partial_t {\bf Q} = & - {\bf u}\cdot\nabla{\bf Q} - v_0 
(\nabla\cdot\overline{{\bf n}{\bf n}{\bf n}})^\text{ST} + D  \nabla^2{\bf Q} 
-\frac{3}{\tau} {\bf Q} \\
&+ 2 ({\bf \Omega} \cdot {\bf Q})^\text{ST} + 2 a_0({\bf \Sigma}\cdot {\bf 
Q})^\text{ST}\\
&+ \frac{2 a_0}{d}{\bf \Sigma} - 2 a_0  ({\bf \Sigma} : \overline{{\bf n}{\bf 
n}{\bf n}{\bf n}})^\text{ST}\\
&+ 2 \gamma_0 v_0\rho (\boldsymbol{J}[{\bf P}]{\bf P})^\text{ST} - 2 \gamma_0 
v_0\rho (\boldsymbol{J}[{\bf P}] \cdot \overline{{\bf n}{\bf n}{\bf 
n}})^\text{ST}\\
&+ 2 \gamma_1 \rho (\boldsymbol{J}[{\bf Q}] \cdot{\bf Q})^\text{ST} + \frac{2 
\gamma_1 \rho}{d} \boldsymbol{J}[{\bf Q}]\\
&- 2 \gamma_1 \rho (\boldsymbol{J}[{\bf Q}] : \overline{{\bf n}{\bf n}{\bf 
n}{\bf n}})^\text{ST}\,.
\end{split}
\end{equation}
Equation~(\ref{nematic1}) shows that the ${\bf Q}$-tensor dynamics involves 
moments up to fourth order. 
We note that a similar problem occurs in the theory of passive liquid 
crystals \cite{ilg1999generating,masao1981molecular,hess1976fokker}. 
At this point we postpone the discussion of suitable closure 
relations for the present microswimmer system to a later stage, but consider 
first the active stress and the Stokes equation. 

\subsubsection{Active stress}
\label{sec: active stress}
The active stress $\boldsymbol \sigma^a$ for a specific \textit{microscopic} 
configuration (depending on the orientation vectors 
${\bf N}^{\sigma}$) of the microswimmers  has already been given in 
Eq.~\eqref{e:active_stress}.
Performing the ensemble and noise average  and using the moment definitions 
(\ref{moments}) together with
Eq.~(\ref{e:def_PDF}), one finds
the \textit{average} active stress 
\be
\label{astress}
\lan{\boldsymbol \sigma}^a({\bf x}, t)\ran
\approx
&& - f_0 \bigl[\zeta_1(\gr \ovl{\bn\bn})+
\zeta_2\;\nabla\cdot  (\gr \ovl{\bn\bn\bn}) \notag \\
&&+\zeta_3\; {\nabla} {\nabla} : (\gr \ovl{\bn\bn\bn\bn}) +
\zeta_4\; \nabla \nabla \nabla : \cdot \, (\gr  
\ovl{\bn\bn\bn\bn\bn})\bigr]\notag. \\
\ee
Equation~(\ref{astress}) features again the need for a closure condition for the 
higher-order moments.

As already discussed in Sec.~\ref{micro_eqs}, there is an additional passive 
stress [see Eq.~(\ref{e:pass_stress})]
due to the elongated shape of the particles, also known in the theory of nematic 
liquid crystals. 
To leading order, the passive stress tensor is proportional to the nematic order 
parameter ${\bf Q}$~\cite{hess2015tensors},
\be
\label{pstress}
\lan{\boldsymbol \sigma}^p({\bf x}, t)\ran
\approx  \rho \vartheta {\bf Q} + ... .
\ee
The negligence of higher-order terms seems justified 
when we assume that the degree of nematic order is small. 
The parameter $\vartheta$, which has units of energy, is concentration 
dependent and drives the isotropic-nematic phase transition of microswimmers.

Combining the active and passive contributions to the average stress, i.e.,
$\lan{\boldsymbol \sigma} \ran= \lan{\boldsymbol \sigma}^a\ran+\lan{\boldsymbol 
\sigma}^p\ran$,
we can set up the averaged Stokes equation~(\ref{SE-b}), which, supplemented by 
the incompressibility condition~(\ref{SE1}), allows for the calculation of the 
average solvent flow field ${\bf u}(\bx,t)$.
We stress at this point that the experimentally observable quantity is not ${\bf 
u}$ but rather the bacterial velocity $\bf w$.
The latter field, which is typically measured through \textit{Particle Image 
Velocimetry} methods, is given by the sum of the
solvent field and the contribution from the swimmers, i.e.,
\be
\label{w_field}
{\bf w}(\bx,t)={\bf u}(\bx,t) + {\bf v}(\bx,t),
\ee
where the swimmer contribution follows
from Eq.~\eqref{LE} and (\ref{polarization}) as
\be
\label{swimmer_velocity}
{\bf v}(\bx,t) = v_0 \ovl{\bn}(\bx,t) = v_0 \bP(\bx,t).
\ee

\subsubsection{Closure relation and final equations}
\label{sec:closurefinal}
We now turn to the closure relation for the hierarchy of moments.
In the main part of this article, we focus on a relatively simple closure 
sufficient to calculate the polarization, $\bP$, and the active stress,
$\lan\boldsymbol\sigma^a\ran$, in a self-consistent manner.
This implies that all moments starting with $k=2$ have to be 
expressed via $\bP$ itself, while ${\bf Q}$ disappears as a dynamical variable. 
We note, however, that an extension
of the theory towards the ${\bf Q}$-dynamics is feasible. We sketch the 
corresponding steps
in Appendix~\ref{A:Q} [see Eq.~(\ref{A:nematic2})].

Considering Eqs.~(\ref{polarl}) and (\ref{astress}) for 
$\bP$ and $\lan\boldsymbol\sigma^a\ran$, respectively, the first step is to 
re-express the moments of order $k\ge 3$.
We here employ the so-called Hand closure \cite{hand1962theory}} (for a 
generalization, see 
Appendix~\ref{A:Closure}), which reads
\be
\label{closure_relations}
(\ovl{\bn\bn\bn})^\text{ST} =(\ovl{\bn\bn\bn \bn})^\text{ST} 
=(\ovl{\bn\bn\bn\bn\bn})^\text{ST}=0.
\ee
where $(...)^\text{ST}$ denotes again the symmetric traceless tensors. We recall 
that these are connected to the full tensorial
moments via Eqs. \eqref{bracket1} to \eqref{bracket2} (see also 
Appendix~\ref{A:Closure}).

As a second step, we have to relate the ${\bf Q}$-tensor (second moment) to the 
polarization.
Following the proposal in our previous study~\cite{heidenreich2016hydrodynamic} 
we use a relation
which is motivated by the Doi theory \cite{RienaeckerHess98, Doi} for the 
shear-induced dynamics of passive rods,
but is adapted for active microswimmer suspensions. 

The original \textit{ansatz} of Doi is given by ${\bf 
Q}   =  q (\bP \bP)^\text{ST}$, where $q$ is the strength (largest 
eigenvalue) of nematic order (for uniaxial order) 
\cite{heidenreich2009orientational}.
However, in the present context the Doi closure neglects the fact that
the active particles permanently generate flow gradients ($\propto \nabla 
\bf{u}$) on the ambient 
fluid. These should, at least locally, affect the orientational order. Indeed, 
already in passive anisotropic fluids, flow gradients (e.g., planar Couette 
flow)
have ordering effects reaching from an induced isotropic-nematic transition to 
tumbling and shear banding \cite{Hess75, Noirez,
OlmstedGoldbart90, Lindner97, Callaghan2001}. %
To include the impact of the \textit{intrinsically} generated flow gradients
in a self-consistent manner we extend the Doi closure according to
\begin{eqnarray}
\label{e:closure_conditions}
 {\bf Q} = q (\bP \bP)^\text{ST} + \lambda_K {\boldsymbol \Sigma},
\end{eqnarray}
where $\lambda_K$ may be considered as tumbling parameter (in analogy to passive 
systems). The tumbling parameter for passive hard rods can be linked to the 
aspect ratio of the 
rods; however, this relation may be different for active particles [see the 
corresponding discussion of Jeffrey tumbling below Eqs.~(\ref{LE1})]. 
In principle, the coefficients $q$ and $\lambda_K$ can be calculated as 
functions of the microscopic parameters; this is described in Appendix 
\ref{A:Q}. Starting point is the dynamical equation~(\ref{A:nematic2}) for ${\bf 
Q}$, which is then considered in the stationary case (along with some other 
approximations). The resulting static equation
for ${\bf Q}$ has exactly the form of the closure condition 
(\ref{e:closure_conditions}).

Next, we apply the closure relations (\ref{closure_relations}) and 
(\ref{e:closure_conditions})
to the active stress 
given in Eq.~(\ref{astress}). 
For a detailed discussion on closure relations see also Appendix~\ref{A:Closure}.
Using, additionally, the condition $\nabla \cdot \bP=0$ and neglecting 
higher-order derivatives (see Appendix~\ref{A:stress}) we find
\be
\label{astress_2}
\lan{\boldsymbol \sigma}^\mathrm{a}({\bf x}, t)\ran
\approx
 &&- f_0\gr \biggl\{
 \xi_1\left[q \left(\bP \bP \right)^\mathrm{(ST)} +\lambda_K \bs
\Sigma  + \f{1}{d} \bI \right]  \notag 
\\
&&+ 2\xi_2(\nabla \bP)^\mathrm{(ST)} + 2\xi_4 \nabla^2(\nabla\bP)^\mathrm{(ST)}
\biggr\},
\ee
where $\xi_1 = b_r+b_f$ and
\be
\label{eq: xi_i 3D}
\xi_2=\f{b_r^2-b_f^2}{10},\quad
\xi_3=\f{b_r^3+b_f^3}{42},\quad 
\xi_4=\f{b_r^4-b_f^4}{280},
\ee
for $d=3$ (for $d=2$ and the general case see Appendix~\ref{A:stress}).
We recall that the active stress, as well as its passive part [see 
Eq.~(\ref{pstress})], enter the average Stokes equation (\ref{SE-b}) via their 
divergence.
Inserting the resulting expressions we obtain from Eq.~(\ref{SE-b})
\be
\label{Delta_u}
\nabla^2 \bu =&&  \f{f_0\rho}{\mu_\mathrm{eff}} \biggl\{
 \xi_1 (1 - \f{\vartheta}{f_0 \xi_1})q \bP \cdot \nabla \bP  \notag \\
 &&+ ( \xi_2 + \xi_4 
\nabla^2)\nabla^2\bP\biggr\} +
\nabla p_\text{eff},
\ee
with the effective bulk viscosity 
\be
\label{mu_eff}
\mu_\text{eff} = \mu^* - \f{1}{2}f_0\rho\xi_1(1 - \f{\vartheta}{f_0 \xi_1}) 
\lambda_K
\ee 
and the effective pressure
\begin{equation}
\label{p_hat}
p_\text{eff} = \frac{1}{\mu_\mathrm{eff}}\left[p+f_0\rho\xi_1\left((1 - \frac{\vartheta}{f_0 
\xi_1}) q |\mathbf{P}|^2\right)/d \right].
\end{equation}

We now turn to the evolution equation for the polarization, Eq.~(\ref{polarl}).
Applying the closure relations Eqs.~(\ref{closure_relations}) and 
(\ref{e:closure_conditions}), neglecting various higher-order derivatives (see 
Appendix~\ref{A:Field}) and utilizing the Stokes equation in the form of 
Eq.~(\ref{Delta_u}), we finally obtain our field equation for the polarization,
\be
\label{polarization_semi_final}
\left(\partial_t + \bu \cdot \nabla  \right) \bP &= & \bs \Omega \cdot \bP + 
\kappa \bs \Sigma \cdot \bP - \lambda_0 \bP \cdot \nabla \bP \notag\\
& &+ \Gamma_2 \nabla^2 \bP + \Gamma_4 \nabla^4 \bP \notag\\
& &- \alpha \bP - \beta |\bP|^2 \bP - \nabla p^*,
\ee 
which contains derivatives up to fourth order as well as nonlinear terms in the 
field $\bP({\bf x},t)$.
Before we comment on the various parameters occurring in 
Eq.~(\ref{polarization_semi_final}),
we first note that it can be rewritten into the more intuitive and compact form
\be
\label{Res_FieldEq}
(\mathfrak{D}_t(\bu) &+& \lambda_0  \bP \cdot \nabla )\bP =  - \f{\delta}{\delta 
\bP} \mathfrak{F} (\bP)- {\nabla} p^* .
\ee 
On the left side we have introduced the derivative
\be
\label{derivative}
\mathfrak{D}_t (\bu)  \bP = (\p_t + \bu\cdot \nabla)\bP - \bs \Omega \cdot \bP - 
\kappa \bs \Sigma \cdot \bP,
\ee
which involves all couplings with the flow field. The parameter $\kappa$ is 
given by
\be
\label{kappa}
\kappa &=& \f{d }{d+2}a_0 - \gc_0 v_0
\lambda_K A_d \gr  + \f{ d}{d+2} \rho \gamma_1 \lambda_K 
A_d.
\ee
Further, the prefactor $\lambda_0$ appearing on the left side of 
Eq.~(\ref{Res_FieldEq}) is given by
\be
\label{lambda_0}
\lambda_0 &=& q v_0 \left( 1 +  \f{  \lambda_K f_0 \rho \xi_1(1 - 
\f{\vartheta}{f_0\xi_1}) }{ 2 \mu_\mathrm{eff}}\right).
\ee

The right side of the field equation~(\ref{Res_FieldEq}) contains a relaxation 
term, that is, a derivative (with respect to $\bP$)
of the functional

\begin{equation}
\label{Landau}
\mathfrak{F} (\mathbf{P},\nabla\mathbf{P},\nabla\nabla\mathbf{P})= \Phi^\mathrm{L} 
+ \frac{1}{2}\Gamma_2 (\nabla \mathbf{P})^2 -\frac{1}{2}\Gamma_4 (\nabla\nabla \mathbf{P})^2,
\end{equation}

where the first term represents the Landau potential 
\be
\label{Landau_hom}
\Phi^\mathrm{L} =\f{1}{2} \alpha |\bP|^2 + \f{1}{4} \beta|\bP|^4
\ee
with the parameters
\begin{subequations}
\label{alpha_beta}
\be
\label{alpha}
\alpha &=&  \frac{1}{\tau} -\f{d-1}{d} \gr\gc_0 v_0 A_d\\
\label{beta}
\beta&=& \left(\f{d-1}{d}\gamma_0 v_0 - \f{d-1}{d+2}\gamma_1 \right) A_d \rho q.
\ee
\end{subequations}
Inspecting these expressions for typical values
of the input constants, it turns out that $\beta$ is positive, 
while $\alpha$ can change its sign. For positive $\alpha$, the function 
$\Phi^\mathrm{L}$ 
has its minimum at $|\bP|=0$, which corresponds to a stable isotropic phase.
In contrast, $\alpha<0$ implies the occurrence of two stable minima at 
$\pm\sqrt{-\alpha/\beta}$,
i.e., a polar (orientationally ordered) phase. From Eq.~(\ref{alpha}) it is seen 
that such a change of sign of $\alpha$
occurs when the "ferromagnetic coupling" ($\propto \gc_0 v_0$) dominates over 
the rotational diffusion ($\sim \tau^{-1}$).
The finite polarization then leads via Eq.~(\ref{swimmer_velocity}) to a 
collective swimmer velocity with magnitude
\be
v_c = v_0|\bP|=v_0 \sqrt{\f{-\alpha}{\beta}}.
\ee

The appearance of the isotropic-to-polar transition is reminiscent of the
Toner-Tu model \cite{Toner_Tu}.
A significant difference, however, becomes clear when we consider the prefactors 
of the gradient terms 
in Eq.~(\ref{Landau}), which are given by
\begin{subequations}
\label{gamma_2_gamma_4}
\be
\Gamma_2&=& D + \f{d-1}{d} \gamma_0 v_0 B_d\rho - v_0 \f{f_0 \rho  \lambda_K}{ 2 
\mu_\mathrm{eff}}\xi_2, \\
\label{gamma_2}
\Gamma_4 &=& \f{d-1}{d} \gamma_0 v_0 C_d\rho - v_0 \f{f_0 \rho  \lambda_K}{ 2 
\mu_\mathrm{eff}}\xi_4.
\label{gamma_4}
\ee
\end{subequations}
The parameter $\Gamma_2$ is positive in the passive limit ($v_0=0$), but can 
become negative for a large active force 
density of the microswimmers, $f_0$, and large values of the self-propulsion 
velocity, $v_0$. 
When this happens, the homogeneous phase, where $\bP$ is a spatially constant 
vector, becomes unstable 
and mesoscale turbulence sets in. This 
behavior can be extracted from a linear stability analysis, as detailed for the 
minimal model [see Eq.~(\ref{GSHE})] in Ref.~\cite{NJP}.
In the vicinity of the threshold, the ratio of the parameters in 
Eqs.~(\ref{gamma_2_gamma_4}), specifically
\be
\Lambda = 2 \pi \sqrt{\f{2 \Gamma_4}{\Gamma_2}}
\ee
then sets the characteristic length scale of the dynamics of the system:
$\Lambda$ determines the size of the vortices.

Finally, the last term on the right side of Eq.~(\ref{Res_FieldEq}) is given by 
\be
\label{p^star}
p^* &=& \f{v_0 \lambda_K}{2\mu_\mathrm{eff}}p + |\bP|^2 \left( -\f{v_0 \lambda_K 
f_0 \rho \xi_1 q}{2\mu_\mathrm{eff}d}(1 - \f{\vartheta}{f_0\xi_1}) - \f{v_0 
q}{d} 
\right).  \notag \\
\ee

From a more universal point of view, the equation for the polarization, Eq. 
\eqref{Res_FieldEq}, may be considered as a generalized Swift-Hohenberg equation
[see Eq.~\eqref{GSHE}],
where the ordinary time derivative was replaced by the generalized derivative 
involving the solvent flow $\bu$ [see Eq.~(\ref{derivative})]. 
The latter is determined by the Stokes equation~(\ref{Delta_u}).
We recall from Eq.~(\ref{w_field}) that the overall bacterial motion is 
characterized by the field  
\begin{equation}
\label{eq: effective velocity}
\bw = v_0\bP + \bu.
\end{equation}

\section{Modelling mesoscale turbulence}

In the last section we have derived continuum equations for the polar order 
parameter and the corresponding Stokes 
equation to describe the collective dynamics of a microswimmer suspension. 
The structure of the order parameter equation is similar to that of the 
phenomenological Eq.~(\ref{GSHE}), but involves,
in addition, a coupling to the dynamics of the surrounding fluid. Indeed, 
recent experiments of \textit{B. subtilis} suspensions and detailed microswimmer 
simulations have shown that this coupling is important for a proper description 
of the system's collective dynamics if boundary conditions play a significant role~\cite{Lushi14}.

In this section, we first write the equations in rescaled form. For the sake 
of clarity, we consider a suspension with purely polar near-field interactions, corresponding to 
$\gamma_1=0$, see Eq.~\eqref{phi_pass}.
To characterize the strength of the flow's response to the activity, we 
introduce a coupling parameter and show that, 
in the limit of very weak coupling, the dynamics of the effective velocity 
$\mathbf{w}$ is adequately described by the phenomenological model 
Eq.~(\ref{GSHE}).

The coupling parameter is a function of the microscopic parameters of the 
microswimmers, 
the microswimmer force density $f_0$ (which is itself proportional to the 
self-propulsion speed $v_0$~\cite{wolgemuth2008collective}) and the effective 
viscosity $\mu_\text{eff}$, see Eq.~\eqref{mu_eff}.

As observed in recent experiments, the effective viscosity of an active 
suspension strongly 
varies with changes in the microscopic parameters and the volume fraction of 
microswimmers. In particular, Aranson and Sokolov found a significant decrease 
of $\mu_{\text{eff}}$ relative to the bare solvent viscosity for intermediate 
volume fractions of the \textit{B. subtilis} suspension~\cite{Sokolov09, 
Heines08}. 
For very high volume fractions the viscosity increases again, yielding values 
larger than the bare solvent viscosity. This surprising behavior is only 
present in active systems and can be reproduced by the present model 
(see~\cite{heidenreich2016hydrodynamic} for details).

Finally, we present numerical results for the full model that illustrate that 
the phenomenological equations can nearly be recovered in the high viscosity 
limit, if effects from boundaries can be neglected. 

\subsection{Scaling of the dynamical equations}

In order to rescale the equations, we introduce the following dimensionless quantities.
First, the persistence number $P_\text{r}$ of the microswimmers' motion 
(persistence length $\ell_\text{P}$ scaled by the microswimmer length $\ell$) is 
given by~\cite{Bechinger2016,Zoettl2016}
\begin{equation}
P_\text{r} = \ell_\text{P}/\ell = v_0 \tau/\ell.
\end{equation}
The strength of the flow's response to the activity is characterized by the dimensionless coupling parameter
\begin{equation}
c_\text{F} = \frac{f_0\rho \ell^2}{10\mu_\text{eff}v_0}.
\end{equation}
The strength of the polar interaction compared to the rotational diffusion time scale $\tau$ yields the dimensionless interaction parameter
\begin{equation}
c_\text{I} = \frac{2}{3}\tau \rho\gamma_0 v_0  A_3.
\end{equation}
For $c_\text{I}<1$ the homogeneous system relaxes to an isotropic state, for $c_\text{I}>1$ to a polar state.

We choose the microswimmers' length $\ell$ as characteristic length and the self-swimming speed $v_0$ as characteristic velocity. The corresponding time scale is given by $\ell/ v_0$. We use a tilde ($\sim$) to denote rescaled, dimensionless quantities and obtain for the gradients 
\begin{equation}
\nabla = \frac{1}{\ell} \tilde{\nabla}, \qquad \nabla^2 = \frac{1}{\ell^2} 
\tilde{\nabla}^2, \qquad \nabla^4 = \frac{1}{\ell^4} \tilde{\nabla}^4.
\end{equation} 
The velocity and its derivatives are rescaled accordingly,
\begin{equation}
\bu = v_0\tilde{\bu}, \qquad \bm{\Omega} = \frac{v_0}{\ell}\tilde{\bm{\Omega}}, \qquad \bm{\Sigma} = \frac{v_0}{\ell}\tilde{\bm{\Sigma}}.
\end{equation}
Using the same scaling for the effective microswimmer velocity [see Eq.~\ref{eq: effective velocity}] we get
\begin{equation}
\label{wScaled}
\tilde{\bw} = \bP + \tilde{\bu},
\end{equation}
which shows that the two fields $\bP$ and $\tilde{\bu}$ are now comparable in magnitude.

To simplify the microswimmer geometry, we set $b_r=\ell$ and $b_f=0$. Additionally, we assume that there is no significant passive nematic contribution to the stress ($\vartheta=0$) and negligible translational diffusion of swimmers ($D=0$). Also, we use the relations obtained for $q$ and $\lambda_K$ [see Eq.~(\ref{A:qAndLambdaK}) in Appendix~\ref{A:Q}].
With the introduced scaling, the Stokes equation [see Eq.~\eqref{Delta_u}] yields
\begin{equation}
\label{scaled_SE}
\tilde{\nabla}^2 \tilde{\mathbf{u}} = c_\text{F} \br{6 c_\text{I}\mathbf{P} 
\cdot 
\tilde{\nabla}\mathbf{P} + \tilde{\nabla}^2 \mathbf{P} + 
\frac{1}{28}\tilde{\nabla^4}\mathbf{P} } + \tilde{\nabla} \tilde{p}_\text{eff}.
\end{equation}
The  polar order parameter equation [see Eq.~\eqref{polarization_semi_final}] is given by
\begin{equation}
\label{scaledP}
\begin{split}
\frac{\partial \mathbf{P}}{\partial \tilde{t}} = 
&- \tilde{\mathbf{u}}\cdot\tilde{\nabla}\mathbf{P} 
+ \tilde{\mathbf{\Omega}} \cdot\mathbf{P} + \kappa
 \tilde{\mathbf{\Sigma}} \cdot \mathbf{P} -\tilde{\lambda}_0 
\mathbf{P}\cdot\tilde{\nabla}\mathbf{P} \\
&- \tilde{\alpha} \mathbf{P} - \tilde{\beta} |\mathbf{P}|^2 \mathbf{P} + \tilde{\Gamma}_2 
\tilde{\nabla}^2 \mathbf{P} 
+ \tilde{\Gamma}_4 \tilde{\nabla}^4 \mathbf{P} - \tilde{\nabla} \tilde{p}^{\ast},\\
\tilde{\nabla} \cdot \bP = &\ 0,
\end{split}
\end{equation}
where
\begin{equation}
\label{rescaledParameters1}
\begin{split}
\tilde{\alpha} &= \left(1 - c_\text{I}\right)/P_\text{r}, 
\quad \tilde{\beta} = \frac{3}{5}c_\text{I}^2/P_\text{r}, \\
\tilde{\Gamma}_2 &= \frac{1}{10}\left(\frac{\epsilon}{\ell}\right)^2 c_\text{I}/P_\text{r} -\frac{a_0}{15}P_\text{r}c_\text{F},
\quad \tilde{\Gamma}_4 = - \frac{a_0}{420}P_\text{r}c_\text{F}, \\
\tilde{\lambda}_0 &= \frac{3}{5}c_\text{I}\left( 1 + \frac{2}{3}a_0 P_\text{r} c_\text{F}\right), 
\quad \kappa = \frac{3}{5} a_0 \left( 1 - \frac{c_\text{I}}{3} \right),\\
\tilde{p}^{\ast} &=\frac{a_0}{15}P_\text{r} \tilde{p} -  \frac{1}{5} |\bP|^2 c_\text{I}\left( 1 + \frac{2}{3}a_0 P_\text{r}c_\text{F} \right).
\end{split}
\end{equation}
In contrast to our previous publication~\cite{heidenreich2016hydrodynamic}, the fourth order gradient term stemming from the expansion of the interaction potential (see Appendix \ref{A:MF}) is neglected here. Thus, the corresponding coefficient $\tilde{\Gamma}_4$ has only one contribution (stemming from the active stress). In doing so, the expansion of the interaction potential ends on an order that does not lead to destabilization on the mesoscopic order parameter level. Also note that, in comparison to \cite{heidenreich2016hydrodynamic}, the parameters given in Eqs.~(\ref{scaledP}) are rescaled. For the definition of the shape parameter $a_0$ see Eq.~(\ref{shape_eq}).

\subsection{Reduction to the phenomenological model}

In the introduction we have mentioned the phenomenological model Eq.~\eqref{GSHE} 
as one key motivation for 
the derivation of the field equations in the present paper. The phenomenological 
model predicts the dynamics in terms of one vector field that describes the 
effective velocity of the microswimmers in the suspension. 
In this paragraph, we will discuss if and under which requirements the 
phenomenological model can be obtained 
from the present one. 

The response of the flow to the activity scales with the dimensionless coupling 
parameter $c_\text{F}$ [see Eq.~(\ref{scaled_SE})]. This parameter is 
proportional to the ratio of active forces (exerted on the solvent by the 
microswimmers) to viscous forces. In the limit
\begin{equation}
c_\text{F} \ll 1,
\end{equation}
the Stokes equation for the net solvent velocity becomes independent of the swimmer orientations, i.e.,
\begin{equation}
\tilde \nabla ^2 \tilde \bu = \tilde \nabla \tilde p.
\end{equation}
In this case, the collective dynamics of the suspension [which is given by 
$\tilde{\bw}$, see Eq.~(\ref{wScaled})] 
depends solely on the orientational dynamics. 
This is also demonstrated later by numerical simulations. 
Following this reasoning, the phenomenological model Eq.~(\ref{GSHE}) 
is rediscovered for a low coupling parameter $c_\text{F}$ in the absence of 
external driving forces (e.g. shear flow) or boundary effects. Note that the product 
$P_\text{r}c_\text{F}$ still has to be sufficiently large compared to 
$c_\text{I}$ to obtain a mesoscale-turbulent state [see 
Eq.~(\ref{rescaledParameters1})].

\begin{figure}
\begin{center}
\begin{tabular}{c}
\includegraphics [width = 9.0 cm]{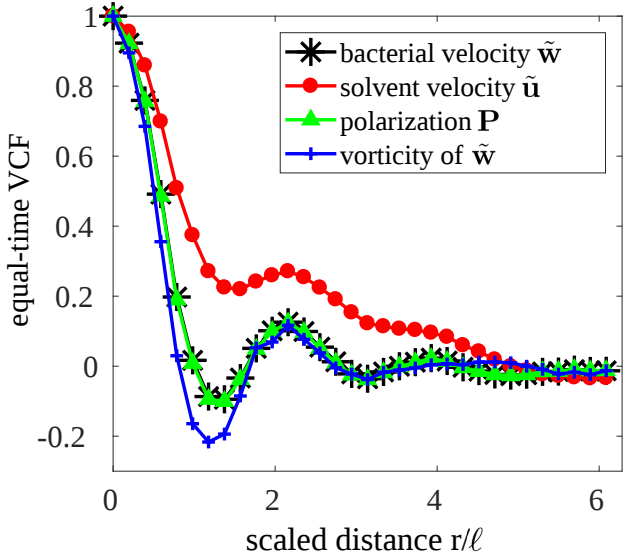}
\end{tabular}
\end{center}
\caption
{ \label{correlation} 
(Color online) Correlation functions for the solvent net flow $\tilde{\bu}$, the 
polarization $\bP$ the effective velocity $\tilde{\bw}$ und the vorticity of $\tilde{\bw}$ 
for small values of the  coupling parameter $c_\text{F} = 0.05$ 
(corresponds to large effective viscosities $\mu_\mathrm{eff}$). 
The remaining parameters are set to $c_\text{I}=2/3,\ \ell_\text{p}= 10\;\mu m,\ 
\epsilon=1\;\mu m$. The length and the diameter of the bacteria are $\ell = 5\;\mu m$ and  $d=0.8 \; \mu m$.
}
\end{figure}

Next, we will show numerical results of the full model in 
spatially 3D for small values of the coupling parameter $c_\text{F}$ (large values 
of the effective viscosity $\mu_\mathrm{eff}$) to demonstrate that in this 
limit the overall collective dynamics of the bacterial suspension is dominated 
by its orientational dynamics. 

For numerical simulations we use periodic boundary conditions and a 
pseudospectral code combined with an operator splitting technique for time 
integration where the linear operator is integrated exactly \cite{PRL}. To 
obtain the equal-time velocity correlation function, we 
consider a 2D slice in the middle of the 3D periodic cube and calculate the 
correlation functions in that plane averaged over 20 time-steps (for details 
see \cite{PRL}). 

Fig.~\ref{correlation} shows equal-time correlation functions for the scaled 
solvent flow $\tilde{\bu}$, the polarization $\bP$, the bacterial collective 
motion $\tilde{\bf w}$ and the vorticity of $\tilde{\bf w}$. The correlation 
functions for the collective bacterial motion and for the polarization collapse. This shows that, for large effective viscosities,
the collective dynamics of the bacteria is dominated by the dynamics of the 
polarization. In this limit the collective dynamics is well described by the 
phenomenological model Eq. \eqref{GSHE}. It is also seen from the correlation 
functions that the typical vortex size (characterized by ranges 
with negative values) is induced by orientational order of the 
bacteria rather than the solvent dynamics. 

To this end, the collective behavior of dense bacterial suspensions can be 
modeled by the one-fluid equation
\be
\label{GSHE1}
\left(\p_{t} +\lambda_0 \bw \cdot \nabla \right) \bw = &-&\nabla p  - \alpha 
\bw - \beta |\bw|^2 \bw \notag  \\ &+& \Gamma_2 \nabla^2
\bw + \Gamma_4 \nabla^4 \bw , \notag \\
\nabla \cdot \bw =&0&,
\ee
with parameters
\begin{equation}
\label{rescaledParameters2}
\begin{split}
\lambda_0 &= \frac{3}{5}c_\text{I}\left( 1 + \frac{2}{3}a_0 P_\text{r} 
c_\text{F}\right), \\
\alpha &= \left(1 - c_\text{I}\right)/P_\text{r}, 
\quad \beta = \frac{3}{5}c_\text{I}^2/P_\text{r}, \\
\Gamma_2 &= \frac{1}{10}\left(\frac{\epsilon}{\ell}\right)^2 
c_\text{I}/P_\text{r} -\frac{a_0}{15}P_\text{r}c_\text{F},
\quad \Gamma_4 = - \frac{a_0}{420}P_\text{r}c_\text{F}
\end{split}
\end{equation}
and abbreviations $c_\text{I} = 8\pi \rho \gamma_0 \epsilon^3 \tau v_0/9$, 
$P_\text{r} = v_0 \tau/\ell$, $c_\text{F} = f_0 \rho \ell^2/(10 
\mu_\mathrm{eff} v_0).$  
Again, time is scaled by $\ell/v_0$, space by $\ell$ and the collective velocity $\bw$ by $v_0$. 

Note that, for the derivation of the field equations, it was presumed that 
densities are large enough such that density fluctuations can be neglected. For 
dilute or semi-dilute microswimmer suspensions correlations or density fluctuations may play a 
significant role and can no longer be neglected \cite{Stenhammer2017}. Near boundaries, such as walls or other obstacles, the assumption of constant swimmer density may also loose its justification due to effects like trapping~\cite{wensink2008aggregation,Hernandez2009}. An extension of the theory to include such density variations will be presented elsewhere.

\section{Conclusions}

The present article is devoted to the systematic derivation of generic field equations for the collective behavior of a large number of microswimmers in a Newtonian fluid.

Following earlier studies \cite{hatwalne2004rheology, Shelley2}, the microswimmers are modeled by moving force dipoles that drive the ambient fluid and generate a net flow of the solvent. 
The new feature of the analysis presented here is the fact that we consider derivatives of the polar order parameter up to fourth order. Furthermore, we use a novel closure condition for the nematic order parameter that includes reorientations from active driving of the solvent flow by the microswimmers not present for passive particle systems. 
The derivation incorporates both, near field (polar and nematic) and far-field hydrodynamic interactions.
The final equations comprise a dynamical equation for the polar order parameter $\bP$ (averaged swimmer orientations) and the dynamics of the solvent flow $\bu$.
We restrict the derivation to sufficiently dense suspensions where density fluctuations play no significant role.

Parameters of the obtained field equations are directly linked to parameters of the microswimmer model. 
For example, the typical vortex size of swarming bacterial suspensions is given by $\Lambda = 2 \pi \sqrt{2 \Gamma_4/\Gamma_2}$. 
In particular, in the limit of large self-swimming speeds, we find $\Lambda \propto \ell$, i.e., the vortex size depends linearly on the microswimmers' length $\ell$.
Indeed, such a dependency was recently found in experiments of \textit{B. subtilis} suspensions \cite{ilkanaiv2017effect}. 

The coupling between the solvent flow and the orientational dynamics 
depends 
strongly on the microscopic parameters, which motivates the introduction of a 
coupling parameter as a function of the parameters of the microswimmer model. We 
distinguish between two limiting cases:

For very weak coupling, the collective dynamics depends solely on the 
orientational dynamics. 
In this limit (and without boundary effects), the phenomenological model Eq. 
\eqref{GSHE} describes the dynamics of the microswimmer suspension very well and 
the field equations derived in this paper can actually be reduced to the simpler 
model.

On the other hand, for a larger coupling parameter, the orientational 
dynamics couples to the solvent flow dynamics.
Here, the intrinsic length scale set by $\Gamma_2$ and $\Gamma_4$ has less 
influence on the dynamical patterns.

It would be very interesting to observe such a dependence of the collective 
behaviour on the coupling (effective viscosity) in experiments.

\begin{acknowledgments}
We thank Igor S. Aranson, Jaume Casademunt, J\"orn Dunkel and Robert Gro{\ss}mann for fruitful discussions.
This work was supported by the Deutsche Forschungsgemeinschaft through GRK 1558 (project B.3) and SFB 910 (projects B2 and B5).
\end{acknowledgments}

\newpage
\begin{widetext}
\appendix
\section{Mean field approximation of the interaction integral}
\label{A:MF}
According to Eq.~(\ref{C2}) in the main text, the active part of the 
"interaction integral" is given by
\be
\mathcal{C}^{(2)}[\Phi^\mathrm{act}] 
&=& (S-1)\;
{\nabla}_{\bf n} \cdot \left\{ \int d{\bf n}'\int d{\bx'}\; {\boldsymbol
\Pi (\bn)} \cdot  
\left[ {\nabla}_\bn  \phi^\mathrm{act}(\bn,\bn',|\bx-\bx'|)\right] 
 \;\mathcal{P}^{(2)}({\bf x},{\bf n};\bx ',{\bf n}',t) 
\right\}
 \notag\\
 &=&
- \gc_0 v_0 (S-1) 
{\nabla}_{\bf n} \cdot \left\{ 
{\boldsymbol \Pi(\bn)} \cdot  \int d{\bf n}'\int_{|\bx-\bx'|<\eps} d{\bx'}\; 
\bn'
 \;\mathcal{P}^{(2)}({\bf x},{\bf n};\bx ',{\bf n}',t) 
 \right\}
\ee
where the second line has been obtained by inserting the expression for the 
activity-induced polar pair 
potential given in Eq.~(\ref{phi_act}). 
We next approximate the two-particle distribution function $\mathcal{P}^{(2)}$, 
defined as
\be
\label{e:def_two_particle_density}
\mathcal{P}^{(2)}({\bf x},{\bf n};\bx ',{\bf n}',t)  =  
\frac{1}{S^2} \sum_{\mu=1}^S\sum_{\nu=1}^S  
\bigg\langle
\delta( {\bf x } - {\bf X}^\mu(t))\,\delta( {\bf n } - {\bf N}^\mu(t)) \delta( 
{\bf x }' - {\bf X}^\nu(t))\,\delta( {\bf n }' - {\bf N}^\nu(t)) 
\bigg\rangle,
\ee  
by factorization (mean-field approximation), i.e.,
$\mathcal{P}^{(2)}({\bf x},{\bf n};\bx ',{\bf n}',t)=\mathcal{P} ({\bf x},{\bf 
n},t) \mathcal{P} ({\bf x}',{\bf n}',t)$.
In the limit $S\gg 1$ the integral then becomes
\be
\label{C2_act_b}
\mathcal{C}^{(2)}[\Phi^\mathrm{act}] 
 &\approx&
- \gc_0 v_0 S {\nabla}_{\bf n} \cdot  \left\{
{\boldsymbol \Pi(\bn)} \;\mathcal{P} ({\bf x},{\bf n},t) \cdot  \int d{\bf
n}'\int_{|\bx-\bx'|<\eps} d{\bx'}\; 
\bn'
 \;\mathcal{P} (\bx ',{\bf n}',t) 
  \right\}
\notag\\
 &=&
- \gc_0 v_0
{\nabla}_{\bf n} \cdot  \left\{
{\boldsymbol \Pi(\bn)} \;\mathcal{P}({\bf x},{\bf n},t) \cdot  
\int_{|\bx-\bx'|<\eps} d{\bx'}\; \gr(\bx ',t) \bP(\bx ',t)
\right\},
 \ee
 where we have employed Eq.~(\ref{rho_def}). 
 To treat the spatial integral over the product $\rho\bP$, we transform to the 
new integration variable $\bs r =\bx-\bx'$ and perform
 a Taylor expansion up to fourth order around $\bs r=0$. This yields
 \be
\bs J[\rho \bP]  = \int_{|\bx-\bx'|<\eps} d{\bx'}\;  \gr(\bx ',t)\bP(\bx ',t) 
&=& \int_{|\bs r|<\eps} d{\bs r}\;  \gr(\bx -\bs r,t)\bP(\bx -\bs r,t) \notag\\
&\approx& \int_{|\bs r|<\eps} d{\bs r}\; \left[1 + \f{1}{2!}\bs r\bs r:\nabla\nabla  +
\f{1}{4!}\bs r\bs r\bs r\bs r::\nabla\nabla\nabla\nabla \right] \gr(\bx ,t) \bP(\bx ,t).
 \ee
 where we have truncated the expansion after the fourth-order term. 
 The gradient operators acting on the product field $\rho\bP$ can be taken out 
of the integral (since the result
 depends on $\bx$ rather than on $\bs r$). Terms
 involving odd powers of $\bs r$ then vanish by symmetry, which has already been 
taken 
 into account in writing the second line.
  The remaining integrals over even powers of $\bs r$ can be carried out 
analytically, yielding 
  the non-zero coefficients in the fourth-order expansion of $\bs J[\rho \bP]$, 
see Eqs.~(\ref{J_act}), (\ref{A2}) and (\ref{A3}) in the main text.
 Inserting the resulting expression into Eq.~(\ref{C2_act_b}) one finally 
obtains Eq.~(\ref{C2_act}).
 
The passive part of the interaction integral is treated in an analogous manner. 
We start from the definition
[from Eq.~(\ref{C2})]
\be
\mathcal{C}^{(2)}[\Phi^\mathrm{pass}] 
&=& (S-1)\;
{\nabla}_{\bf n} \cdot \left\{ \int d{\bf n}'\int d{\bx'}\; {\boldsymbol
\Pi (\bn)} \cdot  
\left[ {\nabla}_\bn  \phi^\mathrm{pass}(\bn,\bn',|\bx-\bx'|)\right] 
 \;\mathcal{P}^{(2)}({\bf x},{\bf n};\bx ',{\bf n}',t) 
\right\}
 \notag\\
 &=&
- \gc_1 (S-1) 
{\nabla}_{\bf n} \cdot \left\{ 
{\boldsymbol \Pi(\bn)} \cdot  \int d{\bf n}'\int_{|\bx-\bx'|<\eps} d{\bx'}\; 
\bn' (\bn'\cdot \bn) \;\mathcal{P}^{(2)}({\bf x},{\bf n};\bx ',{\bf n}',t) 
 \right\}
\ee
where we have used the passive (nematic) pair interaction given in 
Eq.~(\ref{phi_act}).
The mean-field approximation yields in the limit  $S\gg 1$
\be
\mathcal{C}^{(2)}[\Phi^\mathrm{pass}] 
 &\approx&
- \gc_1 S {\nabla}_{\bf n} \cdot  \left\{
{\boldsymbol \Pi(\bn)} \;\mathcal{P} ({\bf x},{\bf n},t) \cdot  \int d{\bf
n}'\int_{|\bx-\bx'|<\eps} d{\bx'}\; 
\bn' (\bn' \cdot \bn)  \;\mathcal{P} (\bx ',{\bf n}',t) 
  \right\}
\notag\\
 &=& - \gc_1
{\nabla}_{\bf n} \cdot  \left\{ {\boldsymbol \Pi(\bn)} \;\mathcal{P}({\bf 
x},{\bf n},t) \cdot  \int_{|\bx-\bx'|<\eps} d{\bx'}\; \gr(\bx ',t) \left({\bf 
Q}(\bx ',t) + \f{1}{d}\bI\right) \cdot \bn
\right\}
\notag\\
 &=& - \gc_1
{\nabla}_{\bf n} \cdot  \left\{ {\boldsymbol \Pi(\bn)} \;\mathcal{P}({\bf 
x},{\bf n},t) \cdot  \int_{|\bx-\bx'|<\eps} d{\bx'}\; \gr(\bx ',t) {\bf 
Q}(\bx ',t)  \cdot \bn
\right\},
 \ee
 where we have used that the contribution from the unit matrix in the second 
line vanishes due to the application of the projector ${\bf \Pi}$. The spatial 
integral is again treated by a Taylor expansion, that is, 
\be
\bs J[\rho {\bf Q}] = \int_{|\bx-\bx'|<\eps} d{\bx'}\;  \gr(\bx ',t){\bf Q}(\bx ',t)
\approx \int_{|\bs r|<\eps} d{\bs r}\; \left[1 + \f{1}{2!}\bs r\bs r:\nabla\nabla  +
\f{1}{4!}\bs r\bs r\bs r\bs r::\nabla\nabla\nabla\nabla \right] \gr(\bx ,t) {\bf Q}(\bx ,t).
 \ee
 The remaining integrals are the same as those calculated for the active 
contribution. 
 From this one obtains Eqs.~(\ref{J_pass}) and (\ref{C2_pass}) in the main text.

\section{Generalized closure conditions}
\label{A:Closure}

In this work we use the Hand closure [Eq.~(\ref{closure_relations})] to truncate the moment hierarchies in the equations for the polarization [see Eq.~(\ref{polarl})], the nematic order parameter [see Eq.~(\ref{nematic1})] and the active stress [see Eq.~(\ref{astress})].
This is the simplest closure that is still consistent with the system's symmetries.
For the sake of completeness, we discuss a somewhat more sophisticated closure relation in the following, the so-called "generalized quadratic closure" \cite{Kroeger2008}, of which the Hand closure and the quadratic closure are special cases.
For nematic suspensions the quadratic closure should be used in order to obtain the Landau-de Gennes potential that ensures a relaxation into the nematic phase. 

By introducing a parameter $s=0,1$ with
$s=0$ referring to the Hand closure and $s=1$ referring to the quadratic closure,
the relation for the moments of order $k\ge 3$ can be compactly written as
\be
\label{e:closure_conditions-1}
(\ovl{n_i n_j n_k})^\text{ST}  &=&    s \left(P_i Q_{jk} 
\right)^\text{ST} 
\\
(\ovl{n_i n_j n_k n_l})^\text{ST} &=&     s\left(Q_{ij} Q_{kl} 
\right)^\text{ST} 
\\
\label{e:closure_conditions-2}
(\ovl{n_i n_j n_k n_l n_m} )^\text{ST} &=&     s\left(Q_{ij} \left[\ovl{n_k 
n_l 
n_m} \right]\right)^\text{ST}.
\ee
Equations~\eqref{e:closure_conditions-1} - 
\eqref{e:closure_conditions-2} are sufficient  
to express \textit{all} higher moments in terms of the mean polarity
$\bP=\ovl{\bn}$ and the nematic order parameter ${\bf Q} = \ovl{\bn \bn} - 
\mathbf{I}/d$. Note that all tensors on the left hand side 
are symmetric and traceless. To obtain from these tensors
the corresponding full versions one may employ Eqs.~(\ref{brackets})
and (\ref{bracket2}). Specifically, for the Hand-Closure ($s=0$), we find for 
the third-order term
\be
\label{e:cr_3}
\ovl{n_i n_jn_k}(\bx,t)
&=& \f{3!}{2d+4} 
(\gd_{ij} \ovl{n_{k}})^\text{SY}
= \f{1}{d+2} 
\left(\gd_{ij} \ovl{n_{k}} +\gd_{ki} \ovl{n_{j}} +\gd_{jk} \ovl{n_{i}}\right)
\ee
and for the fourth-order term
\be
\label{e:cr_4}
\ovl{n_i n_jn_kn_l}(\bx,t)
&=& \f{4!}{4d+16}(\gd_{ij}\ovl{n_k n_l})^\text{SY}
-
\f{4!}{88d+16}(\gd_{ij}\gd_{kl})^\text{SY}
\notag\\
&=&  \f{1}{d+4}\left(
\gd_{ij}\ovl{n_k n_l}+\gd_{ik}\ovl{n_j n_l}+\gd_{il}\ovl{n_k n_j} +
\gd_{kj}\ovl{n_i n_l}+\gd_{lj}\ovl{n_k n_i}+\gd_{kl}\ovl{n_i n_j} 
\right)
-
\notag\\&&
\f{1}{11d+2}\left(
\gd_{ij}\gd_{kl}+\gd_{ik}\gd_{jl}+\gd_{il}\gd_{kj} 
\right).
\ee
Finally, the fifth-order term (which is needed for the calculation of the active 
stress) becomes
\be
\ovl{n_i n_jn_kn_l n_m}(\bx,t)
&=& \f{5!}{12 d+72}  
(\gd_{ij} \ovl{n_{k}n_{l}n_{m}})^\text{SY}
- \f{5!}{120 d+144} 
(\gd_{ij} \gd_{kl}\ovl{n_{m}})^\text{SY}
\notag\\
&=&\f{5!}{(12 d+72)(d+2)} 
(\gd_{ij}\left(\gd_{kl} \ovl{n_{m}} +\gd_{mk} \ovl{n_{l}} +\gd_{lm}
\ovl{n_{k}}\right))^\text{SY}
- \f{5!}{120 d+144} 
(\gd_{ij} \gd_{kl}\ovl{n_{m}})^\text{SY}
\notag\\
&=& \f{5!}{(4 d+24)(d+2)} 
(\gd_{ij}\gd_{kl} \ovl{n_{m}})^\text{SY}
- \f{5!}{120 d+144} 
(\gd_{ij} \gd_{kl}\ovl{n_{m}})^\text{SY}
\notag\\
&=& \left[\f{5!}{(4 d+24)(d+2)} 
- \f{5!}{120 d+144} \right]
(\gd_{ij} \gd_{kl}\ovl{n_{m}})^\text{SY}
\notag\\
&=& \left[\f{1}{(4 d+24)(d+2)} 
- \f{1}{120 d+144} \right]\times
\notag\\&&
4!\left(
\left\{\gd_{ij} \gd_{kl} \right\}_+\ovl{n_{m}}+
\left\{\gd_{mj} \gd_{kl}\right\}_+\ovl{n_{i}}+
\left\{\gd_{im} \gd_{kl}\right\}_+\ovl{n_{j}}+
\left\{\gd_{ij} \gd_{ml}\right\}_+\ovl{n_{k}}+
\left\{\gd_{ij} \gd_{km}\right\}_+\ovl{n_{l}}
\right)
\notag\\
&=& \left[\f{1}{6}\left(
\f{3}{2+ d} -\f{3}{6+ d} -\f{2}{6+5d} 
\right) \right] \times
\notag\\&&
\biggl[
\left( \gd_{ij}\gd_{kl}+\gd_{ik}\gd_{jl}+\gd_{il}\gd_{kj}  \right) 
\ovl{n_{m}} +
\left( \gd_{mj}\gd_{kl}+\gd_{mk}\gd_{jl}+\gd_{ml}\gd_{kj}  \right) 
\ovl{n_{i}} +
\notag\\&&
\left( \gd_{im}\gd_{kl}+\gd_{ik}\gd_{ml}+\gd_{il}\gd_{km}  \right) 
\ovl{n_{j}} +
\left( \gd_{ij}\gd_{ml}+\gd_{im}\gd_{jl}+\gd_{il}\gd_{mj}  \right) 
\ovl{n_{k}} +
\notag\\&&
\left( \gd_{ij}\gd_{km}+\gd_{ik}\gd_{jm}+\gd_{im}\gd_{kj}  \right) 
\ovl{n_{l}}
\biggr].
\ee

\section{Detailed derivation of the field equations}
\label{A:Field}
\subsection{Evolution equation for the polarization}
\label{A:Polarization}
We start from Eq.~(\ref{polarl}), which we display here again for clarity,
\begin{equation}
\label{A:polarl}
\begin{split}
(\partial_t + {\bf u}\cdot\nabla) {\bf P} = & {\bf \Omega} \cdot {\bf P} + a_0 
{\bf \Sigma} \cdot {\bf P} - v_0 \nabla {\bf Q} + D  \nabla^2{\bf P} 
-\frac{1}{\tau} {\bf P} - a_0 {\bf \Sigma} : \overline{{\bf n}{\bf n}{\bf n}} \\
&+\gamma_0 v_0 \rho \frac{d-1}{d} \boldsymbol{J}[{\bf P}] - \gamma_0 v_0 \rho 
{\bf Q} \cdot \boldsymbol{J}[{\bf P}]+ \gamma_1 \rho \boldsymbol{J}[{\bf Q}] 
\cdot {\bf P} - \gamma_1\rho \boldsymbol{J}[{\bf Q}] : \overline{{\bf n}{\bf 
n}{\bf n}}.
\end{split}
\end{equation}
The terms involving third-order moments of ${\bf n}$ can be treated via the Hand 
closure (with $s=0$), where we utilize
the explicit relations between full tensors and symmetric traceless parts given 
in Appendix~\ref{A:Closure}. In particular, we find with
Eq.~(\ref{e:cr_3}) in component form
\begin{equation}
\begin{split}
\Sigma_{ij} \overline{n_i n_j n_k} = \Sigma_{ij} \frac{1}{d+2} (\delta_{ij} P_k 
+ \delta_{ki} P_j + \delta_{jk} P_i)= \frac{1}{d+2} (\Sigma_{kj}  P_j + 
\Sigma_{ik}  P_i) = \frac{2}{d+2} \Sigma_{kj} P_j \,,
\end{split}
\end{equation}

\begin{equation}
\begin{split}
J[{\bf Q}]_{ij} \overline{n_i n_j n_k} = J[{\bf Q}]_{ij} \frac{1}{d+2} 
(\delta_{ij} P_k + \delta_{ki} P_j + \delta_{jk} P_i)= \frac{1}{d+2} (J[{\bf 
Q}]_{kj}  P_j + J[{\bf Q}]_{ik}  P_i)= \frac{2}{d+2} J[{\bf Q}]_{kj} P_j \,.
\end{split}
\end{equation}
Inserting these results into Eq.~(\ref{A:polarl}) yields
\begin{equation}
\label{A:polar_2}
\begin{split}
(\partial_t + {\bf u}\cdot\nabla) {\bf P} = & {\bf \Omega} \cdot {\bf P} + 
\frac{d}{d+2} a_0 {\bf \Sigma} \cdot {\bf P} - v_0 \nabla {\bf Q} + D  
\nabla^2{\bf P} -\frac{1}{\tau} {\bf P} \\
&+\gamma_0 v_0 \rho \frac{d-1}{d} \boldsymbol{J}[{\bf P}] - \gamma_0 v_0 \rho 
{\bf Q} \cdot \boldsymbol{J}[{\bf P}]+  \frac{d}{d+2}\gamma_1 \rho 
\boldsymbol{J}[{\bf Q}] \cdot {\bf P}.
\end{split}
\end{equation}
As a next step, we employ the closure relation for ${\bf Q}$ 
(see Eq.~(\ref{e:closure_conditions}) in the main text), that is,
${\bf Q} = q \left(\bP \bP \right)^\text{ST} + \lambda_K \bs \Sigma$. The 
divergence of ${\bf Q}$ then follows as
\be
\label{divQ}
\nabla \cdot {\bf Q} = q\left( \bP \cdot \nabla \bP - \f{1}{d} 
\nabla |\bP|^2\right) + 
\f{\lambda_K}{2} \nabla^2 \bu.
\ee
Additionally, we insert the explicit forms of the quantities ${{\bs J}[{\bf 
P}]}$ and ${{\bs J}[{\bf Q}]}$ [see Eqs.~(\ref{J_act_pass}] in the main text). 
Further, the products of the ${\bf Q}$-tensor and the quantities $\bs J$ become
\be
\label{Q_times_J}
{\bf Q} \cdot \bs J[\bP] &=& q \; \bP \bP \cdot \bs J[\bP] - \f{q}{d} |\bP|^2 
\bs J[\bP] + \lambda_K \bs \Sigma \cdot \bs J[\bP] \notag \\
&=&  \f{d-1}{d} q A_d |\bP|^2 \bP + q \left(\bP \cdot 
\mathcal{D} \bP\right) \bP - \f{q}{d} |\bP|^2 \left(\mathcal{D} \bP \right) + 
\lambda_K \bs \Sigma \cdot \bs J[\bP],
\ee
and
\be
\label{J_times_P}
 \bs J[{\bf Q}] \cdot \bP = A_d {\bf Q} 
 \cdot \bP +  \bP \cdot \mathcal{D} {\bf Q},
 \ee
where we have introduced the operator $\mathcal{D} := B_d \nabla^2 + C_d 
\nabla^4$ including the constants defined in Eqs.~(\ref{A2}) and (\ref{A3}) for 
$d=2,3$, respectively.
Inserting Eqs.~(\ref{J_act_pass}), (\ref{divQ}), (\ref{Q_times_J}) and 
(\ref{J_times_P}) into Eq.~(\ref{A:polar_2}) we obtain
\be
\label{A:p}
\left(\partial_t + \bu \cdot \nabla  \right) \bP = &-& v_0 \f{\lambda_K}{2} 
\nabla^2 \bu + \bs \Omega \cdot \bP + \kappa \bs \Sigma \cdot \bP - v_0 q \bP 
\cdot \nabla \bP + \widetilde{\Gamma}_2 \nabla^2 \bP + \widetilde{\Gamma}_4 
\nabla^4 \bP - 
\alpha \bP - \beta |\bP|^2 \bP
+ \f{v_0}{d} q \nabla|\bP|^2 \notag \\
&-& \gamma_0 v_0 q \rho \left(\bP \cdot 
\mathcal{D} \bP  \right)^\text{ST}  \bP - \gamma_0 v_0 \lambda_K \rho \bs 
\Sigma \cdot \mathcal{D} \bP 
+ \gamma_1 \f{d}{d+2} q \rho \bP \cdot \left(
\mathcal{D} \bP \bP\right)^\text{ST} 
+ \gamma_1 \f{d}{d+2}  \lambda_K \rho 
\bP \cdot \mathcal{D}\bs \Sigma,
\ee
where
\be
\kappa &=& \f{d}{d+2} a_0 
+ \f{d}{d+2} \gamma_1 A_d \lambda_K \rho 
-\gamma_0 v_0 A_d \lambda_K \rho \\
\alpha &=& \f{1}{\tau} - \f{d-1}{d} \gamma_0 v_0 A_d \rho , \\
\beta &=& \left(\gamma_0 v_0 \f{d-1}{d} - \gamma_1 \f{d-1}{d+2} \right) 
A_d q \rho \\
\widetilde{\Gamma}_2 &=& D + \f{d-1}{d} \gamma_0 v_0 B_d,\\
\widetilde{\Gamma}_4 &=& \f{d-1}{d} \gamma_0 v_0 C_d.
\ee

Equation~(\ref{A:p}) can be somewhat simplified when we neglect the higher-order 
derivatives
contained in the operator $\mathcal{D}$. This yields
\be
\label{A:p_2}
\left(\partial_t + \bu \cdot \nabla  \right) \bP = &-& v_0 \f{\lambda_K}{2} 
\nabla^2 \bu + \bs \Omega \cdot \bP + \kappa \bs \Sigma \cdot \bP - v_0 q \bP 
\cdot \nabla \bP + \widetilde{\Gamma}_2 \nabla^2 \bP + \widetilde{\Gamma}_4 
\nabla^4 \bP + 
\alpha \bP + \beta |\bP|^2 \bP 
+ \f{v_0}{d} q \nabla|\bP|^2. \notag \\
\ee 
Equation~(\ref{A:p_2}) still contains the flow field $\bu$ and its derivatives.
\subsection{Active stress and flow field expressed via the polarization}
\label{A:stress}
To proceed from Eq.~(\ref{A:p_2}), we now consider the average active stress 
given in Eq.~(\ref{astress}).
Using the Hand closure and the fact that, for constant density $\rho$, 
$\bP=\ovl{{\bf n}}$ is source-free [see Eq.~(\ref{divP})], that is, 
$\nabla\cdot\ovl{{\bf n}}=0$, we find
\be
\lan{\boldsymbol \sigma}^a({\bf x}, t)\ran
&\approx 
 - f_0\gr \bigl[&\zeta_1\ovl{\bn\bn}+
\zeta_2\;\nabla\cdot  (\ovl{\bn\bn\bn}) +
\zeta_3\; {\nabla} {\nabla} : (\ovl{\bn\bn\bn\bn}) +
\zeta_4\; \nabla \nabla \nabla : \cdot \, (\ovl{\bn\bn\bn\bn\bn})\bigr]
\notag\\
&=
 - f_0\gr \bigl\{&\xi_1\ovl{\bn\bn}+
\xi_2 \left[\nabla \ovl{\bn} +(\nabla\ovl{\bn})^\top\right]
 +\xi_3 \left[\bI \nabla\nabla :\ovl{\bn\bn}+2 (\nabla \nabla\cdot \ovl{\bn\bn})
 +2(\nabla \nabla\cdot \ovl{\bn\bn})^\top + \nabla^2 \ovl{\bn\bn}
\right]
 \notag\\
 &&+\xi_4 \nabla^2
\left[
\nabla \ovl{\bn} +(\nabla\ovl{\bn})^\top
\right]
\bigr\},
\label{e:sigma_closure}
\ee
where $\xi_1 = \zeta_1=b_r+b_f$ and
\be
\xi_2=\f{\zeta_2}{d+2}
\csp
\xi_3=\f{\zeta_3}{d+4} 
\csp
\xi_4=
\f{\zeta_4\;
}{2}\left(
\f{3}{2+ d} -\f{3}{6+ d} -\f{2}{6+5d} 
\right),
\ee
or explicitly in 2D (for $d=3$ see Eqs.~(\ref{eq: xi_i 3D}) in 
Sec.~\ref{sec:closurefinal})
\be
\xi_2=\f{b_r^2-b_f^2}{8}
\csp
\xi_3=\f{b_r^3+b_f^3}{36} 
\csp
\xi_4=
\f{b_r^4-b_f^4}{192}. 
\ee

Next, we re-express the second moment by the ${\bf Q}$-tensor and apply the
closure relation for ${\bf Q}$ [see Eq.~(\ref{e:closure_conditions})].
We then obtain
\be
\label{A:stress_1}
\lan{\boldsymbol \sigma^a}({\bf x}, t)\ran
&\approx
 - f_0\gr \biggl\{
 &\xi_1 {\bf Q}
+2\xi_2(\nabla \bP)^\text{ST} + 2\xi_4 \nabla^2(\nabla\bP)^\text{ST}
+ \; \xi_3
\biggr\{\nabla^2 {\bf Q} + 4 \left( \nabla \nabla \cdot {\bf Q} 
\right)^\text{ST}  +\mathrm{Tr}\left[\nabla \nabla \cdot {\bf Q} \right]\bI  
\biggr\}\biggl\} \notag \\
 &= - f_0 \rho \biggl\{&\xi_1 \left(q \left(\bP \bP \right)^\text{ST}+ 
\lambda_K \bs \Sigma +\f{1}{d} \bI  \right) + 2\xi_2 \left(\nabla \bP 
\right)^\text{ST}+ 2\xi_4 \nabla^2 \left(\nabla \bP 
\right)^\text{ST} \notag \\
&&+ \xi_3 \left( \f{3}{2} \lambda_K \nabla^2 \bs 
\Sigma + q \nabla^2 \left(\bP \bP  \right)^\text{ST} + q \left( \nabla \nabla 
\cdot (\bP \bP)^\text{ST} \right)^\text{ST}\right) \biggl\}.
\label{e:sigma_closure_simple}
\ee

In what follows we neglect again nonlinear terms involving higher derivatives. 
We further assume 
that the strain rate varies only slowly in space, and thus,
$\bs \nabla^2 \Sigma\approx 0$. 
Equation~(\ref{A:stress_1}) then becomes
\be
\label{A:stress_2}
\bs \sigma^a = -f_0 \rho \biggl\{\xi_1 \left( q \left(\bP \bP 
\right)^\text{ST} + \lambda_K \bs \Sigma + \f{1}{d} \bI \right) + 
2\xi_2 \left(\nabla \bP \right)^\text{ST}+2\xi_4 \nabla^2 \left(\nabla \bP 
\right)^\text{ST}\biggl\}.
\ee
This equation coincides with Eq.~(\ref{astress_2}) in the main text.
We now take the divergence of Eq.~(\ref{A:stress_2}) and do the same for the 
passive stress, Eq.~(\ref{pstress}). Within the latter,
we also use the closure relation for ${\bf Q}$.
Inserting the resulting terms into the Stokes equation and solving for the 
velocity field
we obtain (see Eq.~(\ref{Delta_u}) in the main text)
\be
\label{A:u}
\nabla^2 \bu = \f{ 1}{\mu_\mathrm{eff} } \left( f_0 \rho \xi_1(1 - 
\f{\vartheta}{f_0\xi_1}) q
\bP \cdot \nabla \bP +f_0 \rho \xi_2 \nabla^2 \bP + f_0 \rho \xi_4 \nabla^4 \bP 
\right) + 
\nabla 
p_\text{eff}.
\ee
Inserting Eq.~\eqref{A:u} into \eqref{A:p_2} we finally obtain the field 
equation for the polarization (see Eq.~(\ref{polarization_semi_final}) in the 
main text).

\section{Evolution equation for the $\bf Q$-tensor and generalized Doi closure}
\label{A:Q}

To derive a self-consistent evolution equation for the $\bf Q$-tensor we start 
from Eq.~(\ref{nematic1}), which we display here again for clarity,
\begin{equation}
\label{A:nematic}
\begin{split}
\partial_t {\bf Q} = & - {\bf u}\cdot\nabla{\bf Q} - v_0 
(\nabla\cdot\overline{{\bf n}{\bf n}{\bf n}})^\text{ST} + D  \nabla^2{\bf Q} 
-\frac{3}{\tau} {\bf Q} + 2 ({\bf \Omega} \cdot {\bf Q})^\text{ST} + 2 a_0({\bf 
\Sigma}\cdot {\bf Q})^\text{ST}\\
&+ \frac{2 a_0}{d}{\bf \Sigma} - 2 a_0  ({\bf \Sigma} : \overline{{\bf n}{\bf 
n}{\bf n}{\bf n}})^\text{ST} + 2 \gamma_0 v_0\rho (\boldsymbol{J}[{\bf P}]{\bf 
P})^\text{ST} - 2 \gamma_0 v_0\rho (\boldsymbol{J}[{\bf P}] \cdot \overline{{\bf 
n}{\bf n}{\bf n}})^\text{ST}\\
&+ 2 \gamma_1 \rho (\boldsymbol{J}[{\bf Q}] \cdot{\bf Q})^\text{ST} + \frac{2 
\gamma_1 \rho}{d} \boldsymbol{J}[{\bf Q}]- 2 \gamma_1 \rho (\boldsymbol{J}[{\bf 
Q}] : \overline{{\bf n}{\bf n}{\bf n}{\bf n}})^\text{ST}\,.
\end{split}
\end{equation}
The terms involving third- and fourth-order moments of ${\bf n}$ can be treated 
via the Hand closure (with $s=0$), where we utilize the explicit relations 
between full tensors and symmetric traceless parts given in 
Appendix~\ref{A:Closure}. In particular, we find with
Eq.~(\ref{e:cr_3}) in component form
\be
\left(\f{\partial}{\partial x_i}\ovl{n_i n_j n_k}\right)^\text{ST}
&=& \f{1}{d+2} 
\biggl(\f{\partial}{\partial x_i} (\gd_{ij} P_k +\gd_{ki} P_j +\gd_{jk} P_i 
)\biggr)^\text{ST} = \f{2}{d+2}\left(\f{\partial P_k}{\partial 
x_j}\right)^\text{ST},
\ee
and similarly
\be
\left(J[{\bf P}]_{i}\ovl{n_i n_j n_k}\right)^\text{ST}
&=& \f{1}{d+2} 
\biggl(J[{\bf P}]_{i} (\gd_{ij} P_k +\gd_{ki} P_j +\gd_{jk} P_i 
)\biggr)^\text{ST} = \f{2}{d+2}\left(J[{\bf P}]_{j}P_k\right)^\text{ST}.
\ee
With Eq.~(\ref{e:cr_4}) we further find
\be
\left(\Sigma_{ij}\ovl{n_i n_j n_k n_l}\right)^\text{ST} &=& 
\left(\Sigma_{ij}\biggl[\frac{1}{d+4}
\biggl\{ \delta_{ij} Q_{kl} +
\frac{1}{d} \delta_{ij} \delta_{kl}
 + \delta_{ik} Q_{jl} + \frac{1}{d} \delta_{ik} \delta_{jl} \right. \notag \\
 &&+ \delta_{il} Q_{kj} + \frac{1}{d} \delta_{il}\delta_{kj} + \delta_{kj}
Q_{il} +\frac{1}{d} \delta_{kj}\delta_{il} 
 +   \delta_{lj}Q_{kj} + \frac{1}{d} \delta_{lj} \delta_{ki} \notag\\
 && \left. + \delta_{kl} Q_{ij} + \frac{1}{d} \delta_{kl}\delta_{ij}  \biggl\} 
- \frac{1}{11d+2} (\delta_{ij}\delta_{kl} + \delta_{ik}\delta_{jl} + \delta_{il}
\delta_{kj})\biggl]\right)^\text{ST} \notag \\
&=& \f{4}{d+4} \left(\Sigma_{kj}Q_{jl} \right)^\text{ST} - \left( 
\frac{2}{2+11d} - \frac{4}{d(d+4)}  \right)\Sigma_{kl},
\ee
and similarly
\be
\left(J[{\bf Q}]_{ij}\ovl{n_i n_j n_k n_l}\right)^\text{ST}
&=& \f{4}{d+4} \left(J[{\bf Q}]_{kj}Q_{jl}\right)^\text{ST} - 
\left(\frac{2}{2+11d} - \frac{4}{d(d+4)} \right) J[{\bf Q}]_{kl}.
\ee
Inserting these relations into Eq.~(\ref{A:nematic}) yields
\be
\label{A:nematic2}
\p_t{\bf Q} =  &-& \bu \nabla \cdot {\bf Q} + 2 \bs \Omega \cdot {\bf Q} +D \nabla^2 {\bf Q} - \f{2}{d+2} v_0 \left(\nabla \bP 
\right)^\text{ST}-\f{3}{\tau}{\bf Q} + \frac{2 a_0}{d+2} \bs \Sigma + \f{d}{d+4} 2 a_0 \left(\bs \Sigma \cdot {\bf Q}\right)^\text{ST} \notag \\ 
&+& \frac{d }{d+2} 2 \gamma_0 v_0 \rho \left(\bs J[{\bf P}]{\bf 
P}\right)^\text{ST} + \frac{2\gamma_1 \rho}{d+2} {\bs J[{\bf Q}]} + \f{d}{d+4} 2 \gamma_1 \rho\left(\bs J[{\bf Q}] \cdot {\bf 
Q}\right)^\text{ST} ,
\ee
where we have used the relation
\be
\f{1}{d}+\f{2}{2+11d}-\f{4}{d(d+4)} = \frac{1}{d+2},
\ee
which is valid for $d=2,3$, i.e., only for systems in two- or three-dimensional space.

Eq.~(\ref{A:nematic2}) already provides a self-consistent relation for the time 
evolution of the $\bf Q$-tensor for given polarization and flow field.
For active nematics, higher moments have to be approximated by the quadratic 
closure (see Appendix~\ref{A:Closure}). 

We now proceed towards a derivation of the generalized Doi closure in 
Eq.~(\ref{e:closure_conditions}). To this end we insert the explicit expressions 
for the quantities ${{\bs J}[{\bf P}]}$ and ${{\bs J}[{\bf Q}]}$ [see 
Eqs.~(\ref{J_act_pass})] into Eq.~(\ref{A:nematic2}). Henceforth, we only keep 
terms of linear order in $\bf Q$ and $\bf \Sigma$ and quadratic order in $\bf 
P$. Further, we neglect all gradient terms. We now assume that the dynamics of 
$\bf Q$ is much faster than the dynamics of the polar order parameter $\bf{P}$, 
such that $\p_t{\bf Q}=0$. With these assumptions, we obtain from 
Eq.~(\ref{A:nematic2})
\be
{\bf Q} = q \left(\bP \bP\right)^\text{ST} + \lambda_K \bs \Sigma,
\ee
which has exactly the form of the generalized Doi 
closure~(\ref{e:closure_conditions}). Within this derivation the coefficients 
$q$ and $\lambda_K$ are calculated via
\be
q = \left( 2 v_0 \gamma_0 A_d \rho \f{d}{d+2}\right) \bigg/ \left( \f{3}{\tau} - \frac{2\gamma_1 \rho A_d}{d+2}\right)
\ee
and
\be
\lambda_K = \left( \frac{2 a_0}{d+2} \right) \bigg/ \left( \f{3}{\tau} - \frac{2\gamma_1 \rho A_d}{d+2}\right),
\ee
or explicitly in 3D
\be
\label{A:qAndLambdaK}
q = \f{v_0\gamma_0 A_3\rho\tau}{5/2 - \gamma_1\rho A_3\tau/3}
\qquad \text{and} \qquad
\lambda_K = \f{a_0\tau}{15/2-\gamma_1\rho A_3\tau}.
\ee

\end{widetext}

\bibliographystyle{apsrev4-1}
\bibliography{Derivation_P_Q_resubmitted}

\begin{thebibliography}{96}%
\makeatletter
\providecommand \@ifxundefined [1]{%
 \@ifx{#1\undefined}
}%
\providecommand \@ifnum [1]{%
 \ifnum #1\expandafter \@firstoftwo
 \else \expandafter \@secondoftwo
 \fi
}%
\providecommand \@ifx [1]{%
 \ifx #1\expandafter \@firstoftwo
 \else \expandafter \@secondoftwo
 \fi
}%
\providecommand \natexlab [1]{#1}%
\providecommand \enquote  [1]{``#1''}%
\providecommand \bibnamefont  [1]{#1}%
\providecommand \bibfnamefont [1]{#1}%
\providecommand \citenamefont [1]{#1}%
\providecommand \href@noop [0]{\@secondoftwo}%
\providecommand \href [0]{\begingroup \@sanitize@url \@href}%
\providecommand \@href[1]{\@@startlink{#1}\@@href}%
\providecommand \@@href[1]{\endgroup#1\@@endlink}%
\providecommand \@sanitize@url [0]{\catcode `\\12\catcode `\$12\catcode
  `\&12\catcode `\#12\catcode `\^12\catcode `\_12\catcode `\%12\relax}%
\providecommand \@@startlink[1]{}%
\providecommand \@@endlink[0]{}%
\providecommand \url  [0]{\begingroup\@sanitize@url \@url }%
\providecommand \@url [1]{\endgroup\@href {#1}{\urlprefix }}%
\providecommand \urlprefix  [0]{URL }%
\providecommand \Eprint [0]{\href }%
\providecommand \doibase [0]{http://dx.doi.org/}%
\providecommand \selectlanguage [0]{\@gobble}%
\providecommand \bibinfo  [0]{\@secondoftwo}%
\providecommand \bibfield  [0]{\@secondoftwo}%
\providecommand \translation [1]{[#1]}%
\providecommand \BibitemOpen [0]{}%
\providecommand \bibitemStop [0]{}%
\providecommand \bibitemNoStop [0]{.\EOS\space}%
\providecommand \EOS [0]{\spacefactor3000\relax}%
\providecommand \BibitemShut  [1]{\csname bibitem#1\endcsname}%
\let\auto@bib@innerbib\@empty
\bibitem [{\citenamefont {Sumino}\ \emph {et~al.}(2012)\citenamefont {Sumino},
  \citenamefont {Nagai}, \citenamefont {Shitaka}, \citenamefont {Tanaka},
  \citenamefont {Yoshikawa}, \citenamefont {Chat{\'e}},\ and\ \citenamefont
  {Oiwa}}]{Sumino_Nature}%
  \BibitemOpen
  \bibfield  {author} {\bibinfo {author} {\bibfnamefont {Y.}~\bibnamefont
  {Sumino}}, \bibinfo {author} {\bibfnamefont {K.~H.}\ \bibnamefont {Nagai}},
  \bibinfo {author} {\bibfnamefont {Y.}~\bibnamefont {Shitaka}}, \bibinfo
  {author} {\bibfnamefont {D.}~\bibnamefont {Tanaka}}, \bibinfo {author}
  {\bibfnamefont {K.}~\bibnamefont {Yoshikawa}}, \bibinfo {author}
  {\bibfnamefont {H.}~\bibnamefont {Chat{\'e}}}, \ and\ \bibinfo {author}
  {\bibfnamefont {K.}~\bibnamefont {Oiwa}},\ }\href@noop {} {\bibfield
  {journal} {\bibinfo  {journal} {Nature}\ }\textbf {\bibinfo {volume} {483}},\
  \bibinfo {pages} {448} (\bibinfo {year} {2012})}\BibitemShut {NoStop}%
\bibitem [{\citenamefont {Schaller}\ \emph {et~al.}(2010)\citenamefont
  {Schaller}, \citenamefont {Weber}, \citenamefont {Semmrich}, \citenamefont
  {Frey},\ and\ \citenamefont {Bausch}}]{Schaller_Nature}%
  \BibitemOpen
  \bibfield  {author} {\bibinfo {author} {\bibfnamefont {V.}~\bibnamefont
  {Schaller}}, \bibinfo {author} {\bibfnamefont {C.}~\bibnamefont {Weber}},
  \bibinfo {author} {\bibfnamefont {C.}~\bibnamefont {Semmrich}}, \bibinfo
  {author} {\bibfnamefont {E.}~\bibnamefont {Frey}}, \ and\ \bibinfo {author}
  {\bibfnamefont {A.~R.}\ \bibnamefont {Bausch}},\ }\href@noop {} {\bibfield
  {journal} {\bibinfo  {journal} {Nature}\ }\textbf {\bibinfo {volume} {467}},\
  \bibinfo {pages} {73} (\bibinfo {year} {2010})}\BibitemShut {NoStop}%
\bibitem [{\citenamefont {Nishiguchi}\ and\ \citenamefont
  {Sano}(2015)}]{Nishiguchi2015}%
  \BibitemOpen
  \bibfield  {author} {\bibinfo {author} {\bibfnamefont {D.}~\bibnamefont
  {Nishiguchi}}\ and\ \bibinfo {author} {\bibfnamefont {M.}~\bibnamefont
  {Sano}},\ }\href@noop {} {\bibfield  {journal} {\bibinfo  {journal} {Phys.
  Rev. E}\ }\textbf {\bibinfo {volume} {92}},\ \bibinfo {pages} {052309}
  (\bibinfo {year} {2015})}\BibitemShut {NoStop}%
\bibitem [{\citenamefont {Rabani}\ \emph {et~al.}(2013)\citenamefont {Rabani},
  \citenamefont {Ariel},\ and\ \citenamefont {Be'er}}]{Rabani2013}%
  \BibitemOpen
  \bibfield  {author} {\bibinfo {author} {\bibfnamefont {A.}~\bibnamefont
  {Rabani}}, \bibinfo {author} {\bibfnamefont {G.}~\bibnamefont {Ariel}}, \
  and\ \bibinfo {author} {\bibfnamefont {A.}~\bibnamefont {Be'er}},\
  }\href@noop {} {\bibfield  {journal} {\bibinfo  {journal} {PLOS ONE}\
  }\textbf {\bibinfo {volume} {8}},\ \bibinfo {pages} {1} (\bibinfo {year}
  {2013})}\BibitemShut {NoStop}%
\bibitem [{\citenamefont {Buttinoni}\ \emph {et~al.}(2013)\citenamefont
  {Buttinoni}, \citenamefont {Bialk{\'e}}, \citenamefont {K{\"u}mmel},
  \citenamefont {L{\"o}wen}, \citenamefont {Bechinger},\ and\ \citenamefont
  {Speck}}]{buttinoni2013dynamical}%
  \BibitemOpen
  \bibfield  {author} {\bibinfo {author} {\bibfnamefont {I.}~\bibnamefont
  {Buttinoni}}, \bibinfo {author} {\bibfnamefont {J.}~\bibnamefont
  {Bialk{\'e}}}, \bibinfo {author} {\bibfnamefont {F.}~\bibnamefont
  {K{\"u}mmel}}, \bibinfo {author} {\bibfnamefont {H.}~\bibnamefont
  {L{\"o}wen}}, \bibinfo {author} {\bibfnamefont {C.}~\bibnamefont
  {Bechinger}}, \ and\ \bibinfo {author} {\bibfnamefont {T.}~\bibnamefont
  {Speck}},\ }\href@noop {} {\bibfield  {journal} {\bibinfo  {journal} {Phys.
  Rev. Lett.}\ }\textbf {\bibinfo {volume} {110}},\ \bibinfo {pages} {238301}
  (\bibinfo {year} {2013})}\BibitemShut {NoStop}%
\bibitem [{\citenamefont {Peruani}\ \emph {et~al.}(2012)\citenamefont
  {Peruani}, \citenamefont {Starru{\ss}}, \citenamefont {Jakovljevic},
  \citenamefont {S{\o}gaard-Andersen}, \citenamefont {Deutsch},\ and\
  \citenamefont {B{\"a}r}}]{peruani2012collective}%
  \BibitemOpen
  \bibfield  {author} {\bibinfo {author} {\bibfnamefont {F.}~\bibnamefont
  {Peruani}}, \bibinfo {author} {\bibfnamefont {J.}~\bibnamefont
  {Starru{\ss}}}, \bibinfo {author} {\bibfnamefont {V.}~\bibnamefont
  {Jakovljevic}}, \bibinfo {author} {\bibfnamefont {L.}~\bibnamefont
  {S{\o}gaard-Andersen}}, \bibinfo {author} {\bibfnamefont {A.}~\bibnamefont
  {Deutsch}}, \ and\ \bibinfo {author} {\bibfnamefont {M.}~\bibnamefont
  {B{\"a}r}},\ }\href@noop {} {\bibfield  {journal} {\bibinfo  {journal} {Phys.
  Rev. Lett.}\ }\textbf {\bibinfo {volume} {108}},\ \bibinfo {pages} {098102}
  (\bibinfo {year} {2012})}\BibitemShut {NoStop}%
\bibitem [{\citenamefont {Simha}\ and\ \citenamefont
  {Ramaswamy}(2002)}]{Ramaswamy}%
  \BibitemOpen
  \bibfield  {author} {\bibinfo {author} {\bibfnamefont {R.~A.}\ \bibnamefont
  {Simha}}\ and\ \bibinfo {author} {\bibfnamefont {S.}~\bibnamefont
  {Ramaswamy}},\ }\href@noop {} {\bibfield  {journal} {\bibinfo  {journal}
  {Phys. Rev. Lett.}\ }\textbf {\bibinfo {volume} {89}},\ \bibinfo {pages}
  {058101} (\bibinfo {year} {2002})}\BibitemShut {NoStop}%
\bibitem [{\citenamefont {Zhang}\ \emph {et~al.}(2010)\citenamefont {Zhang},
  \citenamefont {Be'er}, \citenamefont {Florin},\ and\ \citenamefont
  {Swinney}}]{sw10}%
  \BibitemOpen
  \bibfield  {author} {\bibinfo {author} {\bibfnamefont {H.-P.}\ \bibnamefont
  {Zhang}}, \bibinfo {author} {\bibfnamefont {A.}~\bibnamefont {Be'er}},
  \bibinfo {author} {\bibfnamefont {E.-L.}\ \bibnamefont {Florin}}, \ and\
  \bibinfo {author} {\bibfnamefont {H.~L.}\ \bibnamefont {Swinney}},\
  }\href@noop {} {\bibfield  {journal} {\bibinfo  {journal} {Proc. Natl. Acad.
  Sci. U.S.A.}\ }\textbf {\bibinfo {volume} {107}},\ \bibinfo {pages} {13626}
  (\bibinfo {year} {2010})}\BibitemShut {NoStop}%
\bibitem [{\citenamefont {Schwarz-Linek}\ \emph {et~al.}(2012)\citenamefont
  {Schwarz-Linek}, \citenamefont {Valeriani}, \citenamefont {Cacciuto},
  \citenamefont {Cates}, \citenamefont {Marenduzzo}, \citenamefont {Morozov},\
  and\ \citenamefont {Poon}}]{schwarz2012phase}%
  \BibitemOpen
  \bibfield  {author} {\bibinfo {author} {\bibfnamefont {J.}~\bibnamefont
  {Schwarz-Linek}}, \bibinfo {author} {\bibfnamefont {C.}~\bibnamefont
  {Valeriani}}, \bibinfo {author} {\bibfnamefont {A.}~\bibnamefont {Cacciuto}},
  \bibinfo {author} {\bibfnamefont {M.}~\bibnamefont {Cates}}, \bibinfo
  {author} {\bibfnamefont {D.}~\bibnamefont {Marenduzzo}}, \bibinfo {author}
  {\bibfnamefont {A.}~\bibnamefont {Morozov}}, \ and\ \bibinfo {author}
  {\bibfnamefont {W.}~\bibnamefont {Poon}},\ }\href@noop {} {\bibfield
  {journal} {\bibinfo  {journal} {Proc. Natl. Acad. Sci. U.S.A.}\ }\textbf
  {\bibinfo {volume} {109}},\ \bibinfo {pages} {4052} (\bibinfo {year}
  {2012})}\BibitemShut {NoStop}%
\bibitem [{\citenamefont {Wittkowski}\ \emph {et~al.}(2014)\citenamefont
  {Wittkowski}, \citenamefont {Tiribocchi}, \citenamefont {Stenhammar},
  \citenamefont {Allen}, \citenamefont {Marenduzzo},\ and\ \citenamefont
  {Cates}}]{wittkowski2014scalar}%
  \BibitemOpen
  \bibfield  {author} {\bibinfo {author} {\bibfnamefont {R.}~\bibnamefont
  {Wittkowski}}, \bibinfo {author} {\bibfnamefont {A.}~\bibnamefont
  {Tiribocchi}}, \bibinfo {author} {\bibfnamefont {J.}~\bibnamefont
  {Stenhammar}}, \bibinfo {author} {\bibfnamefont {R.~J.}\ \bibnamefont
  {Allen}}, \bibinfo {author} {\bibfnamefont {D.}~\bibnamefont {Marenduzzo}}, \
  and\ \bibinfo {author} {\bibfnamefont {M.~E.}\ \bibnamefont {Cates}},\
  }\href@noop {} {\bibfield  {journal} {\bibinfo  {journal} {Nat. Commun.}\
  }\textbf {\bibinfo {volume} {5}},\ \bibinfo {pages} {4351} (\bibinfo {year}
  {2014})}\BibitemShut {NoStop}%
\bibitem [{\citenamefont {Speck}\ \emph {et~al.}(2014)\citenamefont {Speck},
  \citenamefont {Bialk{\'e}}, \citenamefont {Menzel},\ and\ \citenamefont
  {L{\"o}wen}}]{speck2014effective}%
  \BibitemOpen
  \bibfield  {author} {\bibinfo {author} {\bibfnamefont {T.}~\bibnamefont
  {Speck}}, \bibinfo {author} {\bibfnamefont {J.}~\bibnamefont {Bialk{\'e}}},
  \bibinfo {author} {\bibfnamefont {A.~M.}\ \bibnamefont {Menzel}}, \ and\
  \bibinfo {author} {\bibfnamefont {H.}~\bibnamefont {L{\"o}wen}},\ }\href@noop
  {} {\bibfield  {journal} {\bibinfo  {journal} {Phys. Rev. Lett.}\ }\textbf
  {\bibinfo {volume} {112}},\ \bibinfo {pages} {218304} (\bibinfo {year}
  {2014})}\BibitemShut {NoStop}%
\bibitem [{\citenamefont {Dombrowski}\ \emph {et~al.}(2004)\citenamefont
  {Dombrowski}, \citenamefont {Cisneros}, \citenamefont {Chatkaew},
  \citenamefont {Goldstein},\ and\ \citenamefont
  {Kessler}}]{dombrowski2004self}%
  \BibitemOpen
  \bibfield  {author} {\bibinfo {author} {\bibfnamefont {C.}~\bibnamefont
  {Dombrowski}}, \bibinfo {author} {\bibfnamefont {L.}~\bibnamefont
  {Cisneros}}, \bibinfo {author} {\bibfnamefont {S.}~\bibnamefont {Chatkaew}},
  \bibinfo {author} {\bibfnamefont {R.~E.}\ \bibnamefont {Goldstein}}, \ and\
  \bibinfo {author} {\bibfnamefont {J.~O.}\ \bibnamefont {Kessler}},\
  }\href@noop {} {\bibfield  {journal} {\bibinfo  {journal} {Phys. Rev. Lett.}\
  }\textbf {\bibinfo {volume} {93}},\ \bibinfo {pages} {098103} (\bibinfo
  {year} {2004})}\BibitemShut {NoStop}%
\bibitem [{\citenamefont {Bratanov}\ \emph {et~al.}(2015)\citenamefont
  {Bratanov}, \citenamefont {Jenko},\ and\ \citenamefont
  {Frey}}]{bratanov2015new}%
  \BibitemOpen
  \bibfield  {author} {\bibinfo {author} {\bibfnamefont {V.}~\bibnamefont
  {Bratanov}}, \bibinfo {author} {\bibfnamefont {F.}~\bibnamefont {Jenko}}, \
  and\ \bibinfo {author} {\bibfnamefont {E.}~\bibnamefont {Frey}},\ }\href@noop
  {} {\bibfield  {journal} {\bibinfo  {journal} {Proc. Natl. Acad. Sci.
  U.S.A.}\ }\textbf {\bibinfo {volume} {112}},\ \bibinfo {pages} {15048}
  (\bibinfo {year} {2015})}\BibitemShut {NoStop}%
\bibitem [{\citenamefont {Oza}\ \emph {et~al.}(2016)\citenamefont {Oza},
  \citenamefont {Heidenreich},\ and\ \citenamefont
  {Dunkel}}]{oza2016generalized}%
  \BibitemOpen
  \bibfield  {author} {\bibinfo {author} {\bibfnamefont {A.~U.}\ \bibnamefont
  {Oza}}, \bibinfo {author} {\bibfnamefont {S.}~\bibnamefont {Heidenreich}}, \
  and\ \bibinfo {author} {\bibfnamefont {J.}~\bibnamefont {Dunkel}},\
  }\href@noop {} {\bibfield  {journal} {\bibinfo  {journal} {Eur. Phys. J. E}\
  }\textbf {\bibinfo {volume} {39}},\ \bibinfo {pages} {97} (\bibinfo {year}
  {2016})}\BibitemShut {NoStop}%
\bibitem [{\citenamefont {Heidenreich}\ \emph {et~al.}(2014)\citenamefont
  {Heidenreich}, \citenamefont {Klapp},\ and\ \citenamefont
  {B{\"a}r}}]{heidenreich2014numerical}%
  \BibitemOpen
  \bibfield  {author} {\bibinfo {author} {\bibfnamefont {S.}~\bibnamefont
  {Heidenreich}}, \bibinfo {author} {\bibfnamefont {S.~H.}\ \bibnamefont
  {Klapp}}, \ and\ \bibinfo {author} {\bibfnamefont {M.}~\bibnamefont
  {B{\"a}r}},\ }\href@noop {} {\bibfield  {journal} {\bibinfo  {journal} {J.
  Phys. Conf. Ser.}\ }\textbf {\bibinfo {volume} {490}},\ \bibinfo {pages}
  {012126} (\bibinfo {year} {2014})}\BibitemShut {NoStop}%
\bibitem [{\citenamefont {Ramaswamy}(2010)}]{Ramaswamy2010}%
  \BibitemOpen
  \bibfield  {author} {\bibinfo {author} {\bibfnamefont {S.}~\bibnamefont
  {Ramaswamy}},\ }\href@noop {} {\bibfield  {journal} {\bibinfo  {journal}
  {Annu. Rev. Condens. Matter Phys.}\ }\textbf {\bibinfo {volume} {1}},\
  \bibinfo {pages} {323} (\bibinfo {year} {2010})}\BibitemShut {NoStop}%
\bibitem [{\citenamefont {Romanczuk}\ \emph {et~al.}(2012)\citenamefont
  {Romanczuk}, \citenamefont {B{\"a}r}, \citenamefont {Ebeling}, \citenamefont
  {Lindner},\ and\ \citenamefont {Schimansky-Geier}}]{Romanczuk2012}%
  \BibitemOpen
  \bibfield  {author} {\bibinfo {author} {\bibfnamefont {P.}~\bibnamefont
  {Romanczuk}}, \bibinfo {author} {\bibfnamefont {M.}~\bibnamefont {B{\"a}r}},
  \bibinfo {author} {\bibfnamefont {W.}~\bibnamefont {Ebeling}}, \bibinfo
  {author} {\bibfnamefont {B.}~\bibnamefont {Lindner}}, \ and\ \bibinfo
  {author} {\bibfnamefont {L.}~\bibnamefont {Schimansky-Geier}},\ }\href@noop
  {} {\bibfield  {journal} {\bibinfo  {journal} {Eur. Phys. J. Spec. Top.}\
  }\textbf {\bibinfo {volume} {202}},\ \bibinfo {pages} {1} (\bibinfo {year}
  {2012})}\BibitemShut {NoStop}%
\bibitem [{\citenamefont {Marchetti}\ \emph {et~al.}(2013)\citenamefont
  {Marchetti}, \citenamefont {Joanny}, \citenamefont {Ramaswamy}, \citenamefont
  {Liverpool}, \citenamefont {Prost}, \citenamefont {Rao},\ and\ \citenamefont
  {Simha}}]{Marchetti2013}%
  \BibitemOpen
  \bibfield  {author} {\bibinfo {author} {\bibfnamefont {M.}~\bibnamefont
  {Marchetti}}, \bibinfo {author} {\bibfnamefont {J.}~\bibnamefont {Joanny}},
  \bibinfo {author} {\bibfnamefont {S.}~\bibnamefont {Ramaswamy}}, \bibinfo
  {author} {\bibfnamefont {T.}~\bibnamefont {Liverpool}}, \bibinfo {author}
  {\bibfnamefont {J.}~\bibnamefont {Prost}}, \bibinfo {author} {\bibfnamefont
  {M.}~\bibnamefont {Rao}}, \ and\ \bibinfo {author} {\bibfnamefont {R.~A.}\
  \bibnamefont {Simha}},\ }\href@noop {} {\bibfield  {journal} {\bibinfo
  {journal} {Rev. Mod. Phys.}\ }\textbf {\bibinfo {volume} {85}},\ \bibinfo
  {pages} {1143} (\bibinfo {year} {2013})}\BibitemShut {NoStop}%
\bibitem [{\citenamefont {Elgeti}\ \emph {et~al.}(2015)\citenamefont {Elgeti},
  \citenamefont {Winkler},\ and\ \citenamefont {Gompper}}]{Elgeti2015}%
  \BibitemOpen
  \bibfield  {author} {\bibinfo {author} {\bibfnamefont {J.}~\bibnamefont
  {Elgeti}}, \bibinfo {author} {\bibfnamefont {R.~G.}\ \bibnamefont {Winkler}},
  \ and\ \bibinfo {author} {\bibfnamefont {G.}~\bibnamefont {Gompper}},\
  }\href@noop {} {\bibfield  {journal} {\bibinfo  {journal} {Rep. Prog. Phys.}\
  }\textbf {\bibinfo {volume} {78}},\ \bibinfo {pages} {056601} (\bibinfo
  {year} {2015})}\BibitemShut {NoStop}%
\bibitem [{\citenamefont {Bechinger}\ \emph {et~al.}(2016)\citenamefont
  {Bechinger}, \citenamefont {Di~Leonardo}, \citenamefont {L\"owen},
  \citenamefont {Reichhardt}, \citenamefont {Volpe},\ and\ \citenamefont
  {Volpe}}]{Bechinger2016}%
  \BibitemOpen
  \bibfield  {author} {\bibinfo {author} {\bibfnamefont {C.}~\bibnamefont
  {Bechinger}}, \bibinfo {author} {\bibfnamefont {R.}~\bibnamefont
  {Di~Leonardo}}, \bibinfo {author} {\bibfnamefont {H.}~\bibnamefont
  {L\"owen}}, \bibinfo {author} {\bibfnamefont {C.}~\bibnamefont {Reichhardt}},
  \bibinfo {author} {\bibfnamefont {G.}~\bibnamefont {Volpe}}, \ and\ \bibinfo
  {author} {\bibfnamefont {G.}~\bibnamefont {Volpe}},\ }\href@noop {}
  {\bibfield  {journal} {\bibinfo  {journal} {Rev. Mod. Phys.}\ }\textbf
  {\bibinfo {volume} {88}},\ \bibinfo {pages} {045006} (\bibinfo {year}
  {2016})}\BibitemShut {NoStop}%
\bibitem [{\citenamefont {Z\"{o}ttl}\ and\ \citenamefont
  {Stark}(2016)}]{Zoettl2016}%
  \BibitemOpen
  \bibfield  {author} {\bibinfo {author} {\bibfnamefont {A.}~\bibnamefont
  {Z\"{o}ttl}}\ and\ \bibinfo {author} {\bibfnamefont {H.}~\bibnamefont
  {Stark}},\ }\href@noop {} {\bibfield  {journal} {\bibinfo  {journal} {J.
  Phys. Condens. Matter}\ }\textbf {\bibinfo {volume} {28}},\ \bibinfo {pages}
  {253001} (\bibinfo {year} {2016})}\BibitemShut {NoStop}%
\bibitem [{\citenamefont {Menzel}(2015)}]{Menzel2015}%
  \BibitemOpen
  \bibfield  {author} {\bibinfo {author} {\bibfnamefont {A.~M.}\ \bibnamefont
  {Menzel}},\ }\href@noop {} {\bibfield  {journal} {\bibinfo  {journal} {Phys.
  Rep.}\ }\textbf {\bibinfo {volume} {554}},\ \bibinfo {pages} {1 } (\bibinfo
  {year} {2015})}\BibitemShut {NoStop}%
\bibitem [{\citenamefont {Peshkov}\ \emph
  {et~al.}(2012{\natexlab{a}})\citenamefont {Peshkov}, \citenamefont {Aranson},
  \citenamefont {Bertin}, \citenamefont {Chat{\'e}},\ and\ \citenamefont
  {Ginelli}}]{peshkov2012nonlinear}%
  \BibitemOpen
  \bibfield  {author} {\bibinfo {author} {\bibfnamefont {A.}~\bibnamefont
  {Peshkov}}, \bibinfo {author} {\bibfnamefont {I.~S.}\ \bibnamefont
  {Aranson}}, \bibinfo {author} {\bibfnamefont {E.}~\bibnamefont {Bertin}},
  \bibinfo {author} {\bibfnamefont {H.}~\bibnamefont {Chat{\'e}}}, \ and\
  \bibinfo {author} {\bibfnamefont {F.}~\bibnamefont {Ginelli}},\ }\href@noop
  {} {\bibfield  {journal} {\bibinfo  {journal} {Phys. Rev. Lett.}\ }\textbf
  {\bibinfo {volume} {109}},\ \bibinfo {pages} {268701} (\bibinfo {year}
  {2012}{\natexlab{a}})}\BibitemShut {NoStop}%
\bibitem [{\citenamefont {Peshkov}\ \emph
  {et~al.}(2012{\natexlab{b}})\citenamefont {Peshkov}, \citenamefont {Ngo},
  \citenamefont {Bertin}, \citenamefont {Chat{\'e}},\ and\ \citenamefont
  {Ginelli}}]{peshkov2012continuous}%
  \BibitemOpen
  \bibfield  {author} {\bibinfo {author} {\bibfnamefont {A.}~\bibnamefont
  {Peshkov}}, \bibinfo {author} {\bibfnamefont {S.}~\bibnamefont {Ngo}},
  \bibinfo {author} {\bibfnamefont {E.}~\bibnamefont {Bertin}}, \bibinfo
  {author} {\bibfnamefont {H.}~\bibnamefont {Chat{\'e}}}, \ and\ \bibinfo
  {author} {\bibfnamefont {F.}~\bibnamefont {Ginelli}},\ }\href@noop {}
  {\bibfield  {journal} {\bibinfo  {journal} {Phys. Rev. Lett.}\ }\textbf
  {\bibinfo {volume} {109}},\ \bibinfo {pages} {098101} (\bibinfo {year}
  {2012}{\natexlab{b}})}\BibitemShut {NoStop}%
\bibitem [{\citenamefont {Baskaran}\ and\ \citenamefont
  {Marchetti}(2008)}]{MarchettiP}%
  \BibitemOpen
  \bibfield  {author} {\bibinfo {author} {\bibfnamefont {A.}~\bibnamefont
  {Baskaran}}\ and\ \bibinfo {author} {\bibfnamefont {M.~C.}\ \bibnamefont
  {Marchetti}},\ }\href@noop {} {\bibfield  {journal} {\bibinfo  {journal}
  {Phys. Rev. E}\ }\textbf {\bibinfo {volume} {77}},\ \bibinfo {pages} {011920}
  (\bibinfo {year} {2008})}\BibitemShut {NoStop}%
\bibitem [{\citenamefont {Ahmadi}\ \emph {et~al.}(2005)\citenamefont {Ahmadi},
  \citenamefont {Liverpool},\ and\ \citenamefont {Marchetti}}]{MarchettiQ}%
  \BibitemOpen
  \bibfield  {author} {\bibinfo {author} {\bibfnamefont {A.}~\bibnamefont
  {Ahmadi}}, \bibinfo {author} {\bibfnamefont {T.~B.}\ \bibnamefont
  {Liverpool}}, \ and\ \bibinfo {author} {\bibfnamefont {M.~C.}\ \bibnamefont
  {Marchetti}},\ }\href@noop {} {\bibfield  {journal} {\bibinfo  {journal}
  {Phys. Rev. E}\ }\textbf {\bibinfo {volume} {72}},\ \bibinfo {pages} {060901}
  (\bibinfo {year} {2005})}\BibitemShut {NoStop}%
\bibitem [{\citenamefont {Grossmann}\ \emph {et~al.}(2012)\citenamefont
  {Grossmann}, \citenamefont {Schimansky-Geier},\ and\ \citenamefont
  {Romanczuk}}]{grossmann2012active}%
  \BibitemOpen
  \bibfield  {author} {\bibinfo {author} {\bibfnamefont {R.}~\bibnamefont
  {Grossmann}}, \bibinfo {author} {\bibfnamefont {L.}~\bibnamefont
  {Schimansky-Geier}}, \ and\ \bibinfo {author} {\bibfnamefont
  {P.}~\bibnamefont {Romanczuk}},\ }\href@noop {} {\bibfield  {journal}
  {\bibinfo  {journal} {New J. Phys.}\ }\textbf {\bibinfo {volume} {14}},\
  \bibinfo {pages} {073033} (\bibinfo {year} {2012})}\BibitemShut {NoStop}%
\bibitem [{\citenamefont {Menzel}\ \emph {et~al.}(2016)\citenamefont {Menzel},
  \citenamefont {Saha}, \citenamefont {Hoell},\ and\ \citenamefont
  {L{\"o}wen}}]{menzel2016dynamical}%
  \BibitemOpen
  \bibfield  {author} {\bibinfo {author} {\bibfnamefont {A.~M.}\ \bibnamefont
  {Menzel}}, \bibinfo {author} {\bibfnamefont {A.}~\bibnamefont {Saha}},
  \bibinfo {author} {\bibfnamefont {C.}~\bibnamefont {Hoell}}, \ and\ \bibinfo
  {author} {\bibfnamefont {H.}~\bibnamefont {L{\"o}wen}},\ }\href@noop {}
  {\bibfield  {journal} {\bibinfo  {journal} {J. Chem. Phys.}\ }\textbf
  {\bibinfo {volume} {144}},\ \bibinfo {pages} {024115} (\bibinfo {year}
  {2016})}\BibitemShut {NoStop}%
\bibitem [{\citenamefont {Saintillan}\ and\ \citenamefont
  {Shelley}(2008)}]{Shelley1}%
  \BibitemOpen
  \bibfield  {author} {\bibinfo {author} {\bibfnamefont {D.}~\bibnamefont
  {Saintillan}}\ and\ \bibinfo {author} {\bibfnamefont {M.~J.}\ \bibnamefont
  {Shelley}},\ }\href@noop {} {\bibfield  {journal} {\bibinfo  {journal} {Phys.
  Rev. Lett.}\ }\textbf {\bibinfo {volume} {100}},\ \bibinfo {pages} {178103}
  (\bibinfo {year} {2008})}\BibitemShut {NoStop}%
\bibitem [{\citenamefont {Saintillan}\ and\ \citenamefont
  {Shelley}(2009)}]{Shelley2}%
  \BibitemOpen
  \bibfield  {author} {\bibinfo {author} {\bibfnamefont {D.}~\bibnamefont
  {Saintillan}}\ and\ \bibinfo {author} {\bibfnamefont {M.~J.}\ \bibnamefont
  {Shelley}},\ }\href@noop {} {\bibfield  {journal} {\bibinfo  {journal} {C. R.
  Physique}\ }\textbf {\bibinfo {volume} {14}},\ \bibinfo {pages} {497}
  (\bibinfo {year} {2009})}\BibitemShut {NoStop}%
\bibitem [{\citenamefont {Toner}\ and\ \citenamefont {Tu}(1998)}]{Toner_Tu}%
  \BibitemOpen
  \bibfield  {author} {\bibinfo {author} {\bibfnamefont {J.}~\bibnamefont
  {Toner}}\ and\ \bibinfo {author} {\bibfnamefont {Y.}~\bibnamefont {Tu}},\
  }\href@noop {} {\bibfield  {journal} {\bibinfo  {journal} {Phys. Rev. E}\
  }\textbf {\bibinfo {volume} {58}},\ \bibinfo {pages} {4828} (\bibinfo {year}
  {1998})}\BibitemShut {NoStop}%
\bibitem [{\citenamefont {Gro\ss{}mann}\ \emph {et~al.}(2014)\citenamefont
  {Gro\ss{}mann}, \citenamefont {Romanczuk}, \citenamefont {B\"ar},\ and\
  \citenamefont {Schimansky-Geier}}]{Grossmann2014}%
  \BibitemOpen
  \bibfield  {author} {\bibinfo {author} {\bibfnamefont {R.}~\bibnamefont
  {Gro\ss{}mann}}, \bibinfo {author} {\bibfnamefont {P.}~\bibnamefont
  {Romanczuk}}, \bibinfo {author} {\bibfnamefont {M.}~\bibnamefont {B\"ar}}, \
  and\ \bibinfo {author} {\bibfnamefont {L.}~\bibnamefont {Schimansky-Geier}},\
  }\href@noop {} {\bibfield  {journal} {\bibinfo  {journal} {Phys. Rev. Lett.}\
  }\textbf {\bibinfo {volume} {113}},\ \bibinfo {pages} {258104} (\bibinfo
  {year} {2014})}\BibitemShut {NoStop}%
\bibitem [{\citenamefont {Gro{\ss}mann}\ \emph {et~al.}(2015)\citenamefont
  {Gro{\ss}mann}, \citenamefont {Romanczuk}, \citenamefont {B{\"a}r},\ and\
  \citenamefont {Schimansky-Geier}}]{Grossmann2015}%
  \BibitemOpen
  \bibfield  {author} {\bibinfo {author} {\bibfnamefont {R.}~\bibnamefont
  {Gro{\ss}mann}}, \bibinfo {author} {\bibfnamefont {P.}~\bibnamefont
  {Romanczuk}}, \bibinfo {author} {\bibfnamefont {M.}~\bibnamefont {B{\"a}r}},
  \ and\ \bibinfo {author} {\bibfnamefont {L.}~\bibnamefont
  {Schimansky-Geier}},\ }\href@noop {} {\bibfield  {journal} {\bibinfo
  {journal} {Eur. Phys. J. Spec. Top.}\ }\textbf {\bibinfo {volume} {224}},\
  \bibinfo {pages} {1325} (\bibinfo {year} {2015})}\BibitemShut {NoStop}%
\bibitem [{\citenamefont {Sanchez}\ \emph {et~al.}(2012)\citenamefont
  {Sanchez}, \citenamefont {Chen}, \citenamefont {DeCamp}, \citenamefont
  {Heymann},\ and\ \citenamefont {Dogic}}]{sanchez2012spontaneous}%
  \BibitemOpen
  \bibfield  {author} {\bibinfo {author} {\bibfnamefont {T.}~\bibnamefont
  {Sanchez}}, \bibinfo {author} {\bibfnamefont {D.~T.}\ \bibnamefont {Chen}},
  \bibinfo {author} {\bibfnamefont {S.~J.}\ \bibnamefont {DeCamp}}, \bibinfo
  {author} {\bibfnamefont {M.}~\bibnamefont {Heymann}}, \ and\ \bibinfo
  {author} {\bibfnamefont {Z.}~\bibnamefont {Dogic}},\ }\href@noop {}
  {\bibfield  {journal} {\bibinfo  {journal} {Nature}\ }\textbf {\bibinfo
  {volume} {491}},\ \bibinfo {pages} {431} (\bibinfo {year}
  {2012})}\BibitemShut {NoStop}%
\bibitem [{\citenamefont {Zhou}\ \emph {et~al.}(2014)\citenamefont {Zhou},
  \citenamefont {Sokolov}, \citenamefont {Lavrentovich},\ and\ \citenamefont
  {Aranson}}]{zhou2014living}%
  \BibitemOpen
  \bibfield  {author} {\bibinfo {author} {\bibfnamefont {S.}~\bibnamefont
  {Zhou}}, \bibinfo {author} {\bibfnamefont {A.}~\bibnamefont {Sokolov}},
  \bibinfo {author} {\bibfnamefont {O.~D.}\ \bibnamefont {Lavrentovich}}, \
  and\ \bibinfo {author} {\bibfnamefont {I.~S.}\ \bibnamefont {Aranson}},\
  }\href@noop {} {\bibfield  {journal} {\bibinfo  {journal} {Proc. Natl. Acad.
  Sci. U.S.A.}\ }\textbf {\bibinfo {volume} {111}},\ \bibinfo {pages} {1265}
  (\bibinfo {year} {2014})}\BibitemShut {NoStop}%
\bibitem [{\citenamefont {Genkin}\ \emph {et~al.}(2017)\citenamefont {Genkin},
  \citenamefont {Sokolov}, \citenamefont {Lavrentovich},\ and\ \citenamefont
  {Aranson}}]{genkin2017topological}%
  \BibitemOpen
  \bibfield  {author} {\bibinfo {author} {\bibfnamefont {M.~M.}\ \bibnamefont
  {Genkin}}, \bibinfo {author} {\bibfnamefont {A.}~\bibnamefont {Sokolov}},
  \bibinfo {author} {\bibfnamefont {O.~D.}\ \bibnamefont {Lavrentovich}}, \
  and\ \bibinfo {author} {\bibfnamefont {I.~S.}\ \bibnamefont {Aranson}},\
  }\href@noop {} {\bibfield  {journal} {\bibinfo  {journal} {Phys. Rev. X}\
  }\textbf {\bibinfo {volume} {7}},\ \bibinfo {pages} {011029} (\bibinfo {year}
  {2017})}\BibitemShut {NoStop}%
\bibitem [{\citenamefont {Hemingway}\ \emph {et~al.}(2016)\citenamefont
  {Hemingway}, \citenamefont {Mishra}, \citenamefont {Marchetti},\ and\
  \citenamefont {Fielding}}]{hemingway2016correlation}%
  \BibitemOpen
  \bibfield  {author} {\bibinfo {author} {\bibfnamefont {E.~J.}\ \bibnamefont
  {Hemingway}}, \bibinfo {author} {\bibfnamefont {P.}~\bibnamefont {Mishra}},
  \bibinfo {author} {\bibfnamefont {M.~C.}\ \bibnamefont {Marchetti}}, \ and\
  \bibinfo {author} {\bibfnamefont {S.~M.}\ \bibnamefont {Fielding}},\
  }\href@noop {} {\bibfield  {journal} {\bibinfo  {journal} {Soft Matter}\
  }\textbf {\bibinfo {volume} {12}},\ \bibinfo {pages} {7943} (\bibinfo {year}
  {2016})}\BibitemShut {NoStop}%
\bibitem [{\citenamefont {Doostmohammadi}\ \emph {et~al.}(2016)\citenamefont
  {Doostmohammadi}, \citenamefont {Adamer}, \citenamefont {Thampi},\ and\
  \citenamefont {Yeomans}}]{doostmohammadi2016stabilization}%
  \BibitemOpen
  \bibfield  {author} {\bibinfo {author} {\bibfnamefont {A.}~\bibnamefont
  {Doostmohammadi}}, \bibinfo {author} {\bibfnamefont {M.~F.}\ \bibnamefont
  {Adamer}}, \bibinfo {author} {\bibfnamefont {S.~P.}\ \bibnamefont {Thampi}},
  \ and\ \bibinfo {author} {\bibfnamefont {J.~M.}\ \bibnamefont {Yeomans}},\
  }\href@noop {} {\bibfield  {journal} {\bibinfo  {journal} {Nat. Commun.}\
  }\textbf {\bibinfo {volume} {7}},\ \bibinfo {pages} {10557} (\bibinfo {year}
  {2016})}\BibitemShut {NoStop}%
\bibitem [{\citenamefont {Wensink}\ \emph {et~al.}(2012)\citenamefont
  {Wensink}, \citenamefont {Dunkel}, \citenamefont {Heidenreich}, \citenamefont
  {Drescher}, \citenamefont {Goldstein}, \citenamefont {L{\"o}wen},\ and\
  \citenamefont {Yeomans}}]{PNAS}%
  \BibitemOpen
  \bibfield  {author} {\bibinfo {author} {\bibfnamefont {H.~H.}\ \bibnamefont
  {Wensink}}, \bibinfo {author} {\bibfnamefont {J.}~\bibnamefont {Dunkel}},
  \bibinfo {author} {\bibfnamefont {S.}~\bibnamefont {Heidenreich}}, \bibinfo
  {author} {\bibfnamefont {K.}~\bibnamefont {Drescher}}, \bibinfo {author}
  {\bibfnamefont {R.~E.}\ \bibnamefont {Goldstein}}, \bibinfo {author}
  {\bibfnamefont {H.}~\bibnamefont {L{\"o}wen}}, \ and\ \bibinfo {author}
  {\bibfnamefont {J.~M.}\ \bibnamefont {Yeomans}},\ }\href@noop {} {\bibfield
  {journal} {\bibinfo  {journal} {Proc. Natl. Acad. Sci. U.S.A.}\ }\textbf
  {\bibinfo {volume} {109}},\ \bibinfo {pages} {14308} (\bibinfo {year}
  {2012})}\BibitemShut {NoStop}%
\bibitem [{\citenamefont {Vicsek}\ \emph {et~al.}(1995)\citenamefont {Vicsek},
  \citenamefont {Czir{\'o}k}, \citenamefont {Ben-Jacob}, \citenamefont
  {Cohen},\ and\ \citenamefont {Shochet}}]{vicsek1995novel}%
  \BibitemOpen
  \bibfield  {author} {\bibinfo {author} {\bibfnamefont {T.}~\bibnamefont
  {Vicsek}}, \bibinfo {author} {\bibfnamefont {A.}~\bibnamefont {Czir{\'o}k}},
  \bibinfo {author} {\bibfnamefont {E.}~\bibnamefont {Ben-Jacob}}, \bibinfo
  {author} {\bibfnamefont {I.}~\bibnamefont {Cohen}}, \ and\ \bibinfo {author}
  {\bibfnamefont {O.}~\bibnamefont {Shochet}},\ }\href@noop {} {\bibfield
  {journal} {\bibinfo  {journal} {Phys. Rev. Lett.}\ }\textbf {\bibinfo
  {volume} {75}},\ \bibinfo {pages} {1226} (\bibinfo {year}
  {1995})}\BibitemShut {NoStop}%
\bibitem [{\citenamefont {Dunkel}\ \emph
  {et~al.}(2013{\natexlab{a}})\citenamefont {Dunkel}, \citenamefont
  {Heidenreich}, \citenamefont {Drescher}, \citenamefont {Wensink},
  \citenamefont {B{\"a}r},\ and\ \citenamefont {Goldstein}}]{PRL}%
  \BibitemOpen
  \bibfield  {author} {\bibinfo {author} {\bibfnamefont {J.}~\bibnamefont
  {Dunkel}}, \bibinfo {author} {\bibfnamefont {S.}~\bibnamefont {Heidenreich}},
  \bibinfo {author} {\bibfnamefont {K.}~\bibnamefont {Drescher}}, \bibinfo
  {author} {\bibfnamefont {H.~H.}\ \bibnamefont {Wensink}}, \bibinfo {author}
  {\bibfnamefont {M.}~\bibnamefont {B{\"a}r}}, \ and\ \bibinfo {author}
  {\bibfnamefont {R.~E.}\ \bibnamefont {Goldstein}},\ }\href@noop {} {\bibfield
   {journal} {\bibinfo  {journal} {Phys. Rev. Lett.}\ }\textbf {\bibinfo
  {volume} {110}},\ \bibinfo {pages} {228102} (\bibinfo {year}
  {2013}{\natexlab{a}})}\BibitemShut {NoStop}%
\bibitem [{\citenamefont {Dunkel}\ \emph
  {et~al.}(2013{\natexlab{b}})\citenamefont {Dunkel}, \citenamefont
  {Heidenreich}, \citenamefont {B{\"a}r},\ and\ \citenamefont
  {Goldstein}}]{NJP}%
  \BibitemOpen
  \bibfield  {author} {\bibinfo {author} {\bibfnamefont {J.}~\bibnamefont
  {Dunkel}}, \bibinfo {author} {\bibfnamefont {S.}~\bibnamefont {Heidenreich}},
  \bibinfo {author} {\bibfnamefont {M.}~\bibnamefont {B{\"a}r}}, \ and\
  \bibinfo {author} {\bibfnamefont {R.~E.}\ \bibnamefont {Goldstein}},\
  }\href@noop {} {\bibfield  {journal} {\bibinfo  {journal} {New J. Phys.}\
  }\textbf {\bibinfo {volume} {15}},\ \bibinfo {pages} {045016} (\bibinfo
  {year} {2013}{\natexlab{b}})}\BibitemShut {NoStop}%
\bibitem [{\citenamefont {Zhang}\ \emph {et~al.}(2009)\citenamefont {Zhang},
  \citenamefont {Be'Er}, \citenamefont {Smith}, \citenamefont {Florin},\ and\
  \citenamefont {Swinney}}]{Zhang2009}%
  \BibitemOpen
  \bibfield  {author} {\bibinfo {author} {\bibfnamefont {H.}~\bibnamefont
  {Zhang}}, \bibinfo {author} {\bibfnamefont {A.}~\bibnamefont {Be'Er}},
  \bibinfo {author} {\bibfnamefont {R.~S.}\ \bibnamefont {Smith}}, \bibinfo
  {author} {\bibfnamefont {E.-L.}\ \bibnamefont {Florin}}, \ and\ \bibinfo
  {author} {\bibfnamefont {H.~L.}\ \bibnamefont {Swinney}},\ }\href@noop {}
  {\bibfield  {journal} {\bibinfo  {journal} {EPL}\ }\textbf {\bibinfo {volume}
  {87}},\ \bibinfo {pages} {48011} (\bibinfo {year} {2009})}\BibitemShut
  {NoStop}%
\bibitem [{\citenamefont {Lushi}\ \emph {et~al.}(2014)\citenamefont {Lushi},
  \citenamefont {Wioland},\ and\ \citenamefont {Goldstein}}]{Lushi14}%
  \BibitemOpen
  \bibfield  {author} {\bibinfo {author} {\bibfnamefont {E.}~\bibnamefont
  {Lushi}}, \bibinfo {author} {\bibfnamefont {H.}~\bibnamefont {Wioland}}, \
  and\ \bibinfo {author} {\bibfnamefont {R.~E.}\ \bibnamefont {Goldstein}},\
  }\href@noop {} {\bibfield  {journal} {\bibinfo  {journal} {Proc. Natl. Acad.
  Sci. U.S.A.}\ }\textbf {\bibinfo {volume} {111}},\ \bibinfo {pages} {9733}
  (\bibinfo {year} {2014})}\BibitemShut {NoStop}%
\bibitem [{\citenamefont {Heidenreich}\ \emph {et~al.}(2016)\citenamefont
  {Heidenreich}, \citenamefont {Dunkel}, \citenamefont {Klapp},\ and\
  \citenamefont {B{\"a}r}}]{heidenreich2016hydrodynamic}%
  \BibitemOpen
  \bibfield  {author} {\bibinfo {author} {\bibfnamefont {S.}~\bibnamefont
  {Heidenreich}}, \bibinfo {author} {\bibfnamefont {J.}~\bibnamefont {Dunkel}},
  \bibinfo {author} {\bibfnamefont {S.~H.~L.}\ \bibnamefont {Klapp}}, \ and\
  \bibinfo {author} {\bibfnamefont {M.}~\bibnamefont {B{\"a}r}},\ }\href@noop
  {} {\bibfield  {journal} {\bibinfo  {journal} {Phys. Rev. E}\ }\textbf
  {\bibinfo {volume} {94}},\ \bibinfo {pages} {020601} (\bibinfo {year}
  {2016})}\BibitemShut {NoStop}%
\bibitem [{\citenamefont {Drescher}\ \emph {et~al.}(2011)\citenamefont
  {Drescher}, \citenamefont {Dunkel}, \citenamefont {Cisneros}, \citenamefont
  {Ganguly},\ and\ \citenamefont {Goldstein}}]{Knut}%
  \BibitemOpen
  \bibfield  {author} {\bibinfo {author} {\bibfnamefont {K.}~\bibnamefont
  {Drescher}}, \bibinfo {author} {\bibfnamefont {J.}~\bibnamefont {Dunkel}},
  \bibinfo {author} {\bibfnamefont {L.~H.}\ \bibnamefont {Cisneros}}, \bibinfo
  {author} {\bibfnamefont {S.}~\bibnamefont {Ganguly}}, \ and\ \bibinfo
  {author} {\bibfnamefont {R.~E.}\ \bibnamefont {Goldstein}},\ }\href@noop {}
  {\bibfield  {journal} {\bibinfo  {journal} {Proc. Natl. Acad. Sci. U.S.A.}\
  }\textbf {\bibinfo {volume} {108}},\ \bibinfo {pages} {10940} (\bibinfo
  {year} {2011})}\BibitemShut {NoStop}%
\bibitem [{\citenamefont {Happel}\ and\ \citenamefont
  {Brenner}(2012)}]{Happel}%
  \BibitemOpen
  \bibfield  {author} {\bibinfo {author} {\bibfnamefont {J.}~\bibnamefont
  {Happel}}\ and\ \bibinfo {author} {\bibfnamefont {H.}~\bibnamefont
  {Brenner}},\ }\href@noop {} {\emph {\bibinfo {title} {Low Reynolds number
  hydrodynamics: with special applications to particulate media}}},\
  Vol.~\bibinfo {volume} {1}\ (\bibinfo  {publisher} {Springer Science \&
  Business Media},\ \bibinfo {year} {2012})\BibitemShut {NoStop}%
\bibitem [{\citenamefont {Ginelli}\ \emph {et~al.}(2010)\citenamefont
  {Ginelli}, \citenamefont {Peruani}, \citenamefont {B{\"a}r},\ and\
  \citenamefont {Chat{\'e}}}]{ginelli2010large}%
  \BibitemOpen
  \bibfield  {author} {\bibinfo {author} {\bibfnamefont {F.}~\bibnamefont
  {Ginelli}}, \bibinfo {author} {\bibfnamefont {F.}~\bibnamefont {Peruani}},
  \bibinfo {author} {\bibfnamefont {M.}~\bibnamefont {B{\"a}r}}, \ and\
  \bibinfo {author} {\bibfnamefont {H.}~\bibnamefont {Chat{\'e}}},\ }\href@noop
  {} {\bibfield  {journal} {\bibinfo  {journal} {Phys. Rev. Lett.}\ }\textbf
  {\bibinfo {volume} {104}},\ \bibinfo {pages} {184502} (\bibinfo {year}
  {2010})}\BibitemShut {NoStop}%
\bibitem [{\citenamefont {Yang}\ \emph {et~al.}(2010)\citenamefont {Yang},
  \citenamefont {Marceau},\ and\ \citenamefont {Gompper}}]{yang2010swarm}%
  \BibitemOpen
  \bibfield  {author} {\bibinfo {author} {\bibfnamefont {Y.}~\bibnamefont
  {Yang}}, \bibinfo {author} {\bibfnamefont {V.}~\bibnamefont {Marceau}}, \
  and\ \bibinfo {author} {\bibfnamefont {G.}~\bibnamefont {Gompper}},\
  }\href@noop {} {\bibfield  {journal} {\bibinfo  {journal} {Phys. Rev. E}\
  }\textbf {\bibinfo {volume} {82}},\ \bibinfo {pages} {031904} (\bibinfo
  {year} {2010})}\BibitemShut {NoStop}%
\bibitem [{\citenamefont {Wensink}\ and\ \citenamefont
  {L{\"o}wen}(2008)}]{wensink2008aggregation}%
  \BibitemOpen
  \bibfield  {author} {\bibinfo {author} {\bibfnamefont {H.~H.}\ \bibnamefont
  {Wensink}}\ and\ \bibinfo {author} {\bibfnamefont {H.}~\bibnamefont
  {L{\"o}wen}},\ }\href@noop {} {\bibfield  {journal} {\bibinfo  {journal}
  {Phys. Rev. E}\ }\textbf {\bibinfo {volume} {78}},\ \bibinfo {pages} {031409}
  (\bibinfo {year} {2008})}\BibitemShut {NoStop}%
\bibitem [{\citenamefont {Peruani}\ \emph {et~al.}(2006)\citenamefont
  {Peruani}, \citenamefont {Deutsch},\ and\ \citenamefont
  {B\"ar}}]{Peruani2006}%
  \BibitemOpen
  \bibfield  {author} {\bibinfo {author} {\bibfnamefont {F.}~\bibnamefont
  {Peruani}}, \bibinfo {author} {\bibfnamefont {A.}~\bibnamefont {Deutsch}}, \
  and\ \bibinfo {author} {\bibfnamefont {M.}~\bibnamefont {B\"ar}},\
  }\href@noop {} {\bibfield  {journal} {\bibinfo  {journal} {Phys. Rev. E}\
  }\textbf {\bibinfo {volume} {74}},\ \bibinfo {pages} {030904} (\bibinfo
  {year} {2006})}\BibitemShut {NoStop}%
\bibitem [{\citenamefont {Gro\ss{}mann}\ \emph {et~al.}(2016)\citenamefont
  {Gro\ss{}mann}, \citenamefont {Peruani},\ and\ \citenamefont
  {B\"ar}}]{Grossmann2016}%
  \BibitemOpen
  \bibfield  {author} {\bibinfo {author} {\bibfnamefont {R.}~\bibnamefont
  {Gro\ss{}mann}}, \bibinfo {author} {\bibfnamefont {F.}~\bibnamefont
  {Peruani}}, \ and\ \bibinfo {author} {\bibfnamefont {M.}~\bibnamefont
  {B\"ar}},\ }\href@noop {} {\bibfield  {journal} {\bibinfo  {journal} {Phys.
  Rev. E}\ }\textbf {\bibinfo {volume} {93}},\ \bibinfo {pages} {040102}
  (\bibinfo {year} {2016})}\BibitemShut {NoStop}%
\bibitem [{\citenamefont {Najafi}\ and\ \citenamefont
  {Golestanian}(2004)}]{najafi2004simple}%
  \BibitemOpen
  \bibfield  {author} {\bibinfo {author} {\bibfnamefont {A.}~\bibnamefont
  {Najafi}}\ and\ \bibinfo {author} {\bibfnamefont {R.}~\bibnamefont
  {Golestanian}},\ }\href@noop {} {\bibfield  {journal} {\bibinfo  {journal}
  {Phys. Rev. E}\ }\textbf {\bibinfo {volume} {69}},\ \bibinfo {pages} {062901}
  (\bibinfo {year} {2004})}\BibitemShut {NoStop}%
\bibitem [{\citenamefont {Pooley}\ \emph {et~al.}(2007)\citenamefont {Pooley},
  \citenamefont {Alexander},\ and\ \citenamefont
  {Yeomans}}]{pooley2007hydrodynamic}%
  \BibitemOpen
  \bibfield  {author} {\bibinfo {author} {\bibfnamefont {C.~M.}\ \bibnamefont
  {Pooley}}, \bibinfo {author} {\bibfnamefont {G.~P.}\ \bibnamefont
  {Alexander}}, \ and\ \bibinfo {author} {\bibfnamefont {J.~M.}\ \bibnamefont
  {Yeomans}},\ }\href@noop {} {\bibfield  {journal} {\bibinfo  {journal} {Phys.
  Rev. Lett.}\ }\textbf {\bibinfo {volume} {99}},\ \bibinfo {pages} {228103}
  (\bibinfo {year} {2007})}\BibitemShut {NoStop}%
\bibitem [{\citenamefont {Avron}\ \emph {et~al.}(2005)\citenamefont {Avron},
  \citenamefont {Kenneth},\ and\ \citenamefont
  {Oaknin}}]{avron2005pushmepullyou}%
  \BibitemOpen
  \bibfield  {author} {\bibinfo {author} {\bibfnamefont {J.}~\bibnamefont
  {Avron}}, \bibinfo {author} {\bibfnamefont {O.}~\bibnamefont {Kenneth}}, \
  and\ \bibinfo {author} {\bibfnamefont {D.}~\bibnamefont {Oaknin}},\
  }\href@noop {} {\bibfield  {journal} {\bibinfo  {journal} {New J. Phys.}\
  }\textbf {\bibinfo {volume} {7}},\ \bibinfo {pages} {234} (\bibinfo {year}
  {2005})}\BibitemShut {NoStop}%
\bibitem [{\citenamefont {Maier}\ and\ \citenamefont
  {Saupe}(1958)}]{maier1958einfache}%
  \BibitemOpen
  \bibfield  {author} {\bibinfo {author} {\bibfnamefont {W.}~\bibnamefont
  {Maier}}\ and\ \bibinfo {author} {\bibfnamefont {A.}~\bibnamefont {Saupe}},\
  }\href@noop {} {\bibfield  {journal} {\bibinfo  {journal} {Z. Naturforsch.
  A}\ }\textbf {\bibinfo {volume} {13}},\ \bibinfo {pages} {564} (\bibinfo
  {year} {1958})}\BibitemShut {NoStop}%
\bibitem [{\citenamefont {Maier}\ and\ \citenamefont
  {Saupe}(1959)}]{maier1959einfache}%
  \BibitemOpen
  \bibfield  {author} {\bibinfo {author} {\bibfnamefont {W.}~\bibnamefont
  {Maier}}\ and\ \bibinfo {author} {\bibfnamefont {A.}~\bibnamefont {Saupe}},\
  }\href@noop {} {\bibfield  {journal} {\bibinfo  {journal} {Z. Naturforsch.
  A}\ }\textbf {\bibinfo {volume} {14}},\ \bibinfo {pages} {882} (\bibinfo
  {year} {1959})}\BibitemShut {NoStop}%
\bibitem [{\citenamefont {Maier}\ and\ \citenamefont
  {Saupe}(1960)}]{maier1960einfache}%
  \BibitemOpen
  \bibfield  {author} {\bibinfo {author} {\bibfnamefont {W.}~\bibnamefont
  {Maier}}\ and\ \bibinfo {author} {\bibfnamefont {A.}~\bibnamefont {Saupe}},\
  }\href@noop {} {\bibfield  {journal} {\bibinfo  {journal} {Z. Naturforsch.
  A}\ }\textbf {\bibinfo {volume} {15}},\ \bibinfo {pages} {287} (\bibinfo
  {year} {1960})}\BibitemShut {NoStop}%
\bibitem [{\citenamefont {Jeffery}(1922)}]{Jeffery}%
  \BibitemOpen
  \bibfield  {author} {\bibinfo {author} {\bibfnamefont {G.~B.}\ \bibnamefont
  {Jeffery}},\ }\href@noop {} {\bibfield  {journal} {\bibinfo  {journal} {Proc.
  R. Soc. Lond. A}\ }\textbf {\bibinfo {volume} {102}},\ \bibinfo {pages} {161}
  (\bibinfo {year} {1922})}\BibitemShut {NoStop}%
\bibitem [{\citenamefont {Hinch}\ and\ \citenamefont {Leal}(1979)}]{HinchLeal}%
  \BibitemOpen
  \bibfield  {author} {\bibinfo {author} {\bibfnamefont {E.~J.}\ \bibnamefont
  {Hinch}}\ and\ \bibinfo {author} {\bibfnamefont {L.~G.}\ \bibnamefont
  {Leal}},\ }\href@noop {} {\bibfield  {journal} {\bibinfo  {journal} {J. Fluid
  Mech.}\ }\textbf {\bibinfo {volume} {92}},\ \bibinfo {pages} {591 } (\bibinfo
  {year} {1979})}\BibitemShut {NoStop}%
\bibitem [{\citenamefont {Pedley}\ and\ \citenamefont
  {Kessler}(1992)}]{PedleyKessler}%
  \BibitemOpen
  \bibfield  {author} {\bibinfo {author} {\bibfnamefont {T.}~\bibnamefont
  {Pedley}}\ and\ \bibinfo {author} {\bibfnamefont {J.~O.}\ \bibnamefont
  {Kessler}},\ }\href@noop {} {\bibfield  {journal} {\bibinfo  {journal} {Annu.
  Rev. Fluid Mech.}\ }\textbf {\bibinfo {volume} {24}},\ \bibinfo {pages} {313}
  (\bibinfo {year} {1992})}\BibitemShut {NoStop}%
\bibitem [{\citenamefont {Rafa{\"i}}\ \emph {et~al.}(2010)\citenamefont
  {Rafa{\"i}}, \citenamefont {Jibuti},\ and\ \citenamefont
  {Peyla}}]{Peyla2010}%
  \BibitemOpen
  \bibfield  {author} {\bibinfo {author} {\bibfnamefont {S.}~\bibnamefont
  {Rafa{\"i}}}, \bibinfo {author} {\bibfnamefont {L.}~\bibnamefont {Jibuti}}, \
  and\ \bibinfo {author} {\bibfnamefont {P.}~\bibnamefont {Peyla}},\
  }\href@noop {} {\bibfield  {journal} {\bibinfo  {journal} {Phys. Rev. Lett.}\
  }\textbf {\bibinfo {volume} {104}},\ \bibinfo {pages} {098102} (\bibinfo
  {year} {2010})}\BibitemShut {NoStop}%
\bibitem [{\citenamefont {Jibuti}\ \emph {et~al.}(2017)\citenamefont {Jibuti},
  \citenamefont {Zimmermann}, \citenamefont {Rafa{\"\i}},\ and\ \citenamefont
  {Peyla}}]{Jibuti2017}%
  \BibitemOpen
  \bibfield  {author} {\bibinfo {author} {\bibfnamefont {L.}~\bibnamefont
  {Jibuti}}, \bibinfo {author} {\bibfnamefont {W.}~\bibnamefont {Zimmermann}},
  \bibinfo {author} {\bibfnamefont {S.}~\bibnamefont {Rafa{\"\i}}}, \ and\
  \bibinfo {author} {\bibfnamefont {P.}~\bibnamefont {Peyla}},\ }\href@noop {}
  {\bibfield  {journal} {\bibinfo  {journal} {Phys. Rev. E}\ }\textbf {\bibinfo
  {volume} {96}},\ \bibinfo {pages} {052610} (\bibinfo {year}
  {2017})}\BibitemShut {NoStop}%
\bibitem [{\citenamefont {Hinch}(2010)}]{Hinch10}%
  \BibitemOpen
  \bibfield  {author} {\bibinfo {author} {\bibfnamefont {J.}~\bibnamefont
  {Hinch}},\ }\href@noop {} {\bibfield  {journal} {\bibinfo  {journal} {J.
  Fluid Mech.}\ }\textbf {\bibinfo {volume} {663}},\ \bibinfo {pages} {8}
  (\bibinfo {year} {2010})}\BibitemShut {NoStop}%
\bibitem [{\citenamefont {Einstein}(1906)}]{Einstein}%
  \BibitemOpen
  \bibfield  {author} {\bibinfo {author} {\bibfnamefont {A.}~\bibnamefont
  {Einstein}},\ }\href@noop {} {\bibfield  {journal} {\bibinfo  {journal} {Ann.
  Phys.}\ }\textbf {\bibinfo {volume} {324}},\ \bibinfo {pages} {289} (\bibinfo
  {year} {1906})}\BibitemShut {NoStop}%
\bibitem [{\citenamefont {Batchelor}(1974)}]{Batchelor}%
  \BibitemOpen
  \bibfield  {author} {\bibinfo {author} {\bibfnamefont {G.}~\bibnamefont
  {Batchelor}},\ }\href@noop {} {\bibfield  {journal} {\bibinfo  {journal}
  {Annu. Rev. Fluid Mech.}\ }\textbf {\bibinfo {volume} {6}},\ \bibinfo {pages}
  {227} (\bibinfo {year} {1974})}\BibitemShut {NoStop}%
\bibitem [{\citenamefont {Haines}\ \emph {et~al.}(2008)\citenamefont {Haines},
  \citenamefont {Aranson}, \citenamefont {Berlyand},\ and\ \citenamefont
  {Karpeev}}]{Heines08}%
  \BibitemOpen
  \bibfield  {author} {\bibinfo {author} {\bibfnamefont {B.~M.}\ \bibnamefont
  {Haines}}, \bibinfo {author} {\bibfnamefont {I.~S.}\ \bibnamefont {Aranson}},
  \bibinfo {author} {\bibfnamefont {L.}~\bibnamefont {Berlyand}}, \ and\
  \bibinfo {author} {\bibfnamefont {D.~A.}\ \bibnamefont {Karpeev}},\
  }\href@noop {} {\bibfield  {journal} {\bibinfo  {journal} {Phys. Biol.}\
  }\textbf {\bibinfo {volume} {5}},\ \bibinfo {pages} {046003} (\bibinfo {year}
  {2008})}\BibitemShut {NoStop}%
\bibitem [{\citenamefont {Krieger}\ and\ \citenamefont
  {Dougherty}(1959)}]{Krieger1959}%
  \BibitemOpen
  \bibfield  {author} {\bibinfo {author} {\bibfnamefont {I.~M.}\ \bibnamefont
  {Krieger}}\ and\ \bibinfo {author} {\bibfnamefont {T.~J.}\ \bibnamefont
  {Dougherty}},\ }\href@noop {} {\bibfield  {journal} {\bibinfo  {journal}
  {Trans. Soc. Rheol.}\ }\textbf {\bibinfo {volume} {3}},\ \bibinfo {pages}
  {137} (\bibinfo {year} {1959})}\BibitemShut {NoStop}%
\bibitem [{\citenamefont {Brinkmann}(1949)}]{Brinkmann49}%
  \BibitemOpen
  \bibfield  {author} {\bibinfo {author} {\bibfnamefont {H.~C.}\ \bibnamefont
  {Brinkmann}},\ }\href@noop {} {\bibfield  {journal} {\bibinfo  {journal}
  {Appl. Sci. Res.}\ }\textbf {\bibinfo {volume} {1}},\ \bibinfo {pages} {27}
  (\bibinfo {year} {1949})}\BibitemShut {NoStop}%
\bibitem [{\citenamefont {Hatwalne}\ \emph {et~al.}(2004)\citenamefont
  {Hatwalne}, \citenamefont {Ramaswamy}, \citenamefont {Rao},\ and\
  \citenamefont {Simha}}]{hatwalne2004rheology}%
  \BibitemOpen
  \bibfield  {author} {\bibinfo {author} {\bibfnamefont {Y.}~\bibnamefont
  {Hatwalne}}, \bibinfo {author} {\bibfnamefont {S.}~\bibnamefont {Ramaswamy}},
  \bibinfo {author} {\bibfnamefont {M.}~\bibnamefont {Rao}}, \ and\ \bibinfo
  {author} {\bibfnamefont {R.~A.}\ \bibnamefont {Simha}},\ }\href@noop {}
  {\bibfield  {journal} {\bibinfo  {journal} {Phys. Rev. Lett.}\ }\textbf
  {\bibinfo {volume} {92}},\ \bibinfo {pages} {118101} (\bibinfo {year}
  {2004})}\BibitemShut {NoStop}%
\bibitem [{\citenamefont {Baskaran}\ and\ \citenamefont
  {Marchetti}(2009)}]{Baskaran2009}%
  \BibitemOpen
  \bibfield  {author} {\bibinfo {author} {\bibfnamefont {A.}~\bibnamefont
  {Baskaran}}\ and\ \bibinfo {author} {\bibfnamefont {M.~C.}\ \bibnamefont
  {Marchetti}},\ }\href@noop {} {\bibfield  {journal} {\bibinfo  {journal}
  {Proc. Natl. Acad. Sci. U.S.A.}\ }\textbf {\bibinfo {volume} {106}},\
  \bibinfo {pages} {15567} (\bibinfo {year} {2009})}\BibitemShut {NoStop}%
\bibitem [{\citenamefont {Spagnolie}\ and\ \citenamefont
  {Lauga}(2012)}]{Spagnolie2012}%
  \BibitemOpen
  \bibfield  {author} {\bibinfo {author} {\bibfnamefont {S.~E.}\ \bibnamefont
  {Spagnolie}}\ and\ \bibinfo {author} {\bibfnamefont {E.}~\bibnamefont
  {Lauga}},\ }\href@noop {} {\bibfield  {journal} {\bibinfo  {journal} {J.
  Fluid Mech.}\ }\textbf {\bibinfo {volume} {700}},\ \bibinfo {pages} {105}
  (\bibinfo {year} {2012})}\BibitemShut {NoStop}%
\bibitem [{\citenamefont {Dhont}\ and\ \citenamefont
  {Briels}(2002)}]{Dhont2002}%
  \BibitemOpen
  \bibfield  {author} {\bibinfo {author} {\bibfnamefont {J.~K.~G.}\
  \bibnamefont {Dhont}}\ and\ \bibinfo {author} {\bibfnamefont {W.~J.}\
  \bibnamefont {Briels}},\ }\href@noop {} {\bibfield  {journal} {\bibinfo
  {journal} {J. Chem. Phys.}\ }\textbf {\bibinfo {volume} {117}},\ \bibinfo
  {pages} {3992} (\bibinfo {year} {2002})}\BibitemShut {NoStop}%
\bibitem [{\citenamefont {McCourt}\ \emph {et~al.}(1990)\citenamefont
  {McCourt}, \citenamefont {Beenakker}, \citenamefont {K{\"o}hler},\ and\
  \citenamefont {Ku\u{s}\u{c}er}}]{McCourt}%
  \BibitemOpen
  \bibfield  {author} {\bibinfo {author} {\bibfnamefont {F.~R.~W.}\
  \bibnamefont {McCourt}}, \bibinfo {author} {\bibfnamefont {J.~J.~M.}\
  \bibnamefont {Beenakker}}, \bibinfo {author} {\bibfnamefont {W.}~\bibnamefont
  {K{\"o}hler}}, \ and\ \bibinfo {author} {\bibfnamefont {I.}~\bibnamefont
  {Ku\u{s}\u{c}er}},\ }\href@noop {} {\emph {\bibinfo {title} {Nonequilibrium
  Phenomena in Polyatomic Gases}}}\ (\bibinfo  {publisher} {Clarendon Press},\
  \bibinfo {year} {Oxford 1990})\BibitemShut {NoStop}%
\bibitem [{\citenamefont {Turzi}(2011)}]{Turzi2011}%
  \BibitemOpen
  \bibfield  {author} {\bibinfo {author} {\bibfnamefont {S.}~\bibnamefont
  {Turzi}},\ }\href@noop {} {\bibfield  {journal} {\bibinfo  {journal} {J.
  Math. Phys.}\ }\textbf {\bibinfo {volume} {52}},\ \bibinfo {pages} {053517}
  (\bibinfo {year} {2011})}\BibitemShut {NoStop}%
\bibitem [{\citenamefont {Kr{\"o}ger}\ \emph {et~al.}(2008)\citenamefont
  {Kr{\"o}ger}, \citenamefont {Ammar},\ and\ \citenamefont
  {Chinesta}}]{Kroeger2008}%
  \BibitemOpen
  \bibfield  {author} {\bibinfo {author} {\bibfnamefont {M.}~\bibnamefont
  {Kr{\"o}ger}}, \bibinfo {author} {\bibfnamefont {A.}~\bibnamefont {Ammar}}, \
  and\ \bibinfo {author} {\bibfnamefont {F.}~\bibnamefont {Chinesta}},\
  }\href@noop {} {\bibfield  {journal} {\bibinfo  {journal} {J. Non-Newtonian
  Fluid Mech.}\ }\textbf {\bibinfo {volume} {149}},\ \bibinfo {pages} {40}
  (\bibinfo {year} {2008})}\BibitemShut {NoStop}%
\bibitem [{\citenamefont {Hess}(2015)}]{hess2015tensors}%
  \BibitemOpen
  \bibfield  {author} {\bibinfo {author} {\bibfnamefont {S.}~\bibnamefont
  {Hess}},\ }\href@noop {} {\emph {\bibinfo {title} {Tensors for physics}}}\
  (\bibinfo  {publisher} {Springer},\ \bibinfo {year} {2015})\BibitemShut
  {NoStop}%
\bibitem [{\citenamefont {Risken}(1984)}]{Risken}%
  \BibitemOpen
  \bibfield  {author} {\bibinfo {author} {\bibfnamefont {H.}~\bibnamefont
  {Risken}},\ }\href@noop {} {\emph {\bibinfo {title} {The Fokker-Planck
  Equation}}}\ (\bibinfo  {publisher} {Springer},\ \bibinfo {year}
  {1984})\BibitemShut {NoStop}%
\bibitem [{\citenamefont {Jacobs}(2010)}]{Jacobs}%
  \BibitemOpen
  \bibfield  {author} {\bibinfo {author} {\bibfnamefont {K.}~\bibnamefont
  {Jacobs}},\ }\href@noop {} {\emph {\bibinfo {title} {Stochastic processes for
  physicists: understanding noisy systems}}}\ (\bibinfo  {publisher} {Cambridge
  University Press},\ \bibinfo {year} {2010})\BibitemShut {NoStop}%
\bibitem [{\citenamefont {Ilg}\ \emph {et~al.}(1999)\citenamefont {Ilg},
  \citenamefont {Karlin},\ and\ \citenamefont
  {{\"O}ttinger}}]{ilg1999generating}%
  \BibitemOpen
  \bibfield  {author} {\bibinfo {author} {\bibfnamefont {P.}~\bibnamefont
  {Ilg}}, \bibinfo {author} {\bibfnamefont {I.~V.}\ \bibnamefont {Karlin}}, \
  and\ \bibinfo {author} {\bibfnamefont {H.~C.}\ \bibnamefont {{\"O}ttinger}},\
  }\href@noop {} {\bibfield  {journal} {\bibinfo  {journal} {Phys. Rev. E}\
  }\textbf {\bibinfo {volume} {60}},\ \bibinfo {pages} {5783} (\bibinfo {year}
  {1999})}\BibitemShut {NoStop}%
\bibitem [{\citenamefont {Masao}(1981)}]{masao1981molecular}%
  \BibitemOpen
  \bibfield  {author} {\bibinfo {author} {\bibfnamefont {D.}~\bibnamefont
  {Masao}},\ }\href@noop {} {\bibfield  {journal} {\bibinfo  {journal} {J.
  Polym. Sci. Part B Polym. Phys.}\ }\textbf {\bibinfo {volume} {19}},\
  \bibinfo {pages} {229} (\bibinfo {year} {1981})}\BibitemShut {NoStop}%
\bibitem [{\citenamefont {Hess}(1976)}]{hess1976fokker}%
  \BibitemOpen
  \bibfield  {author} {\bibinfo {author} {\bibfnamefont {S.}~\bibnamefont
  {Hess}},\ }\href@noop {} {\bibfield  {journal} {\bibinfo  {journal} {Z.
  Naturforsch. A}\ }\textbf {\bibinfo {volume} {31}},\ \bibinfo {pages} {1034}
  (\bibinfo {year} {1976})}\BibitemShut {NoStop}%
\bibitem [{\citenamefont {Hand}(1962)}]{hand1962theory}%
  \BibitemOpen
  \bibfield  {author} {\bibinfo {author} {\bibfnamefont {G.~L.}\ \bibnamefont
  {Hand}},\ }\href@noop {} {\bibfield  {journal} {\bibinfo  {journal} {J. Fluid
  Mech.}\ }\textbf {\bibinfo {volume} {13}},\ \bibinfo {pages} {33} (\bibinfo
  {year} {1962})}\BibitemShut {NoStop}%
\bibitem [{\citenamefont {Rin{\"a}cker}\ and\ \citenamefont
  {Hess}(1998)}]{RienaeckerHess98}%
  \BibitemOpen
  \bibfield  {author} {\bibinfo {author} {\bibfnamefont {G.}~\bibnamefont
  {Rin{\"a}cker}}\ and\ \bibinfo {author} {\bibfnamefont {S.}~\bibnamefont
  {Hess}},\ }\href@noop {} {\bibfield  {journal} {\bibinfo  {journal} {Physica
  A}\ }\textbf {\bibinfo {volume} {267}},\ \bibinfo {pages} {294} (\bibinfo
  {year} {1998})}\BibitemShut {NoStop}%
\bibitem [{\citenamefont {Doi}\ and\ \citenamefont {Edwards}(1988)}]{Doi}%
  \BibitemOpen
  \bibfield  {author} {\bibinfo {author} {\bibfnamefont {M.}~\bibnamefont
  {Doi}}\ and\ \bibinfo {author} {\bibfnamefont {S.~F.}\ \bibnamefont
  {Edwards}},\ }\href@noop {} {\emph {\bibinfo {title} {The theory of polymer
  dynamics}}},\ Vol.~\bibinfo {volume} {73}\ (\bibinfo  {publisher} {Oxford
  University Press},\ \bibinfo {year} {1988})\BibitemShut {NoStop}%
\bibitem [{\citenamefont {Heidenreich}(2009)}]{heidenreich2009orientational}%
  \BibitemOpen
  \bibfield  {author} {\bibinfo {author} {\bibfnamefont {S.}~\bibnamefont
  {Heidenreich}},\ }\emph {\bibinfo {title} {Orientational dynamics and flow
  properties of polar and non-polar hard-rod fluids}},\ \href@noop {} {Ph.D.
  thesis},\ \bibinfo  {school} {Technische Universit\"at Berlin} (\bibinfo
  {year} {2009})\BibitemShut {NoStop}%
\bibitem [{\citenamefont {Hess}(1975)}]{Hess75}%
  \BibitemOpen
  \bibfield  {author} {\bibinfo {author} {\bibfnamefont {S.}~\bibnamefont
  {Hess}},\ }\href@noop {} {\bibfield  {journal} {\bibinfo  {journal} {Z.
  Naturforsch. A}\ }\textbf {\bibinfo {volume} {30}},\ \bibinfo {pages} {1224}
  (\bibinfo {year} {1975})}\BibitemShut {NoStop}%
\bibitem [{\citenamefont {Pujolle-Robic}\ and\ \citenamefont
  {Noirez}(2001)}]{Noirez}%
  \BibitemOpen
  \bibfield  {author} {\bibinfo {author} {\bibfnamefont {C.}~\bibnamefont
  {Pujolle-Robic}}\ and\ \bibinfo {author} {\bibfnamefont {L.}~\bibnamefont
  {Noirez}},\ }\href@noop {} {\bibfield  {journal} {\bibinfo  {journal}
  {Nature}\ }\textbf {\bibinfo {volume} {409}},\ \bibinfo {pages} {167}
  (\bibinfo {year} {2001})}\BibitemShut {NoStop}%
\bibitem [{\citenamefont {Olmsted}\ and\ \citenamefont
  {Goldbart}(1990)}]{OlmstedGoldbart90}%
  \BibitemOpen
  \bibfield  {author} {\bibinfo {author} {\bibfnamefont {P.~D.}\ \bibnamefont
  {Olmsted}}\ and\ \bibinfo {author} {\bibfnamefont {P.}~\bibnamefont
  {Goldbart}},\ }\href@noop {} {\bibfield  {journal} {\bibinfo  {journal}
  {Phys. Rev. A}\ }\textbf {\bibinfo {volume} {41}},\ \bibinfo {pages}
  {4578(R)} (\bibinfo {year} {1990})}\BibitemShut {NoStop}%
\bibitem [{\citenamefont {Cappelaere}\ \emph {et~al.}(1997)\citenamefont
  {Cappelaere}, \citenamefont {Berret}, \citenamefont {Decruppe}, \citenamefont
  {Cressely},\ and\ \citenamefont {Lindner}}]{Lindner97}%
  \BibitemOpen
  \bibfield  {author} {\bibinfo {author} {\bibfnamefont {E.}~\bibnamefont
  {Cappelaere}}, \bibinfo {author} {\bibfnamefont {J.~F.}\ \bibnamefont
  {Berret}}, \bibinfo {author} {\bibfnamefont {J.~P.}\ \bibnamefont
  {Decruppe}}, \bibinfo {author} {\bibfnamefont {R.}~\bibnamefont {Cressely}},
  \ and\ \bibinfo {author} {\bibfnamefont {P.}~\bibnamefont {Lindner}},\
  }\href@noop {} {\bibfield  {journal} {\bibinfo  {journal} {Phys. Rev. E}\
  }\textbf {\bibinfo {volume} {56}},\ \bibinfo {pages} {1869} (\bibinfo {year}
  {1997})}\BibitemShut {NoStop}%
\bibitem [{\citenamefont {Fischer}\ and\ \citenamefont
  {Callaghan}(2001)}]{Callaghan2001}%
  \BibitemOpen
  \bibfield  {author} {\bibinfo {author} {\bibfnamefont {E.}~\bibnamefont
  {Fischer}}\ and\ \bibinfo {author} {\bibfnamefont {P.~T.}\ \bibnamefont
  {Callaghan}},\ }\href@noop {} {\bibfield  {journal} {\bibinfo  {journal}
  {Phys. Rev. E}\ }\textbf {\bibinfo {volume} {64}},\ \bibinfo {pages} {011501}
  (\bibinfo {year} {2001})}\BibitemShut {NoStop}%
\bibitem [{\citenamefont {Wolgemuth}(2008)}]{wolgemuth2008collective}%
  \BibitemOpen
  \bibfield  {author} {\bibinfo {author} {\bibfnamefont {C.~W.}\ \bibnamefont
  {Wolgemuth}},\ }\href@noop {} {\bibfield  {journal} {\bibinfo  {journal}
  {Biophys. J.}\ }\textbf {\bibinfo {volume} {95}},\ \bibinfo {pages} {1564}
  (\bibinfo {year} {2008})}\BibitemShut {NoStop}%
\bibitem [{\citenamefont {Sokolov}\ and\ \citenamefont
  {Aranson}(2009)}]{Sokolov09}%
  \BibitemOpen
  \bibfield  {author} {\bibinfo {author} {\bibfnamefont {A.}~\bibnamefont
  {Sokolov}}\ and\ \bibinfo {author} {\bibfnamefont {I.~S.}\ \bibnamefont
  {Aranson}},\ }\href@noop {} {\bibfield  {journal} {\bibinfo  {journal} {Phys.
  Rev. Lett.}\ }\textbf {\bibinfo {volume} {103}},\ \bibinfo {pages} {148101}
  (\bibinfo {year} {2009})}\BibitemShut {NoStop}%
\bibitem [{\citenamefont {Stenhammar}\ \emph {et~al.}(2017)\citenamefont
  {Stenhammar}, \citenamefont {Nardini}, \citenamefont {Nash}, \citenamefont
  {Marenduzzo},\ and\ \citenamefont {Morozov}}]{Stenhammer2017}%
  \BibitemOpen
  \bibfield  {author} {\bibinfo {author} {\bibfnamefont {J.}~\bibnamefont
  {Stenhammar}}, \bibinfo {author} {\bibfnamefont {C.}~\bibnamefont {Nardini}},
  \bibinfo {author} {\bibfnamefont {R.~W.}\ \bibnamefont {Nash}}, \bibinfo
  {author} {\bibfnamefont {D.}~\bibnamefont {Marenduzzo}}, \ and\ \bibinfo
  {author} {\bibfnamefont {A.}~\bibnamefont {Morozov}},\ }\href {\doibase
  10.1103/PhysRevLett.119.028005} {\bibfield  {journal} {\bibinfo  {journal}
  {Phys. Rev. Lett.}\ }\textbf {\bibinfo {volume} {119}},\ \bibinfo {pages}
  {028005} (\bibinfo {year} {2017})}\BibitemShut {NoStop}%
\bibitem [{\citenamefont {Hernandez-Ortiz}\ \emph {et~al.}(2009)\citenamefont
  {Hernandez-Ortiz}, \citenamefont {Underhill},\ and\ \citenamefont
  {Graham}}]{Hernandez2009}%
  \BibitemOpen
  \bibfield  {author} {\bibinfo {author} {\bibfnamefont {J.~P.}\ \bibnamefont
  {Hernandez-Ortiz}}, \bibinfo {author} {\bibfnamefont {P.~T.}\ \bibnamefont
  {Underhill}}, \ and\ \bibinfo {author} {\bibfnamefont {M.~D.}\ \bibnamefont
  {Graham}},\ }\href@noop {} {\bibfield  {journal} {\bibinfo  {journal} {J.
  Phys. Condens. Matter}\ }\textbf {\bibinfo {volume} {21}},\ \bibinfo {pages}
  {204107} (\bibinfo {year} {2009})}\BibitemShut {NoStop}%
\bibitem [{\citenamefont {Ilkanaiv}\ \emph {et~al.}(2017)\citenamefont
  {Ilkanaiv}, \citenamefont {Kearns}, \citenamefont {Ariel},\ and\
  \citenamefont {Be'er}}]{ilkanaiv2017effect}%
  \BibitemOpen
  \bibfield  {author} {\bibinfo {author} {\bibfnamefont {B.}~\bibnamefont
  {Ilkanaiv}}, \bibinfo {author} {\bibfnamefont {D.~B.}\ \bibnamefont
  {Kearns}}, \bibinfo {author} {\bibfnamefont {G.}~\bibnamefont {Ariel}}, \
  and\ \bibinfo {author} {\bibfnamefont {A.}~\bibnamefont {Be'er}},\
  }\href@noop {} {\bibfield  {journal} {\bibinfo  {journal} {Phys. Rev. Lett.}\
  }\textbf {\bibinfo {volume} {118}},\ \bibinfo {pages} {158002} (\bibinfo
  {year} {2017})}\BibitemShut {NoStop}%
\end{thebibliography}%

\end{document}